\documentclass[useAMS,usenatbib]{mn2e}
\usepackage{times}
\usepackage{xcolor}
\usepackage{graphicx}
\usepackage{amssymb}
\usepackage{epstopdf}
\usepackage{subcaption}
\usepackage{placeins}
\usepackage{float}
\newcommand{\mytilde}{\raise.17ex\hbox{$\scriptstyle\mathtt{\sim}$}}

\def\dlos{\langle d_{\rm LOS}\rangle}

\def\lya{Ly$\alpha$ }
\def\hmpc{h^{-1}{\rm Mpc}}

\title[Large--scale 3D mapping of the intergalactic medium using the Lyman $\alpha$ forest]{Large--scale 3D mapping of the intergalactic medium using the Lyman $\alpha$ forest}
\author [M. Ozbek, R. A. C. Croft and N. Khandai]{Melih Ozbek${^1}$, Rupert A. C. Croft$^{1}$ and Nishikanta Khandai$^{2}$\\
${}^1$ McWilliams Center for Cosmology, Department of Physics, Carnegie Mellon University, 5000 Forbes Avenue, Pittsburgh, PA 15213, USA\\
${}^2$ School of Physical Sciences, National Institute of Science Education and Research, Bhubaneswar 751005, India \\
}

\begin{document}

\maketitle

\begin{abstract}

 Maps of the large-scale structure of the Universe at redshifts 2--4 can be made with the Lyman-$\alpha$ forest which are complementary to low redshift galaxy surveys. We apply the Wiener interpolation method of \citeauthor{caucci2008} to construct three-dimensional maps from sets of Lyman-$\alpha$ forest spectra taken from cosmological hydrodynamic simulations. We mimic some current and future quasar redshift surveys (BOSS, eBOSS and MS--DESI) by choosing similar sightline densities. We use these appropriate subsets of the Lyman-$\alpha$ absorption sightlines to reconstruct the full three dimensional Lyman-$\alpha$ flux field and perform comparisons between the true and the reconstructed fields. We study global statistical properties of the intergalactic medium (IGM) maps with auto--correlation and cross--correlation analysis, slice plots, local peaks and point by point scatter. We find that both the density field and the statistical properties of the IGM are recovered well enough that the resulting IGM maps can be meaningfully considered to represent 
large-scale maps of the Universe in agreement with  \citeauthor{caucci2008}, on larger scales and for sparser sightlines than had been tested previously. Quantitatively, for sightline parameters comparable to current and near future surveys the correlation coefficient between true and 
reconstructed fields is $ r > 0.9 $ on scales \textgreater $\, 30 \, \hmpc$. 
The properties of the maps are relatively insensitive
to the precise form of the covariance matrix used. The final BOSS quasar Lyman-$\alpha$ forest sample will 
allow maps to be made with a resolution of $\sim30 \, \hmpc$ over a volume of $\sim15 \, h^{-3}{\rm Gpc^3}$ between redshifts 1.9 and 2.3.

\end{abstract}



\section{INTRODUCTION}
 The structure of the Intergalactic Medium (IGM) can be studied using the  Lyman-$\alpha$ forest, the absorption features due to neutral hydrogen
seen in quasar spectra \citep{lynds1971}. 
Besides quasars, other background sources can be used to probe
absorption along the line of sight. These include Lyman-break galaxies (LBGs) at redshifts higher than 2
\citep{steidel1992deep, steidel1998lyman, steidel2001lyman, adelberger2005spatial, lee2014a}, although their much fainter magnitudes makes this
significantly more difficult. Gamma Ray Bursts (GRB) can also be
used \citep{totani2013}. At $z=3$, each individual spectrum provides one dimensional 
information for a $\sim$ 400 $\hmpc$ skewer in length
if spectral coverage is obtained for the  Lyman-$\alpha$ --
Lyman-$\beta$ region.
This 1D information can be used to constrain the matter fluctuation amplitude \citep{weinberg1999closing, kim2004power} and the temperature-density relation \citep{rollinde2001, lee2015igm}.
 Transverse information can be obtained with pairs of quasars \citep{dinshaw1994common, dinshaw1995large, fang1995size, petry1998small, hennawi2007}. 
Using this transverse information, the three dimensional correlation 
function of the Lyman-$\alpha$ forest on large scales (up to 100 $\hmpc$)
 was  measured for the first time, using data from 
the Baryon Oscillation Spectroscopic Survey (BOSS) Data Release 9 
(DR9) \citep{slosar2011}. BOSS has also enabled the first measurement
of Baryon Acoustic Oscillations using the \lya forest on 
even larger scales \citep{boss2616ng, slosar2013measurement, delubac2014baryon, aubourg2014cosmological}. Such BAO measurements at z $\sim$ 2 with the Lyman-$\alpha$ forest are currently the only way to put constraints 
on dark energy at these redshifts.

If the line of sight density is 
sufficiently high, inversion methods can be utilized in order to
directly make
three-dimensional maps of the intergalactic medium from collections
of 1D spectra \citep{pichon2001}. \citet{caucci2008} and \citet{lee2014a, lee2014b} have shown that the topological and statistical 
properties of the IGM can be reconstructed accurately 
even on small scales that correspond to mean line of sight separations of 
the order of 1 arcmin. In this paper, we explore how well maps can be made on much larger scales, corresponding to mean quasar angular
separations of tens of arc minutes. Our work, focusing on large-scale structure, is therefore complementary to the studies where individual structures like protoclusters and voids in the Lyman-$\alpha$ forest have been identified \citep{stark2014, stark2015}. In a companion paper \citep{ozbek2015} we present
maps made from the BOSS final data release \citep{alam2015eleventh}, which contains such a
sample of quasars covering a wide area.

 The Lyman-$\alpha$ forest provides a means of studying the large 
scale structure in the redshift range  $ 2 \lesssim z \lesssim 3.5 $. 
Before recent large-scale structure surveys
such as BOSS \citep{dawson2013baryon}, the 
sky density of known 
background quasars in this redshift range over most of the sky
was of the order of 1 per square degree (e.g., from the 2dF quasar survey \citep{miller2002possible, outram20032df, croom20042df,   miller2004200},
and from SDSS I and II \citep{schneider2002sloan, schneider2003sloan, schneider2005sloan, richards2006sloan}. Except for some small areas with higher
observed densities of objects (e.g., \citealt{rollinde2003correlation}) the Lyman-$\alpha$ forest was treated as a collection of discrete 1D individual quasar sightlines.

Recently, however, the increasing number of discovered quasars 
with suitable redshifts ($z>2$) for ground based study 
has made it possible to correlate information over large scales in three dimensions. BOSS features a high QSO density of $\sim15$ $\deg^{-2}$ ($\sim180,000$ QSOs in the redshift range $2.15 < z < 4$ over 10,000 $\deg^{2}$). Each QSO provides Lyman-$\alpha$ forest information along a skewer of length $\sim$ 400 $\hmpc$, and the typical mean separation for spectra in BOSS is $\sim$ 20 comoving  $\hmpc$ \citep{lee2013boss}. This is what has enabled clustering statistics of the Lyman-$\alpha$ forest
to be measured in three dimensions as mentioned above. 
In the future, one can expect yet higher densities of sightlines and more
precise measurements, in view of the fact that even more quasars will be available for analysis (e.g., 45 $\deg^{-2}$ proposed for MS--DESI \citep{DESI2013}, see Table \ref{Table:ThreeExps}). In this paper, we use the quantity $n_{\rm LOS}$ for the areal quasar density in observational surveys, as given in Table \ref{Table:ThreeExps}, first row. For the number of sightlines chosen at random to carry out the interpolation in the simulation cube, we define a new quantity $N_{\rm LOS}$, whose appropriate values are chosen according to $n_{\rm LOS}$ values from observational surveys, as shown with the red markers in Fig. \ref{Fig:LOSDensityforExps}. 
 
 
 
At the redshifts relevant to the work we present here ($z=$ 2--4), 
most of the volume of the IGM is in photoionization equilibrium. The thermal state of the IGM is determined by photoionization heating and adiabatic cooling 
\citep{hui1997equation}.
Lya forest traces weakly overdense regions at these redshifts \citep{bi1993lyman}, and so we can use it to study the large scale structure in this diffuse medium.
The IGM density fluctuations
follow those of the  total matter potential,  tracing dark matter at large scales ($\ge 0.1$ $\hmpc$) and respond to pressure smoothing on small 
scales ($< 100$ $h^{-1} {\rm kpc}$) \citep{sargent1980distribution, ikeuchi1986baryon, rees1986lyman}.
 
 The Lyman-$\alpha$ forest is useful for studying  large 
scale structure at high redshift ($2 < z <4$), in a fashion complementary
 to galaxy surveys at low redshift. The redshift range accessible for 
surveying large samples of
the latter is limited by the surface brightness of galaxies scaling 
as $(1+z)^{-4}$. The space density
of tracer objects in current large-scale maps of the Universe 
above redshift $z=0.5$ declines rapidly (e.g., in the CMASS sample of BOSS DR9 \citep{BOSS_CMASS_DR9},
the comoving space density is $\sim 3.6 \times 10^{-4}$ $h^3 Mpc^{-3}$ at $ \sim z = 0.5$ but falls off rapidly with increasing redshift). 
Here we define large-scale maps as those which cover an appreciable fraction
of the sky area.
At redshifts $z>2$, quasars are the only tracers that can be 
used to make such maps, and the space density
of objects declines to the order of $10^{-6}$ $h^3 Mpc^{-3}$ \citep{dawson2013baryon}. With a mean 3D separation of $ \sim$ 100 $\hmpc$ or larger, the 
shot noise in maps made using quasars is extremely high. Using absorption
lines, for example metal lines, where multiple lines are available for each quasar offers a way to increase the space density of tracers \citep{vikas2013moderate, zhu2013jhu}. The Lyman-$\alpha$ forest, being a continuous
field offers a way of reducing the level of noise in  maps even
further.
  
When one has saturation in the Lyman-$\alpha$ forest, the Lyman-$\beta$ transition should provide  information due to its lower cross section, which makes it a potentially better probe at high overdensities \citep{shull2000} as high as 10 times the mean density at $z=2-3$.
Since Lyman-$\beta$ absorption occurs at a lower rest wavelength (1026 \AA), the Lyman-$\alpha$ forest overlaps the Lyman-$\beta$ 
forest, and so statistical techniques are needed to make use
of the Lyman-$\beta$ information \citep{dijkstra2003, irvsivc2013detection}.
In principle higher order transitions could be used  together 
with the Lyman-$\alpha$ forest in mapmaking also.

 
The structure of this paper is as follows: We first describe the properties of the cosmological hydrodynamic simulation we 
use and our choices of simulated data samples at redshifts 2 and 3 with varying line of sight densities and noise levels in accordance with current and future sky surveys in \S 2. 
In \S 3, we review Wiener filtering and how it is used to carry out the recovery the flux field of the IGM. We look at point to point comparisons of
true and reconstructed fields in \S 4.1, and then provide correlation plots to test the validity of the reconstruction as a function of scale, followed by non-Gaussianity analysis via Kolmogorov--Smirnov tests in the following subsections. In \S 4.4, we look for local peaks in the true field and the reconstructed field and examine slice images to make a visual comparison between the true and reconstructed fields. Finally, we test the use of an alternative input 
correlation matrix in the Wiener filtering.
We summarize our results in \S 5 and also provide closing comments.

\section{SIMULATION}

In order to evaluate the expected performance of
map-making reconstruction on Lyman-$\alpha$ forest data from BOSS and other
observational surveys, we make use of a large hydrodynamic cosmological 
simulation of the $\Lambda$CDM model. We use the smoothed particle 
hydrodynamics code P--GADGET (see \citealt{springel2005cosmological, di2012cold}) to evolve a distribution of $2 \times 4096^2 = 137$ billion particles in a cubical
periodic volume of side length 400 $\hmpc$. 
The simulation cosmological parameters were $h = 0.702,\, \Omega_{\Lambda} = 0.725,\, \Omega_m = 0.275, \, \Omega_b = 0.046, \, n_s = 0.968\; $and$ \; \sigma_8 = 0.82.$ The mass per particle was $1.19 \times 10^7$ $h^{-1} M_{\odot}$ (gas) and $5.92 \times 10^7$ $h^{-1} M_{\odot}$ (dark matter). 
A gravitational force resolution of 3.25 kpc/h comoving was used. 
The power spectrum of the simulation initial conditions was taken from CAMB 
\citep{lewis2000efficient}.
The simulation was run with 
an ultraviolet background radiation field consistent with 
\citet{haardt1995radiative}. Cooling and star formation were included. However the latter used a 
lower density threshold than usual (for example in \citealt{springel2002cosmological}) so that gas particles are rapidly converted to collisionless gas particles. 
This was done to speed up execution of the simulation. As a result the stellar properties of galaxies in
the simulation are not predicted reliably but this has no significant effect on the diffuse IGM that gives rise to the Lyman-$\alpha$ forest. Black hole 
formation and feedback from stars were also switched off in the simulation.

\subsection{Data}

\begin{table}
\begin{center}
\begin{tabular}{cccc}
{\bf } & {\bf BOSS} & {\bf eBOSS} & {\bf MS--DESI} \\
\hline
\textbf{$n_{\rm LOS} / \deg^{2}$}            &   16	               &  25               &  45 \\
Sky coverage ($\deg^{2}$)       &  10400        & 7500              &  14000 \\
\lya QSOs (thousands)     &  180      	          & 250               &  1000  \\
Spectral Resolution                &  $\sim$ 2000  & $\sim$ 2000       & $\sim$ 3500 \\
\hline
\end{tabular} \caption{LOS density, sky coverage, targeted Lyman-$\alpha$ QSOs and spectral resolution parameters for three sky surveys.}\label{Table:ThreeExps}
\end{center}
\end{table}

\begin{figure}
\centering
\includegraphics[scale=0.26]{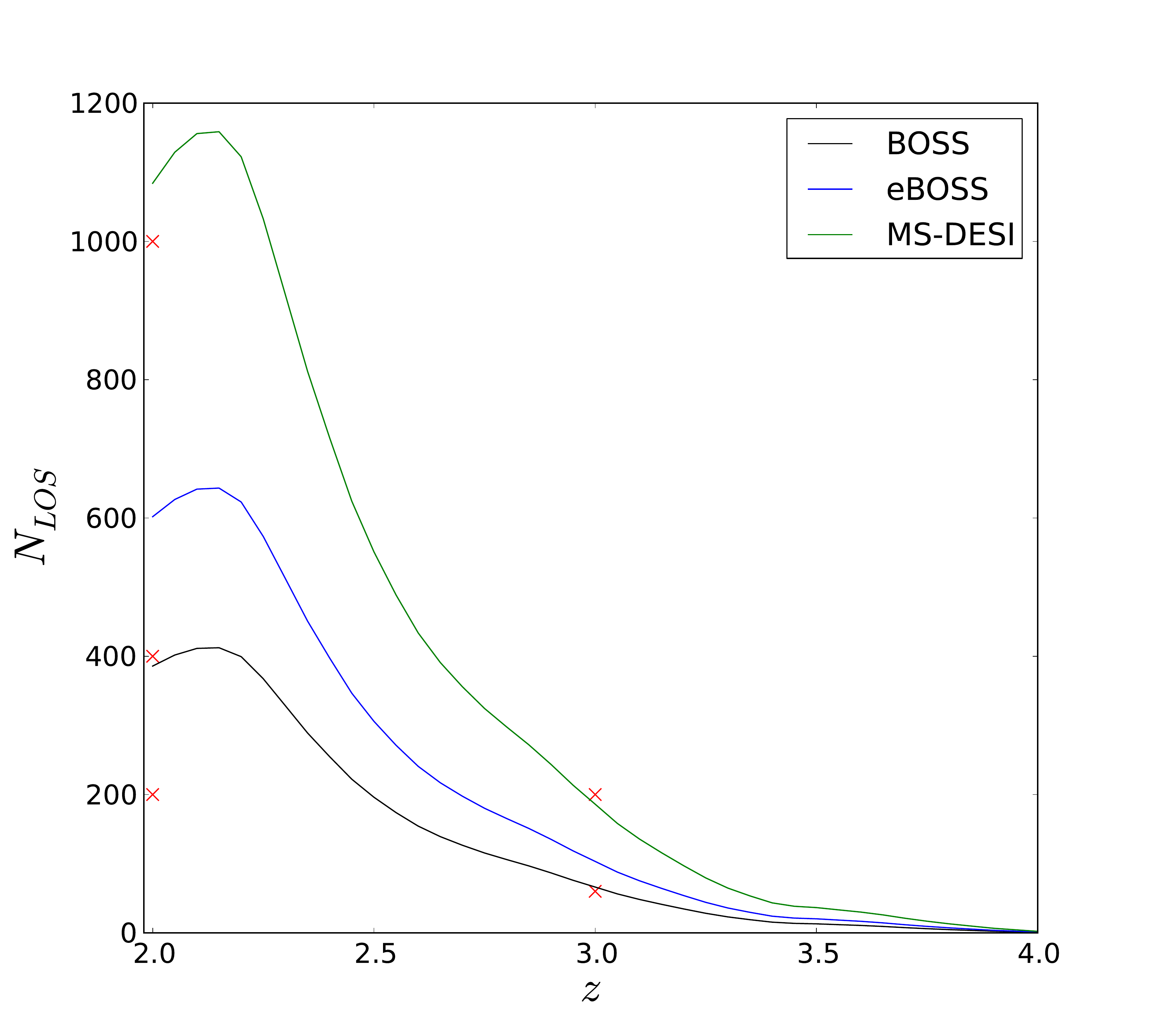}
\caption{$N_{\rm LOS}$ passing through our simulation volume at different redshifts according to BOSS, eBOSS and MS--DESI (assuming the other two experiments have the same distribution of QSOs with respect to redshift as BOSS). The red markers show the fiducial choices for our work with simulation data. The BOSS quasar catalog indicates that the number of quasars peaks at z $\sim$ 2.25 and decreases rapidly at higher redshifts.}
\label{Fig:LOSDensityforExps}
\end{figure} 

We use two simulation snapshots at redshifts of $z = 2$ and $z = 3$
to generate two sets of Lyman-$\alpha$ spectra using information
from the particle distribution \citep{hernquist1996lyman}.
We make spectra set out on a grid with $176^2 = 30,976$ 
evenly spaced sightlines, resulting in 2.27 $\hmpc$ spacing.
This can be compared in the line of sight direction
 with BOSS pixels of width $\Delta{v} = 69.02$ km $\rm{s^{-1}}$ 
\citep{lee2013boss}, which is $\sim$ 0.6 $\hmpc$ at $z=3$.
Each simulation sightline was generated with high resolution, 10,560 pixels, 
in order to resolve the thermal broadening when computing the optical depth. 
The spectra were then downsampled (by averaging the transmitted flux 
over 60 pixels) to 176 pixels. The full set of simulation 
data sets therefore consist of $176^3$ data values each.

In order to roughly approximate the noise which will be expected in 
observational data, we add random uncorrelated Gaussian pixel noise 
to the data sets with a signal to noise ratio S/N = 1 or 2 per unit simulation pixel in 176 pixels per 400 $\hmpc$ sightline.
This is similar to BOSS which has S/N of order unity 
\citep{lee2015igm}. If BOSS Lyman-$\alpha$ data were binned to 9.3 $\hmpc$ pixels, the mean S/N ratio would be 2.5. Similarly,
the simulation data with S/N = 1, when binned to pixels of the same size results in an S/N ratio of 4. Therefore, we use the S/N = 1 case as a close match for the BOSS noise level, whereas the other noise level, S/N = 2, is given as an example with less noise.
A random subset of the $176^2$ sightlines was chosen, according to LOS area
densities from the experiments BOSS \citep{alam2015eleventh}, eBOSS \citep{raichoor2015sdss} and MS--DESI \citep{DESI2013}
(e.g., $N_{\rm LOS}$ $\sim$ 400 LOS passing through the simulation volume
for BOSS) from our simulation box to carry 
out the reconstruction (see Table \ref{Table:ThreeExps} and Fig. \ref{Fig:LOSDensityforExps}). We define $N_{\rm LOS}$ as the number of lines of sight chosen to reconstruct the entire volume in the simulation box, whereas $n_{\rm LOS}$ (see, e.g., Table \ref{Table:ThreeExps}) denotes the total number of sightlines along the entire redshift range from the observer to quasars.

Our data points derived from the simulation are optical depths, ($\tau = -\log_{e} F / F_0$), where $F$ is the flux received at a certain location in 
space and $F_0$ is the unabsorbed flux. The data are convolved with the peculiar velocity, therefore we work in redshift space. In \citep{caucci2008} the authors worked with the density field directly. We will however work with the flux, and our maps will be reconstructions of  the three dimensional
flux field, in redshift space.
The relation between the 
gas density and the optical depth is

\begin{equation}
\delta(x) = \frac{1}{\alpha} \log(\frac{\tau(x)}{A(\bar{z})})
\end{equation}

\noindent
where $\delta(x) \approx \frac{\rho - \bar{\rho}} {\bar{\rho}}$ is the density contrast, and $\alpha$ and $A(\bar{z})$ are redshift dependent factors.
 We present our results in terms of flux contrast,
 $\delta_{F} = (F/\left<F\right>) -1$, where $\left<F\right>$ is the mean
transmitted flux computed from all spectra.

\subsection{Simulated datasets}

We have made 12 simulated data sets with different sightline densities and
noise levels. These are summarised in 
 Table \ref{Table:Simulated Data Sets}. Sightline density choices were made to mimic those of current or future observational surveys, as shown in Fig. \ref{Fig:LOSDensityforExps} with red markers. Some data sets have a LOS density that is even lower than that of BOSS (e.g. the data set labelled "z2\textunderscore N200") but still allow an accurate recovery of the flux field, as we will see in \S 4. Data sets with higher sightline densities (e.g. "z2\textunderscore N1000", which is comparable to that of MS-DESI) result in an even better inference of the field. Therefore, as observational surveys find more quasars, even more accurate density maps will be available with the \lya forest. 

\begin{table}
\begin{center}
\begin{tabular}{ccccc}
{\bf Sample} & {\bf Redshift} & {$\mathbf{N_{\rm LOS}}$} & {\bf Noise} &  $\mathbf{\dlos}$($\hmpc$) \\
\hline
z2\textunderscore N200 & 2 & 200 & Noiseless & 28.28 \\
z2\textunderscore N200\textunderscore SN2 & 2 & 200 & S/N=2 & 28.28  \\
z2\textunderscore N200\textunderscore SN1 & 2 & 200 & S/N=1 & 28.28 \\
z2\textunderscore N400 & 2 & 400 & Noiseless & 20.00 \\
z2\textunderscore N400\textunderscore SN2 & 2 & 400 & S/N=2 & 20.00  \\
z2\textunderscore N400\textunderscore SN1 & 2 & 400 & S/N=1 & 20.00 \\
z2\textunderscore N1000 & 2 & 1000 & Noiseless & 12.65 \\
z2\textunderscore N1000\textunderscore SN2 & 2 & 1000 & S/N=2 & 12.65  \\
z2\textunderscore N1000\textunderscore SN1 & 2 & 1000 & S/N=1 & 12.65 \\
z3\textunderscore N60 & 3 & 60 & Noiseless & 51.64\\
z3\textunderscore N200 & 3 & 200 & Noiseless & 28.28 \\
z3\textunderscore N200\textunderscore SN2 & 3 & 200 & S/N=2 & 28.28 \\
\hline
\end{tabular} \caption{Our choices of simulated data sets at redshifts 2 and 3 with different LOS density and noise levels.} \label{Table:Simulated Data Sets}
\end{center}
\end{table}

\section{RECONSTRUCTION}  

There are several methods which can be used to interpolate between the sparse
absorption skewers in the Lyman-$\alpha$ forest. For example, recent work by \citet{cisewski2014} used  local polynomial smoothing for this purpose. 
The method we choose in this paper is Wiener filtering, pioneered
in this context by \citet{pichon2001}, and used by \citet{caucci2008},
and \citet{lee2014b} to make maps from simulated data, and by 
\citet{lee2014a} 
to make the first 3 dimensional maps from observations.

\mathversion{bold}

We consider the values of the flux contrast in the reconstructed field to 
be entries in a column vector $\mathsf{M}$, and the values of the flux contrast
in the absorption skewer data to be entries in a column vector $\mathsf{D}$.
In general the entries of $\mathsf{M}$ will represent values on a uniform grid
of voxels as we are constructing a map which covers all space within the map
boundary. In our simulation tests, they will be covering
the cubical simulation volume uniformly. We choose not to make use of the
simulation periodic boundary conditions, in order to mimic some aspects
of real data.  
Using Wiener filtering, the reconstructed 3D field $\mathsf{M}$ can be inferred from the absorption skewer data $\mathsf{D}$ by computing

\begin{equation}
    {\mathsf{M}}={\mathsf{ C_{\rm MD}}} \cdot ({\mathsf{C_{\rm DD}}} + {\mathsf{{N)^{-1}}}}
    \cdot {\mathsf D}\,$,$
\end{equation}

\noindent
where ${\mathsf{C_{\rm MD}}}$ and ${\mathsf{C_{\rm DD}}}$ are the map--data and data--data covariance matrices and ${\mathsf{N}}$ is the diagonal noise matrix.
In the present work, we assume the  noise to be uncorrelated, so that the
entries of ${\mathsf{N}}$ are inversely proportional to the square root of the 
number of pixels in each cell.
The covariance matrices encode the expected correlation
structure of the field. In most of our work we use the 
following simple form
advocated by \citet{pichon2001} and \citet{caucci2008},

\mathversion{normal}

\begin{eqnarray}
    \textbf{\textsf{C}}(x_1,x_2,{\bf x_{1\perp}},{\bf x_{2\perp}}) =
    \sigma^2\times\exp\Big(-\frac{(x_1-x_2)^2}{L_{||}^2}\Big)\times
    \nonumber\\ 
    \exp\Big(-\frac{|{\bf x_{1\perp}}-
      {\bf x_{2\perp}}|^2}{L_{\perp}^2} \Big)\,,  
    \label{Eq:correl_matrix}
\end{eqnarray}

\noindent
where  $(x_1-x_2)$ and $|{\bf x_{1\perp}}-{\bf x_{2\perp}}|$ represent the 
distances between two pixels, parallel and perpendicular to the LOSs
respectively, $L_{||}$ and $L_{\perp}$ are correlation lengths parallel 
and perpendicular to the LOSs, while the variance $\sigma^2$ is calculated 
directly from the field.\mathversion{bold} The ${\mathsf{C_{\rm DD}}}$ covariance matrix contains 
correlation information between the initial data points only 
(the ${\mathsf{D}}$ array), whereas ${\mathsf{C_{\rm MD}}}$ contains information of 
the 3D pixel locations of the map to be inferred and the ${\mathsf{D}}$ array.

\mathversion{normal}

 In order to test how well the reconstruction works as a function of
line of sight density, we make several different
data samples by choosing a subset of our lines of sight at 
random. The areal density of the sightlines are 
those that correspond to 
some current and planned experiments, e.g. BOSS \citep{dawson2013baryon} and DESI 
\citep{levi2013desi} 
(see Table \ref{Table:ThreeExps} and Fig. \ref{Fig:LOSDensityforExps}). 

Our numerical code to carry out the reconstruction
splits the simulation volume into ``subcubes''.
The interpolation is then carried out separately for each 
subcube in parallel with the others and in the
final step the results are combined to form the whole 
reconstructed simulation cube. In order to make the calculations more 
computationally tractable, we decrease the resolution of the field from 
$176^3$ to $44^3$ pixels.

We introduce a buffer volume 
on the edges of the subcubes, allowing them to overlap, in order to
avoid edge 
artefacts. In our fiducial
reconstruction of the simulation 
we use 64 subcubes overall and a buffer of 40 $\hmpc$ on each side 
for each subcube. Each subcube therefore has a side length of 
180 $\hmpc$, including the buffer regions. 
We have tested and checked that 
adjusting the number of subcubes or changing the number of pixels does not 
significantly alter the results. 

The code used does not take into account 
the periodic boundary conditions of the whole simulation box, in order
to approximate the situation which will occur for observational data. 
This means that the reconstruction will be less accurate
near the edges of the cubical simulation volume.
For this reason, when choosing slice images to compare
real and reconstructed fields, we choose the middle planes of the cube
rather than the edges. 
We find by visual inspection that there is no significant difference
in the quality of reconstructions when increasing the separation
from the box edge by greater than 50 $\hmpc$.
We repeat some 
statistical calculations after truncating the cube by 50 $\hmpc$ from each 
edge in order to test the importance of edge artefacts.

 The resolution of the maps is determined by the mean 
separation between quasar lines of sight:

\begin{equation}
\dlos = {\frac{L_{Box}}{\sqrt{N_{\rm LOS}}}} 
\label{Eq:dlos}
\end{equation} 

\noindent 
For $N_{\rm LOS} = 200$, $\dlos$ is equal to $28.28$ $\hmpc$.
For a study of the map resolution as a function of exposure time, 
the reader is referred to \citealt{lee2014b}.
\citealt{pichon2001}
have shown that typical values for the correlation lengths $L_{||}$ and 
$L_{\perp}$ should be of the order of $\dlos$ in order to avoid numerical
instabilities in the matrix inversion leading to
spurious structures. 
We smooth both true and reconstructed fields with an isotropic 
Gaussian filter with a standard deviation $\sigma_{S} = 1.4 \dlos$, in the 
latter case after carrying out the reconstruction.

\section{Analysis}

After the reconstruction of the field
we analyse the relationship between the true and reconstructed fields,
bearing in mind that both have been smoothed, as stated above.

\subsection{Scatter Plots}

\begin{figure*}
\centering
\begin{subfigure}{0.32\textwidth}
\includegraphics[scale=0.18]{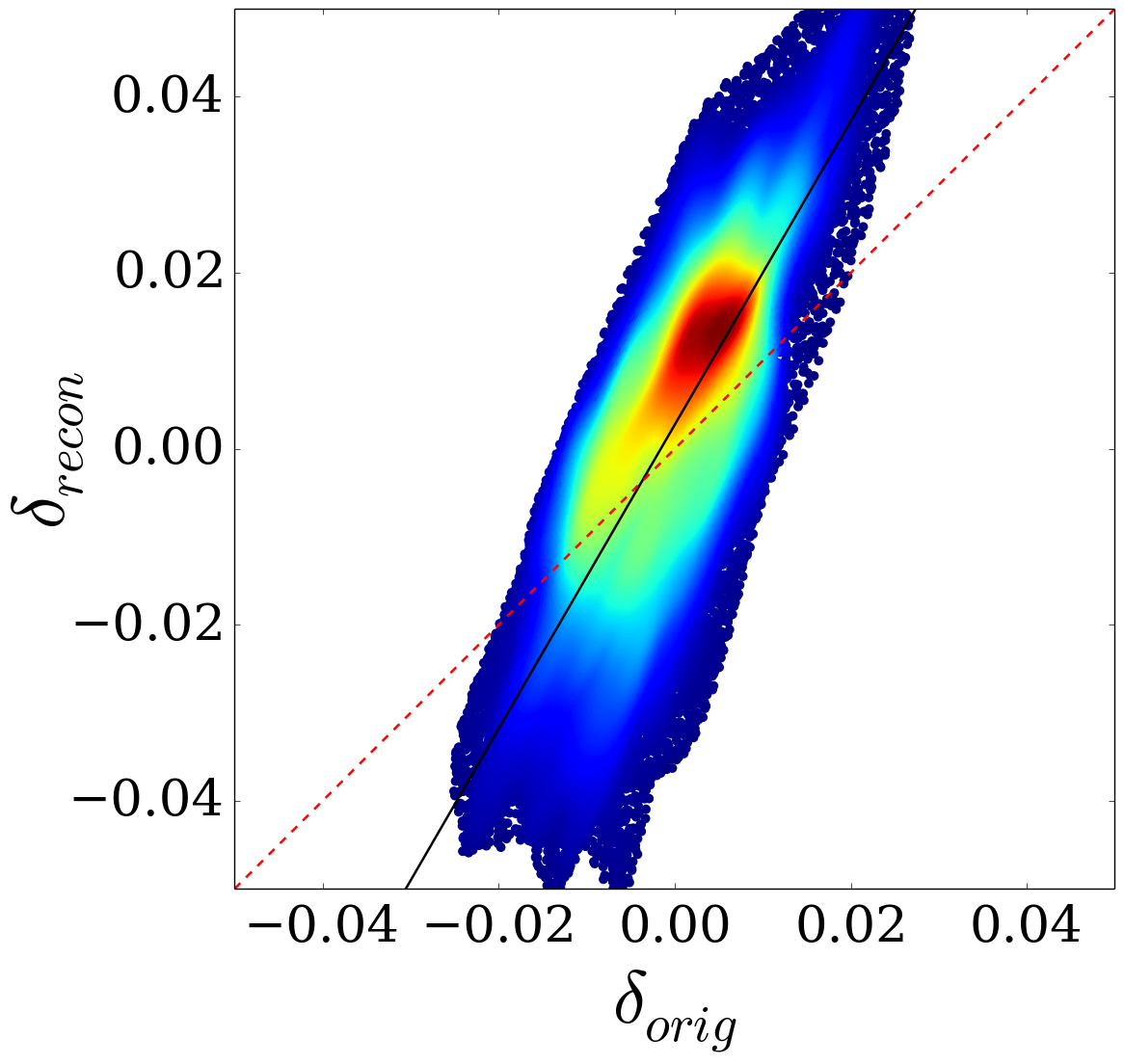}
\caption{$z=2$, $N_{\rm LOS} = 200$, Noiseless}
\end{subfigure}
\begin{subfigure}{0.32\textwidth}
\includegraphics[scale=0.18]{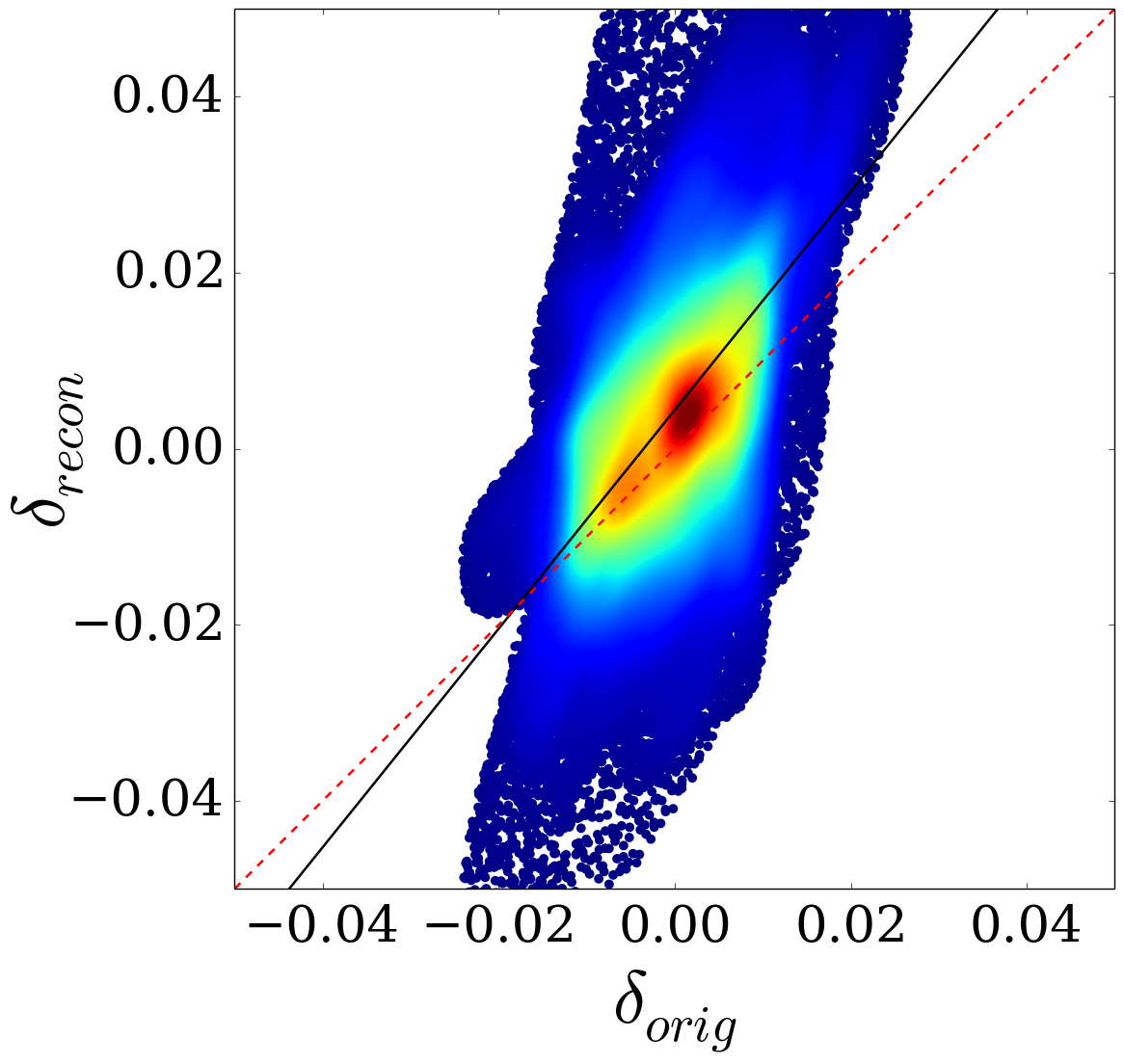}
\caption{$z=2$, $N_{\rm LOS} = 200$, S/N=2}
\end{subfigure}
\begin{subfigure}{0.32\textwidth}
\includegraphics[scale=0.18]{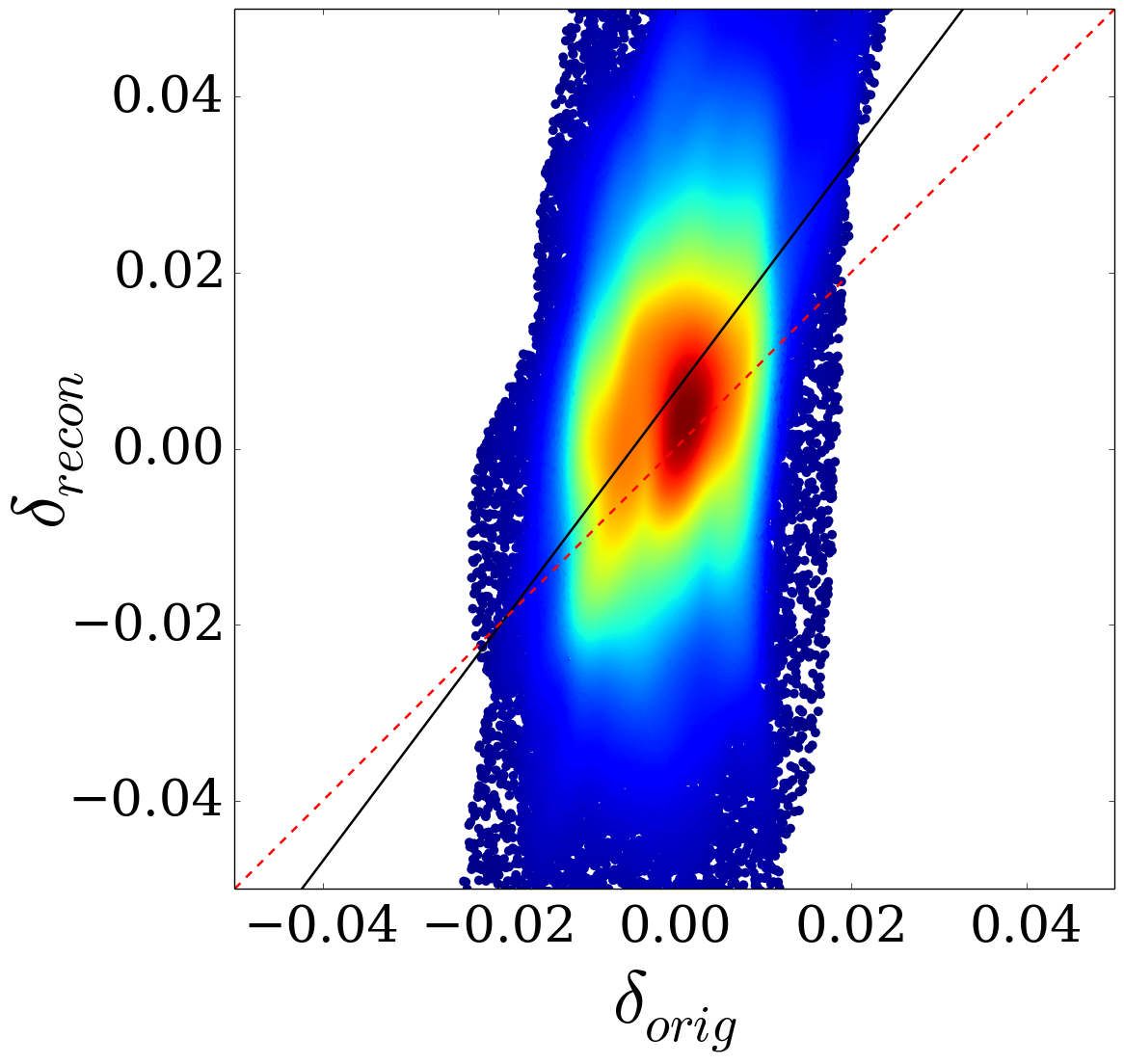}
\caption{$z=2$, $N_{\rm LOS} = 200$, S/N=1}
\end{subfigure}

\vspace*{0.6cm}

\begin{subfigure}{0.32\textwidth}
\includegraphics[scale=0.18]{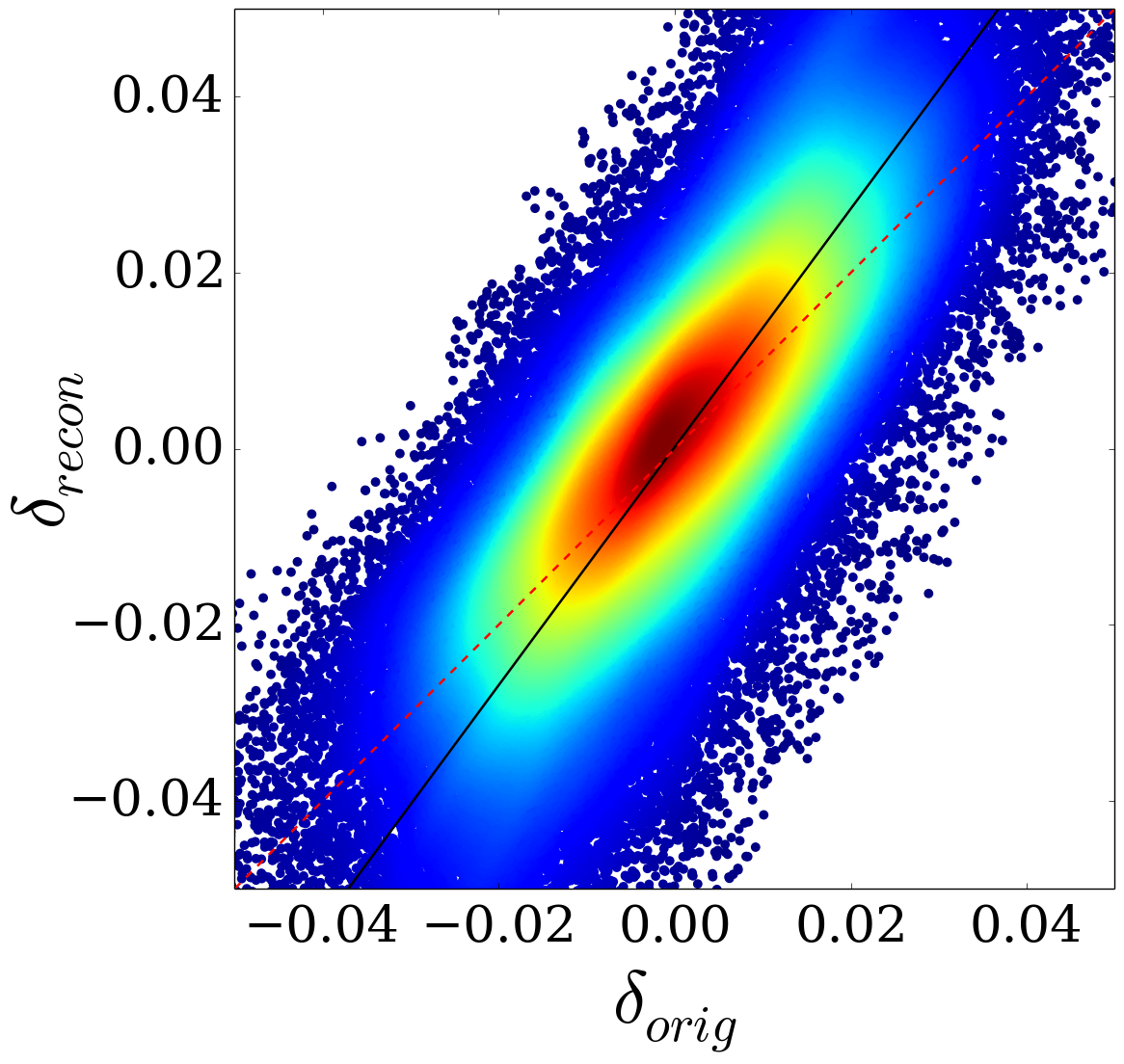}
\caption{$z=2$, $N_{\rm LOS} = 1000$, Noiseless}
\end{subfigure}
\begin{subfigure}{0.32\textwidth}
\includegraphics[scale=0.18]{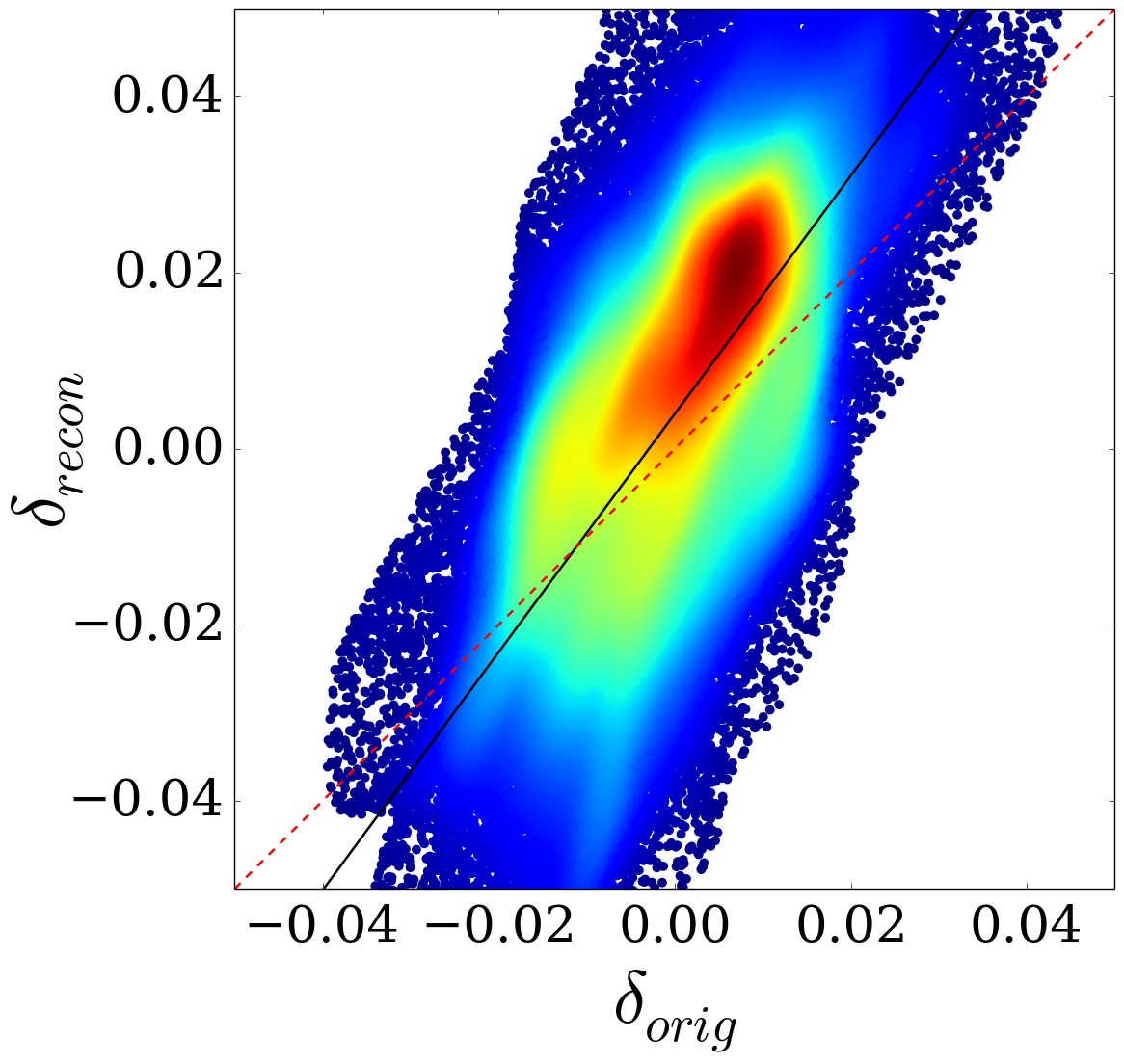}
\caption{$z=3$, $N_{\rm LOS} = 200$, Noiseless}
\end{subfigure}

\vspace*{0.2cm}

\caption{
Scatter plots of the true flux contrast ($\delta_{orig}$)
in the simulated maps compared to the reconstructed flux contrast ($\delta_{recon}$). We show reconstructions at different redshifts, sightline densities
and signal to noise ratios, as follows:
The top row, which is from an analysis at $z=2$ with $N_{\rm LOS} = 200$,
demonstrates the effect of adding noise to our data and carrying out the reconstruction in order to mimic observational data. From left to right, the analyses for data which are noiseless, S/N=2 and S/N=1 are shown. The bottom left is from an analysis at $z=2$ with $N_{\rm LOS} = 1000$. Bottom right plot shows our results at $z=3$ with $N_{\rm LOS} = 200$. The black lines indicate linear regression fits and the red dashed lines show the $y=x$ line. Colours show the density of the points, with red being the densest and blue denoting the most sparse. For each plot, the red area contains 68 per cent of the data points.} 
\label{Fig: Scatter Plots}
\end{figure*}

We first show point by point scatter plots in Fig. \ref{Fig: Scatter Plots}.
We plot the reconstructed flux contrast, $\delta_{recon}$
against the true flux contrast, $\delta_{orig}$. The results of
the point to point comparison of the fields are summarised in 
Table \ref{Table:RMS Errors and Smoothing Level}.

Throughout the paper, we use "original field" (or "true field") with the meaning that we keep all of the Lyman-$\alpha$ skewers in the cube, while "recovered field" or "reconstructed field" means the flux field inferred with the given LOS density with the quasars located at redshift 2 or 3. In the top left panel of Fig. \ref{Fig: Scatter Plots},
we show the results for the the $N_{\rm LOS}=200$ dataset with no noise. We can
see the slope of the relation between  $\delta_{recon}$
and $\delta_{orig}$ is biased (this was also found 
by \citealt{lee2014a}), in the sense that the recovered field has more contrast 
than the original field. After fitting a linear regression we find that the
slope is 1.73, whereas the y--intercept is consistent with zero.
This bias depends on the interplay between the Wiener filter smoothing 
scales and fluctuations in the field that are missed in the sparse sampling.
The bias is larger when the number density of sightlines is low (compare the 
top left panel of Fig. \ref{Fig: Scatter Plots} 
which has a fitted slope of 1.73 and a sightline density 
5 times less than the bottom left panel, which has a fitted slope of 1.35).
Any correction for this bias is likely to be empirical, and therefore 
in the rest of our analysis we apply the simplest correction, by 
renormalizing the $\delta_{recon}$ according to the slope of the regression.

\begin{table*}
\begin{center}
\begin{tabular}{ccccccc}
{\bf Sample} & {$\mathbf{RMS({\delta_{orig}})}$} & {$\mathbf{RMS({\delta_{recon}})}$} & {$\mathbf{RMS({\delta_{diff}})}$} & {$\mathbf{\% \; Error}$} & {\bf Pearson Coefficient ($\mathbf{r}$)} & {$\mathbf{\sigma_S}$}($\hmpc$) \\
\hline

z2\textunderscore N200 & 0.00817 & 0.0104 & 0.00666 & 20.4 & 0.783 & 39.6 \\
z2\textunderscore N200\textunderscore SN2 & 0.00809 & 0.0139 & 0.0120 & 37.0 & 0.578 & 39.6 \\
z2\textunderscore N200\textunderscore SN1 & 0.00805 & 0.0195 & 0.0185 & 57.3 & 0.412 & 39.6 \\
z2\textunderscore N400 & 0.0122 & 0.0154 & 0.00957 & 19.6 & 0.790 & 28.0 \\
z2\textunderscore N400\textunderscore SN2 & 0.0121 & 0.0228 & 0.0199 & 41.0 & 0.530 & 28.0 \\
z2\textunderscore N400\textunderscore SN1 & 0.0121 & 0.0452 & 0.0417 & 86.1 & 0.421 & 28.0 \\
z2\textunderscore N1000 & 0.0193 & 0.0234 & 0.0133 & 17.2 & 0.824 & 17.8 \\
z2\textunderscore N1000\textunderscore SN2 & 0.0192 & 0.0342 & 0.0284 & 36.9 & 0.563 & 17.8  \\
z2\textunderscore N1000\textunderscore SN1 & 0.0195 & 0.0364 & 0.0307 & 39.3 & 0.537 & 17.8  \\
z3\textunderscore N60 & 0.00432 & 0.00695 & 0.00599 & 34.7 & 0.622 & 72.3 \\
z3\textunderscore N200 & 0.0128 & 0.0187 & 0.0139 & 27.1 & 0.686 & 39.6 \\
z3\textunderscore N200\textunderscore SN2 & 0.0127 & 0.0379 & 0.0370 & 72.8 & 0.334 & 39.6 \\

\hline
\end{tabular} \caption{RMS values, percentage error, Pearson coefficient and the standard deviation for the isotropic Gaussian filter size for different samples. The reader is referred to Table \ref{Table:Simulated Data Sets} for
the sample definitions.} 
\label{Table:RMS Errors and Smoothing Level}
\end{center}
\end{table*}

In order to quantitatively test the quality of the reconstruction, we compute the error by 
calculating the ratio of the root mean square (RMS) of the pixel by pixel difference to
the RMS of the true field ($\delta_{orig}$) which only includes 95 per cent of the true pixels ($\pm$2$\sigma$ from the mean, which is 0), therefore avoiding outlier points:


\begin{equation}
e_{\%}=100 \frac{\sqrt{\sum(\delta_{\rm orig} - \delta_{\rm recon})^{2}}} {4 \sqrt{\sum \delta_{\rm orig}^{2}}}
\end{equation}

From Fig. \ref{Fig: Scatter Plots}, we can see from the top row
of panels that the addition of noise to the input field does affect the 
reconstruction. The RMS error (after bias correction)
is 20.4 per cent for the noiseless case
and 37.0 per cent and 57.3 per cent for the cases with S/N=2 and S/N=1 respectively in the pixels
of size 9.30 $\hmpc$ used in our analysis. We remind the reader that the
BOSS \lya forest data when rebinned in this way has a mean
S/N ratio of 2.5. Therefore, we estimate that the RMS error for
the reconstructed BOSS data will be 36.7 per cent. 

We also provide the Pearson coefficient ($r$) as a measure 
of the linear correlation between $\delta_{orig}$ and $\delta_{recon}$. Total positive correlation between the original field and the reconstruction would correspond to $r=1$, while no correlation would be $r=0$ and total negative correlation would be $r=-1$. At redshift $z=2$ with $N_{\rm LOS} = 200$, $r=0.783$, and it is even higher when the number of sightlines is increased, as expected.


Increasing the number of sightlines, as shown in the bottom 
left panel of Fig. \ref{Fig: Scatter Plots}
allows the smoothing scale to be reduced and the resolution
of finer features in the
flux contrast field. The RMS flux contrast fluctuations
increase to 0.0193 for this sample (z2\textunderscore N1000, see Table \ref{Table:RMS Errors and Smoothing Level})
and the percentage error on the reconstruction stays approximately
the same as the lower resolution reconstruction in the top left panel.

Finally we show results for the higher redshift, $z=3$ in the bottom
right panel of Fig. \ref{Fig: Scatter Plots}. For the sample z3\textunderscore N200,
we use the same number of sightlines, as the top left ($z=2$) panel, but the
RMS accuracy of the reconstruction is lower by a factor of 0.75. The quality of the reconstruction is low for the sample z3\textunderscore N60, as can
be seen in Fig. \ref{Fig:Slices_z3_nonoise_60LOS}. However, the percentage error in Table \ref{Table:RMS Errors and Smoothing Level} is comparable to those from
other samples due to the fact that the dynamic range in the true field is low because of the much higher
$\sigma_S$ value for that sample. The flux 
contrast fluctuations are larger
at the higher redshift, because of the greater overall
optical depth in the \lya forest, but this does not translate
into a better reconstruction.

\subsection{Cross Correlation}

In the previous section we have seen how the true and reconstructed fields
compare on a point--by--point basis for one specific value of the smoothing
scale, given by $\sigma=\sqrt{2} \dlos$. We can also study how 
similar the fields are at different scales. Instead of
 using different filter scales, we make use of the correlation functions
of the field, measuring the correlation between points separated by 
a distance $r$. We compute the auto--correlation of the true field, 
the cross correlation and the standardized cross correlation between the true
 and the recovered fields. In general, the correlation function of two 
fields 1 and 2 is defined by:

\begin{equation}
\xi_{12}(r)=<\delta_{1}(x)\delta_{2}(x+r)>
\end{equation}

\noindent
For the auto--correlation, 1 and 2 represent the same fields. 
The standardized cross correlation is defined by: 

{
\begin{equation}
C_{12}(r)=\xi_{12}(r)/\sqrt{\xi_{11}(r)\cdot \xi_{22}(r)}, 
\end{equation}
}

\noindent
where $\xi_{12}$ is the cross-correlatin
and $\xi_{11}$ and $\xi_{22}$ are the auto--correlations.
$C_{12}$ enables us to quantify the accuracy of the reconstruction as a 
function of scale, with a value of unity indicating perfect 
fidelity. It should be noted that we do not expect good agreement 
for scales smaller than the fiducial smoothing length, the natural 
resolution of the map. According to Equation \ref{Eq:dlos}, the fields 
with $N_{\rm LOS} = 200$ are smoothed with $\sigma_{S}=39.6$ $\hmpc$, 
while those with $N_{\rm LOS} = 1000$ are smoothed with $\sigma_{S}=17.7$ $\hmpc$. 

\begin{center}
\begin{figure*}
\centering
\begin{subfigure}{0.49\textwidth}
\includegraphics[scale=0.42]{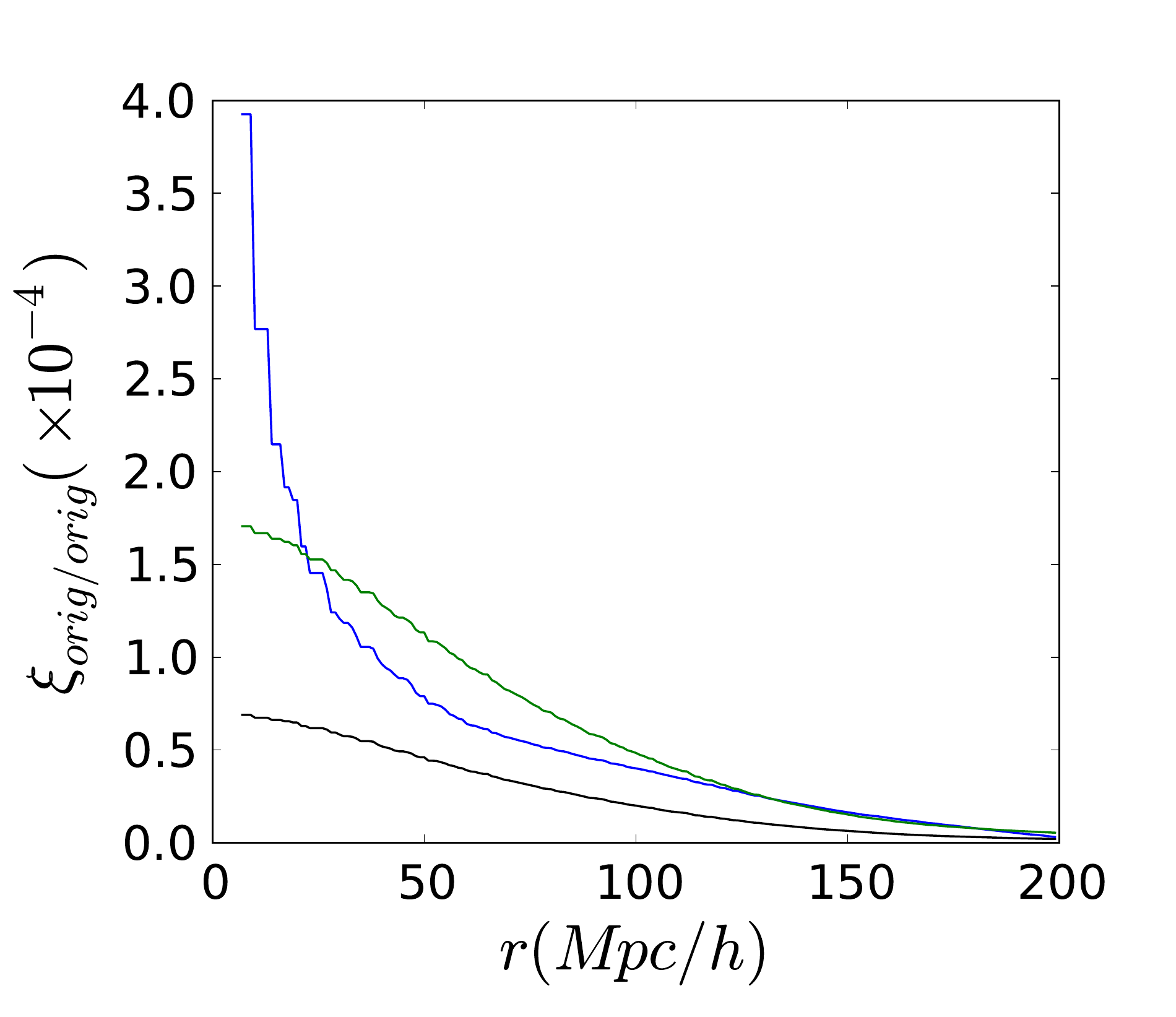}
\caption{Auto--correlation}
\end{subfigure}
\begin{subfigure}{0.49\textwidth}
\includegraphics[scale=0.42]{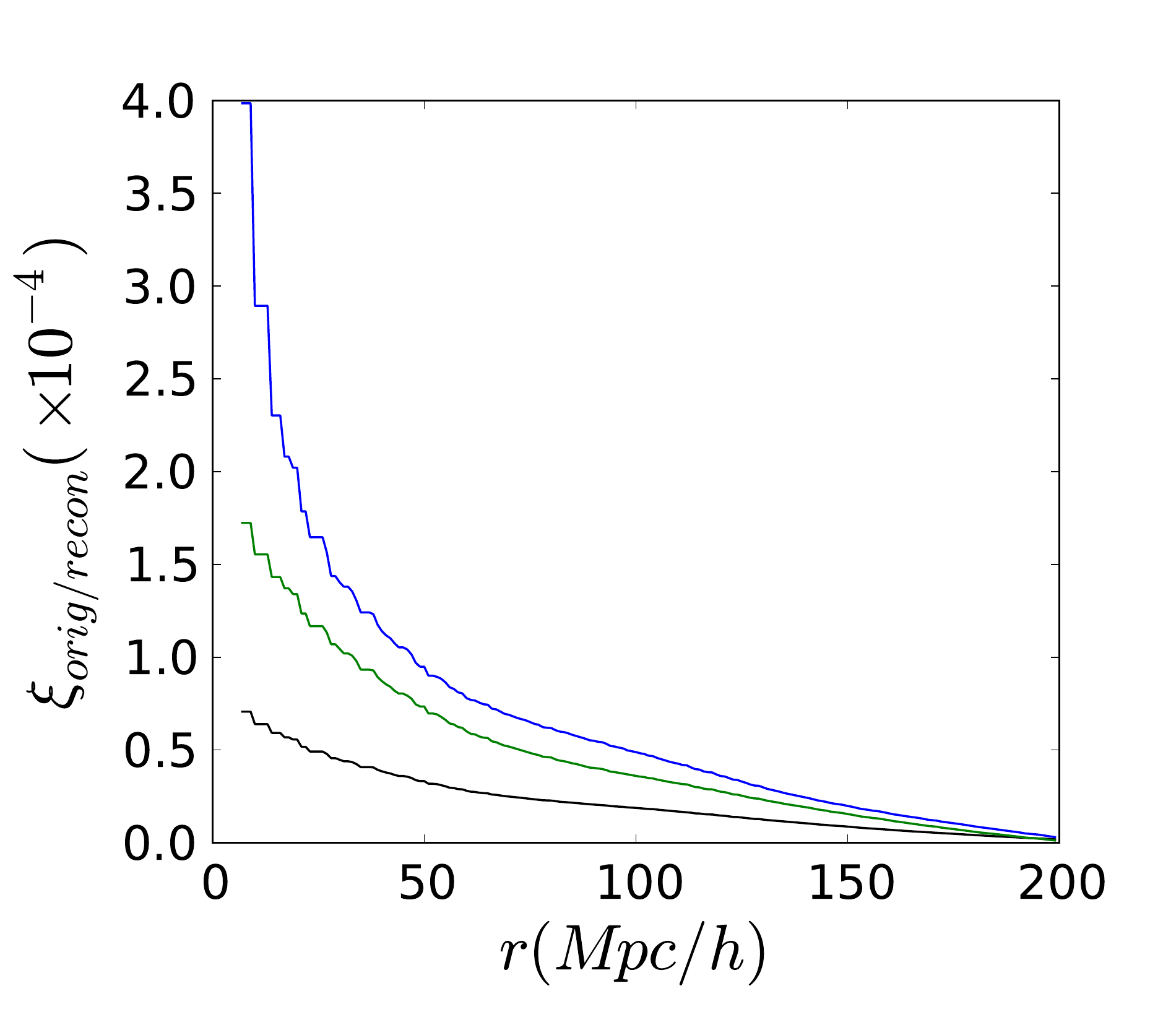}
\caption{Cross--correlation}
\end{subfigure}
\begin{subfigure}{0.49\textwidth}
\includegraphics[scale=0.42]{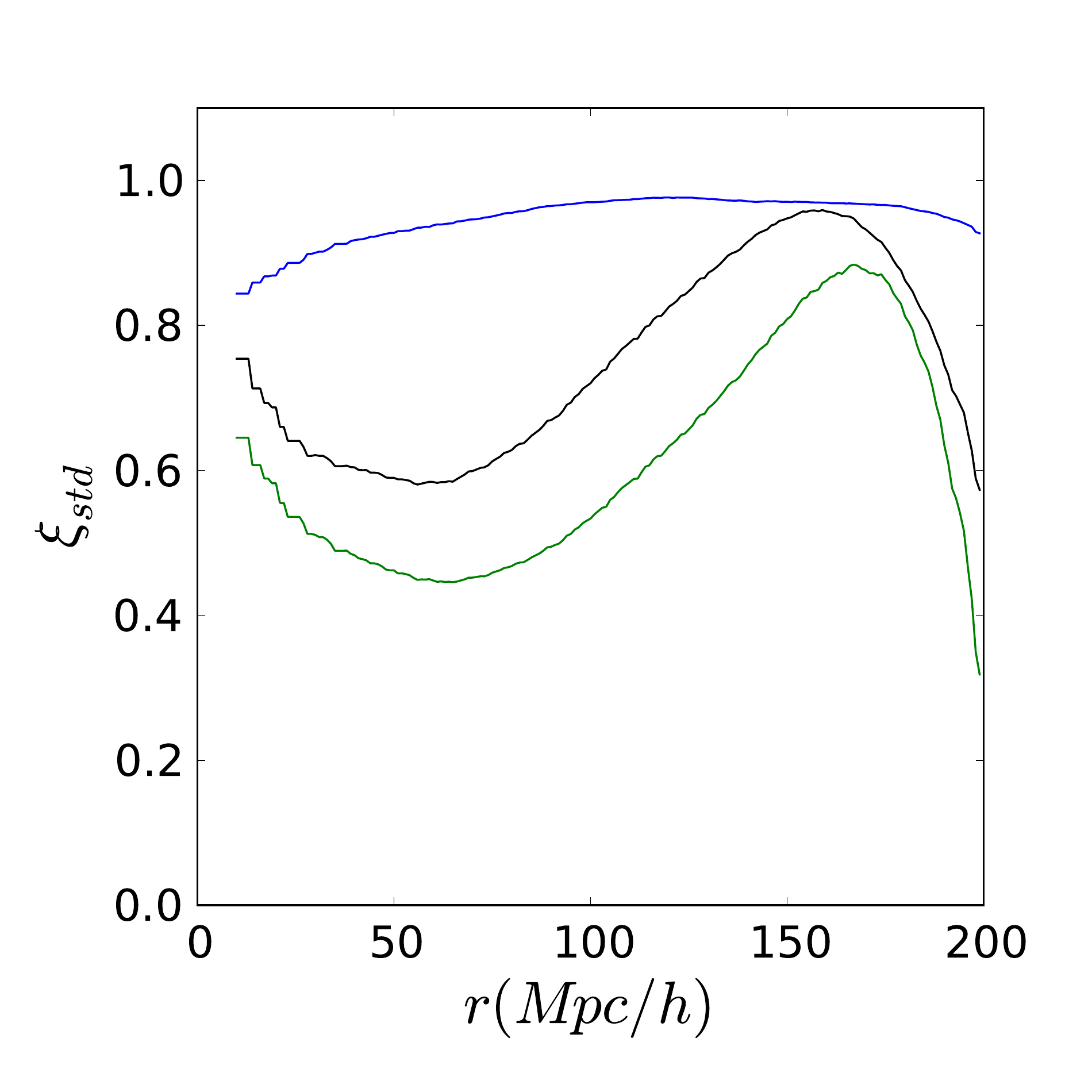}
\caption{Standardized cross--correlation}
\end{subfigure}
\begin{subfigure}{0.49\textwidth}
\includegraphics[scale=0.42]{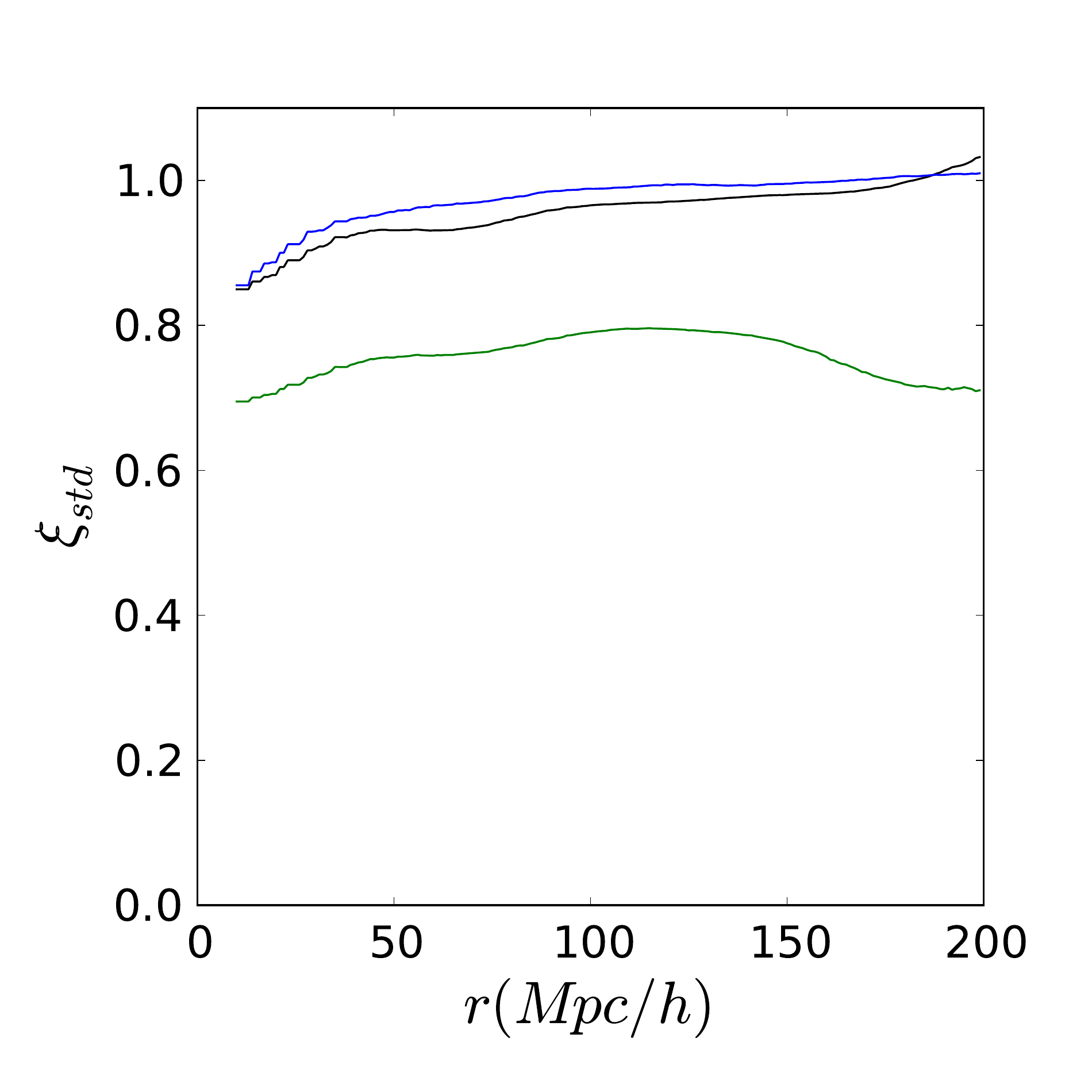}
\caption{Standardized cross--correlation (truncated cube)}
\end{subfigure}

\vspace*{0.4cm}

\caption{Correlation functions of the true and reconstructed fields
as a function of scale. In each panel, the 
black colour curves show the results for $z=2$ with $N_{\rm LOS} = 200$, 
the blue colour $z=2$ with $N_{\rm LOS} = 1000$ and green $z=3$ with 
$N_{\rm LOS} = 200$. Panel (a) shows the auto--correlation function of the
true field and panel (b)
the cross--correlation function of the true and reconstructed fields. Panel 
(c) shows the standardized cross--correlation function 
for the entire simulation volume computed using the equation $C_{12}/\sqrt{A_1\cdot A_2}$. Panel (d) shows the standardized cross--correlation function 
computed only for the part of the simulation volume that is at least
50 $\hmpc$ from an edge.
}
\label{Fig:Correlation Plots}
\end{figure*}
\end{center}

If we look at the case of low sampling density, $N_{\rm LOS} = 200$, we see that 
$C_{12}$  can be as low as 0.45 on scales which are approximately 1.5--2 times
the smoothing scale. 
Repeating the same analysis for unsmoothed fields with this LOS density
does give 
better agreement at low scales -- clearly, some structure is erased due to 
smoothing. For larger values of $N_{\rm LOS}$, on the other hand, we get good 
agreement  despite the smoothing (Fig. \ref{Fig:Correlation Plots}). 
After smoothing both fields with the fiducial smoothing length, the recovery improves with distance until very large distances,
where it decreases again. This decrease is due to edge effects, as on
large scales, much of the volume is close to an edge.  In panel (d)
of Fig. \ref{Fig:Correlation Plots} we restrict the
measurements of $\xi_{12}$ to the volume of the simulation cube
left after eliminating all regions within 50 $\hmpc$ of an edge. 
Comparing with panel (c) of Fig. \ref{Fig:Correlation Plots}, we can 
see that the agreement on small scales has also been improved,
 showing the positive effects of getting rid of the edge artefacts.
In a large survey such as BOSS
which spans a contiguous volume of several gigaparsecs,
 most of the volume will be much  further from an edge than 50 $\hmpc$,
so that edge effects should be a small issue.

When the edge effects have been removed, we can see that the $C_{12}$ 
measurement is close to 1 for all scales greater
than the smoothing filter scale at $z=2$, indicating essentially
perfect statistical agreement. At redshift $z=3$, $C_{12}$ is never 
greater than 0.8, which may indicate that the less evolved structures
at higher redshift make accurate reconstruction more difficult.

\subsection{Non--Gaussianity and the Flux Probability Distribution}

The density field probed by the \lya forest is 
expected to be in the mildly non-linear regime. When smoothed on
large scales, which we necessarily must do in order to 
construct our interpolated maps, we expect that the flux probability
distribution should be quite close to Gaussian. Indeed this Gaussian
assumption underlies the reconstruction carried out with the
Wiener filter in Equation \ref{Eq:correl_matrix}. It is therefore of
interest to compare the reconstructed and true flux probability
density functions with each other and with a normal distribution.

\begin{table}
\begin{tabular}{cccc}

{\bf DATA SET} & {\bf D value} & {\bf p value}\\

\hline

$\mathbf{z=2, N_{\rm LOS}=200}$, {\bf Noiseless}            &                &               \\
Original                   &      0.067	          &           0.63     \\
Reconstruction      &      	0.060          &    0.77          \\

$\mathbf{z=2, N_{\rm LOS}=200}$, {\bf S/N=2}              &      	          &                \\
Original             &      0.068         &        0.60       \\
Reconstruction             &      	0.14          &         0.011     \\

$\mathbf{z=2, N_{\rm LOS}=200}$, {\bf S/N=1}              &      	          &                \\
Original             &     0.058 	          &   0.81            \\
Reconstruction            &    0.23 	          &       4.4$\times10^{-6}$         \\

$\mathbf{z=2, N_{\rm LOS}=400}$, {\bf Noiseless}              &      	          &                \\
Original            &    0.044  	          &   0.97           \\
Reconstruction             &      0.069          &      0.57         \\

$\mathbf{z=2, N_{\rm LOS}=1000}$, {\bf Noiseless}              &      	          &                \\
Original             &    0.088  	          &     0.28           \\
Reconstruction          &     0.057	          &     0.83           \\

$\mathbf{z=3, N_{\rm LOS}=200}$, {\bf Noiseless}              &      	          &                \\
Original             &     0.065         &    0.68            \\
Reconstruction           &     0.098	          &      0.17         \\
            
\hline
\end{tabular}
\caption{Kolmogorov--Smirnov test results giving 
the probability ($p$ value) of the flux pdfs of the
real and the reconstructed data being
drawn from a Gaussian distribution. We show results for different
numbers of quasar sightlines through our simulation
volume, $N_{\rm LOS}$, redshifts and signal to noise
ratios.} \label{Table:Kolmogorov-Smirnov}
\end{table}

In order to look for deviations in the flux
pdfs, we use the Kolmogorov--Smirnov (KS) test. 
We first compute the mean and the 
standard deviation $\sigma$ of a Gaussian fit 
to the true field. We construct a cumulative probability distribution
from this and compare it to the cumulative probability distributions of
the true field and the reconstructions, for various $N_{\rm LOS}$ values,
redshifts and levels of noise.

With the KS test, we compute quantitative measures
of the similarity of the flux pdfs to
normal distributions.  The test statistic (D value) is the maximum of the 
difference in the cumulative distribution functions of the particular
field being tested and the 
Gaussian.  The closer this value is to 0, the more likely it is
that the data sets are 
have been drawn from the same distribution. Furthermore, the 
p value, which is computed from the test statistic, 
represents the significance level threshold below which the null 
hypothesis (that the data sets come from the same distribution) 
will be accepted.

When computing the flux pdfs, we would like the data points to be as
independent as possible, and so our data points should at least be
separated by distances greater than the smoothing scale, because smoothing
would correlate the measurements.
Because of this, we downsample each data set,
picking only a $5^{3}$ grid of values (i.e., data points separated by 
80 $\hmpc$ 
in each direction in the 400 $\hmpc$ volume). 
Our KS test results are shown in Table \ref{Table:Kolmogorov-Smirnov},
for different line of sight densities, redshifts and signal to noise
ratios.

In all cases for the true field we find high p values and low D
values, which means that the original field was approximately
Gaussian to start with. The recovered field also shows the same property as well  
in Table \ref{Table:Kolmogorov-Smirnov}, for recovery from data samples
with no added noise. When noise was added, however, the reconstructed
maps became significantly non--Gaussian, with the p value decreasing
as the signal to noise ratio decreased. Furthermore, the fact that the D and p values between
samples with different LOS density are significantly different can be attributed to the fact that the smoothing filter size
depends on the LOS density itself. As we have mentioned above, the 
noisiest data, which has S/N=1, is significantly worse than the majority of 
BOSS data, for example, but the effect of noise in changing
the pdf shape of the reconstructed field should still be borne in mind
in an analysis of observed data.

\subsection{Peaks In the Density Field}

Searching for local maxima in the reconstructed flux density field
offers one means of defining objects and finding them. 
Such peaks are likely to correspond to the locations
of forming clusters or superclusters of galaxies.
The properties of these density maxima can be used to constrain the cosmological 
model \citep{bardeen1986statistics, croft1998space, de2007peaks, de2010peaks}. It is therefore of interest to compare the peaks of the reconstructed
flux density field with those in the true flux density field in the simulation.

\begin{figure*}
\centering
\begin{subfigure}{0.32\textwidth}
\includegraphics[scale=0.24]{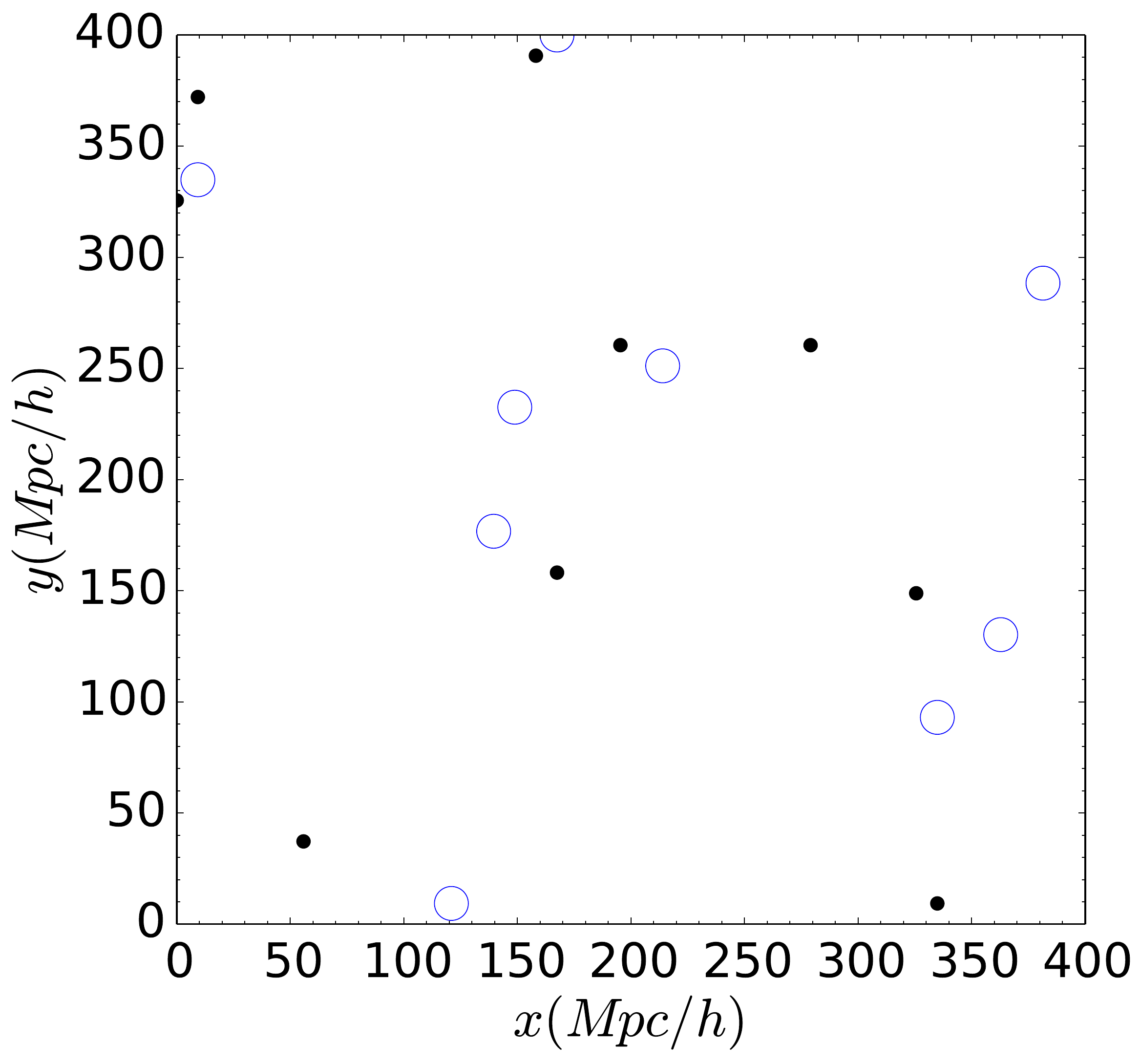}
\caption{$z=2$, $N_{\rm LOS} = 200$}
\end{subfigure}
\begin{subfigure}{0.32\textwidth}
\includegraphics[scale=0.24]{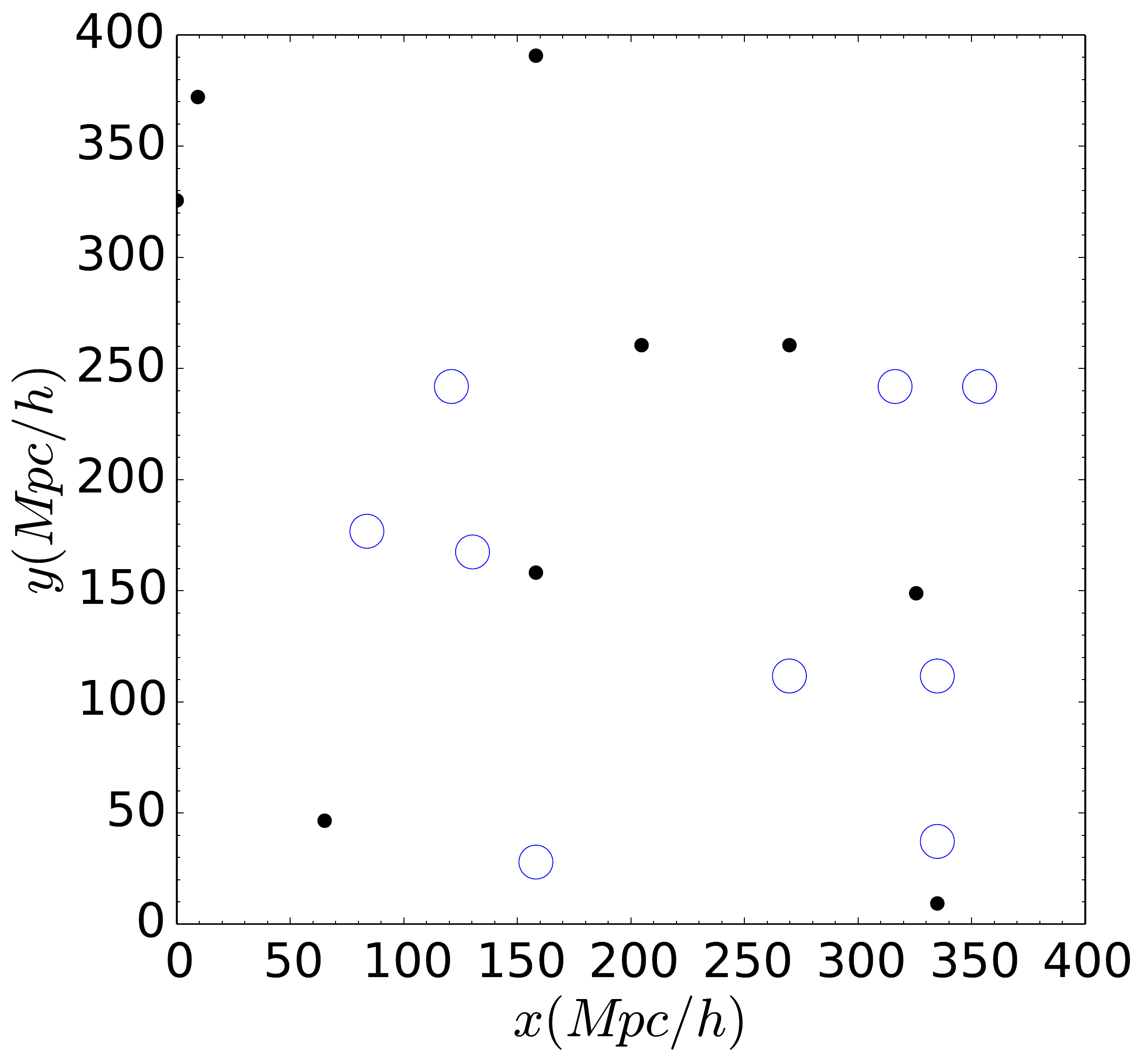}
\caption{$z=3$, $N_{\rm LOS} = 200$}
\end{subfigure}
\begin{subfigure}{0.32\textwidth}
\includegraphics[scale=0.24]{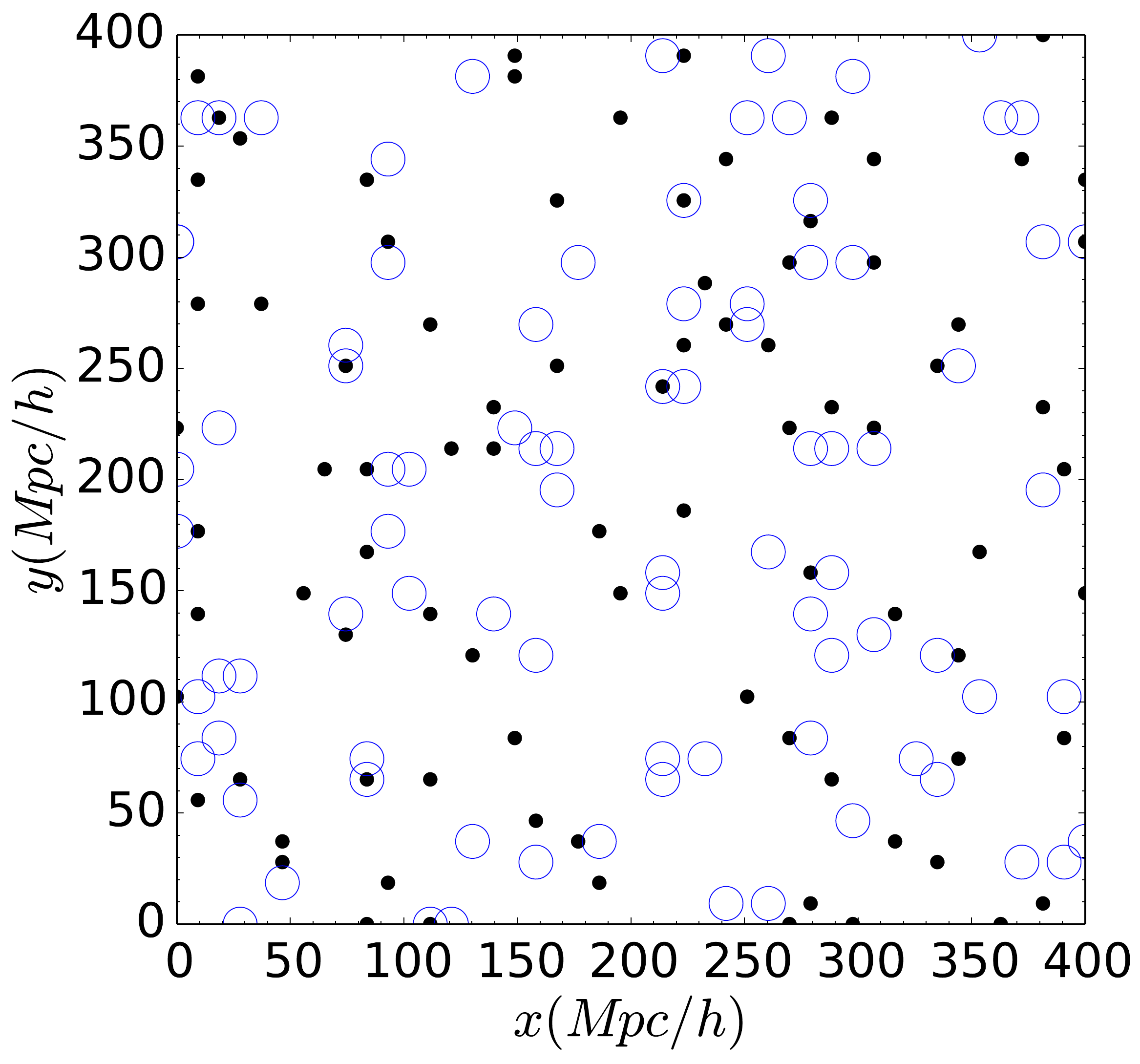}
\caption{$z=2$, $N_{\rm LOS} = 1000$}
\end{subfigure}
\caption{Coordinates of potential superclusters (local peaks) from noiseless analysis. Black dots show the results from the true field, while blue empty circles show the results from the recovered field. Most black dots are enclosed by or neighbouring a blue circle, indicating accurate statistics of the recovered field -- two potential superclusters at the top right are on top of each other.}
\label{Fig:Superclusters}
\end{figure*}

We search for density peaks in the three dimensional volume of the simulation.
Density and flux are inversely related, therefore we identify a simulation 3D pixel as a local peak if its flux value 
is the smallest amongst the 26 neighbouring 3D pixels surrounding it. 
As expected, we find that the number density of local peaks is 
strongly dependent on the smoothing filter size.
We find that for a filter size of 39.6 $ \hmpc$, appropriate for 
$N_{\rm LOS}=200$, we find 9 local peaks in the simulation volume
at $z=2$ (Fig. \ref{Fig:Superclusters}a), and for a filter size of 17.8 $\hmpc$, appropriate for 
$N_{\rm LOS}=1000$, we find 87 peaks (Fig. \ref{Fig:Superclusters}c), where both of these figures
are for the true field. Furthermore, $z=2$ and $z=3$ data samples have very similar true peak locations for the same filter size. When noise is added to mimic observational data, we discover 8 local peaks for the real simulation flux field, while the reconstructed field contains 10 local peaks for the data set z2\textunderscore N200\textunderscore SN1.

In Fig. \ref{Fig:Superclusters} we show a comparison between the true 
field local peaks and those of the reconstructed field. The number of peaks
in both cases match exactly for the three different
combinations of $N_{\rm LOS}$ and redshift shown. The positions of the
peaks are  visually a reasonable match, with better agreement for $z=2$ than
$z=3$. The structures  traced out in the plot
with $N_{\rm LOS}=1000$ by the reconstructed peaks do seem visually
to trace out those in the real peaks.

 There is not a one to one correspondence however. We can quantify the
level of agreement by counting the number of peaks in the true field which
have a peak in the reconstructed field within one smoothing length. 
This is 33.3 per cent for $z=2$, $N_{\rm LOS} = 200$.
The expectation from randomly positioned peaks with the same number density (computed using
1000 Monte Carlo trials) is 11.5 per cent. This means that the reconstruction
is a factor of 2.89 better than
random. The equivalent measures for $z=3$, $N_{\rm LOS}=200$ and
$z=2$, $N_{\rm LOS}=1000$ are 11.1 per cent and 32.1 per cent peaks within 1 
smoothing length respectively and factors of 1.13 and 10.4 better than random.


This type of analysis could be potentially extended to look at the structures
that are enclosed within isodensity contours. This would reveal
the morphology of the IGM on large scales.
At the scales we are probing here (smoothing scales $> 10$ $\hmpc$),
sheet and filament-like topologies are relatively difficult to see
 (as we shall
see in our visualisations in the next section).
On smaller scales, these characteristics are readily apparent in simulated
maps (e.g. \citealt{pichon2001}) and the first maps made from
observational data with these techniques \citep{lee2014a}. The most straightforward cosmological constraints will come from the peak density, and the reconstruction technique does very well: We get perfect agreement between the real and reconstructed fields for noiseless comparison, and for the noisy case (with z2\textunderscore N200\textunderscore SN1), the number of peaks agrees within 25 per cent.


\subsection{Slice Image Comparisons}

\begin{figure}
\begin{subfigure}{0.23\textwidth}
\includegraphics[scale=0.24, angle=-90]
{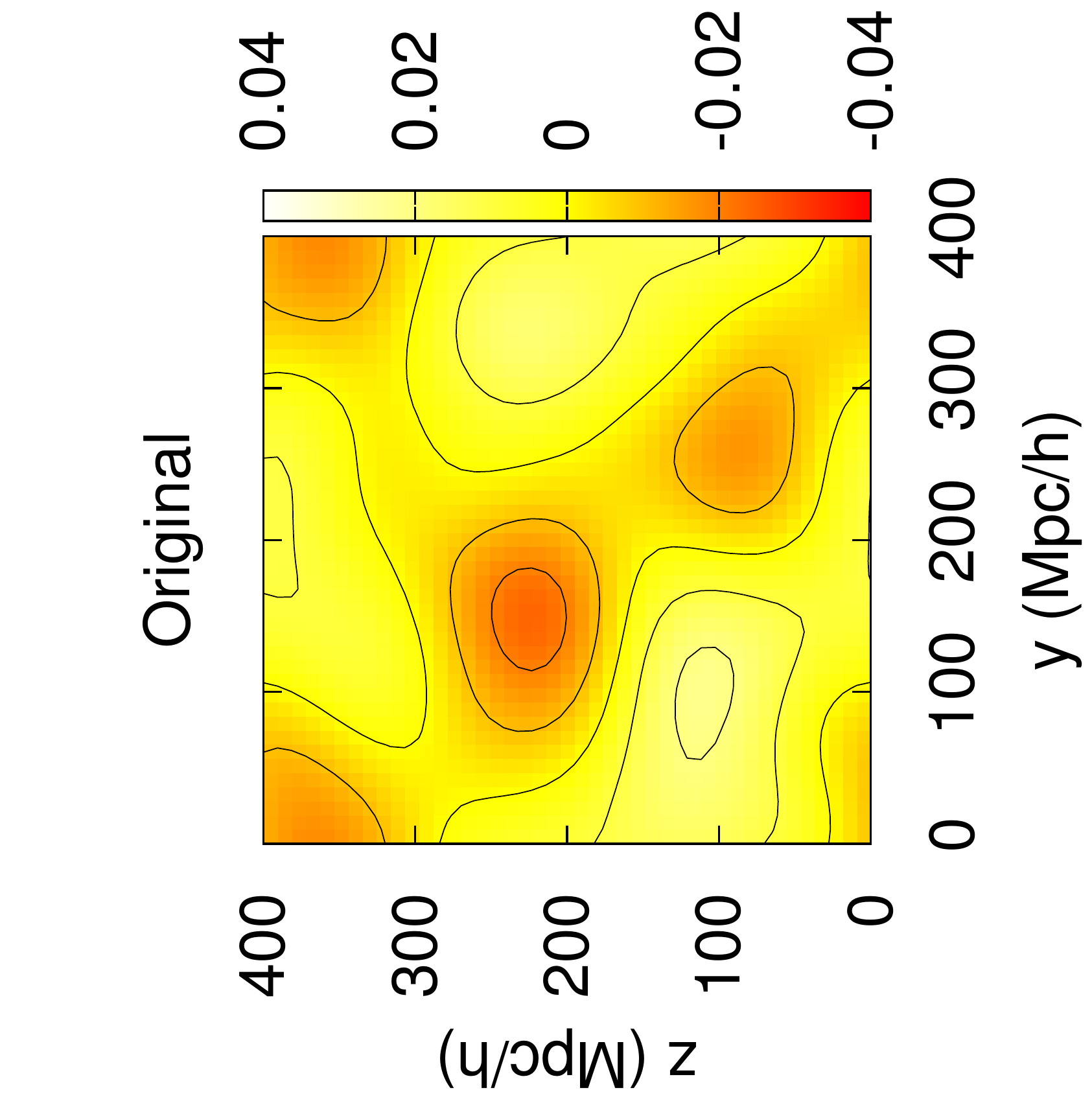}
\captionsetup{justification=centering}
\caption{$z=2$, $N_{\rm LOS}=200$, perpendicular to LOS}
\end{subfigure}
\begin{subfigure}{0.23\textwidth}
\includegraphics[scale=0.24, angle=-90]
{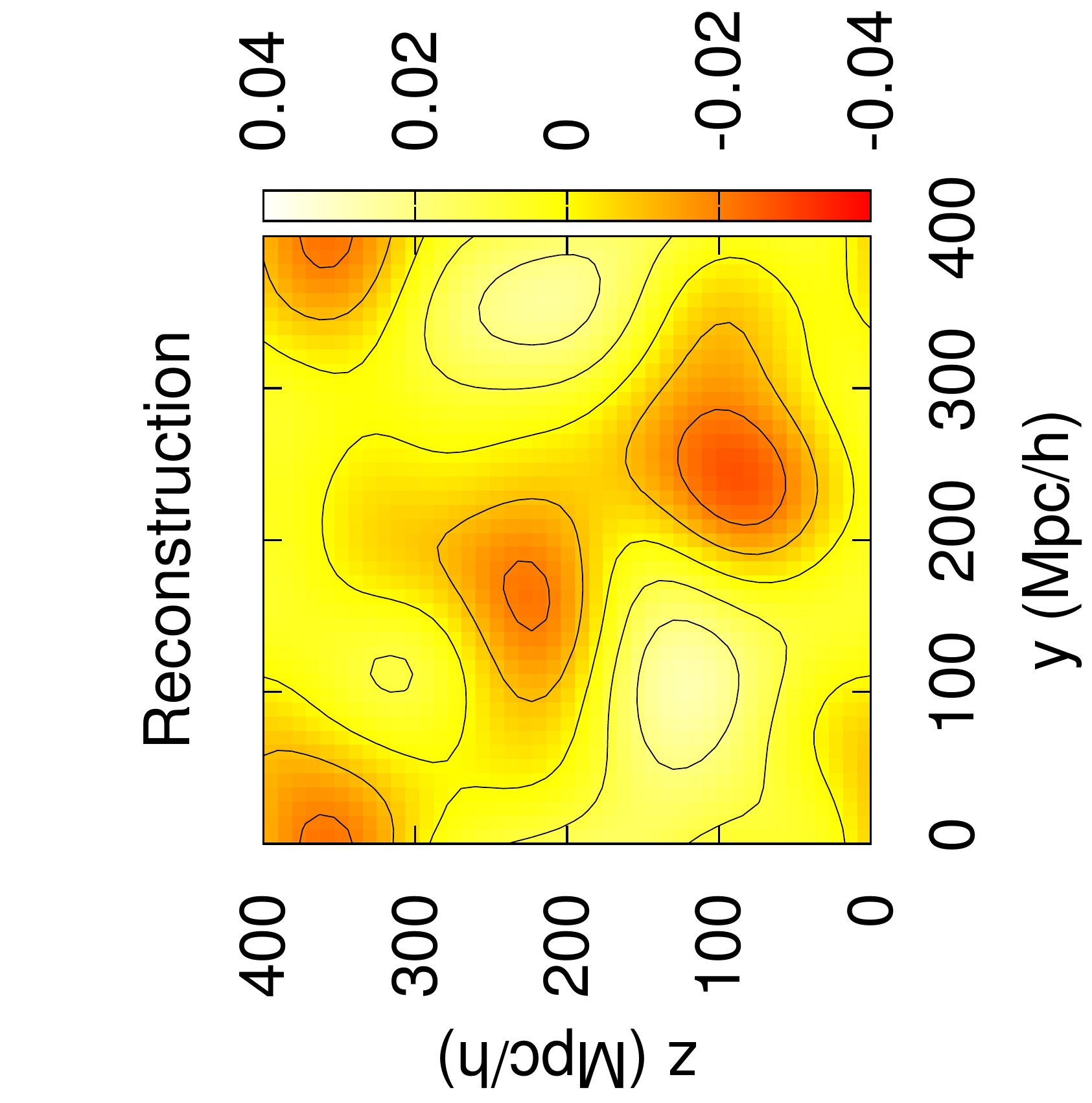}
\captionsetup{justification=centering}
\caption{$z=2$, $N_{\rm LOS}=200$, perpendicular to LOS}
\end{subfigure}

\vspace{0.1mm}

\begin{subfigure}{0.23\textwidth}
\includegraphics[scale=0.24, angle=-90]
{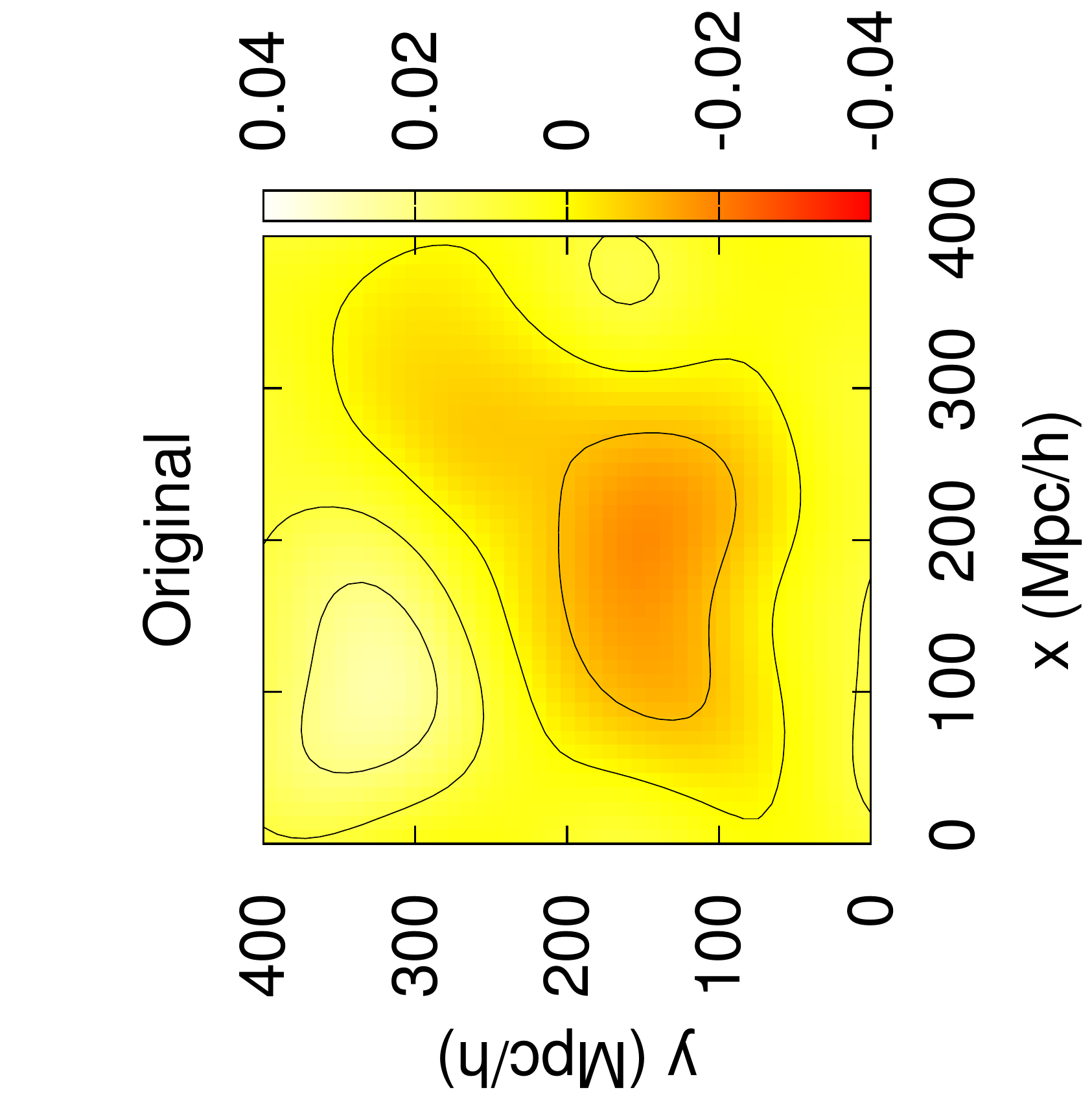}
\captionsetup{justification=centering}
\caption{$z=2$, $N_{\rm LOS}=200$, parallel to LOS}
\end{subfigure}
\begin{subfigure}{0.23\textwidth}
\includegraphics[scale=0.24, angle=-90]
{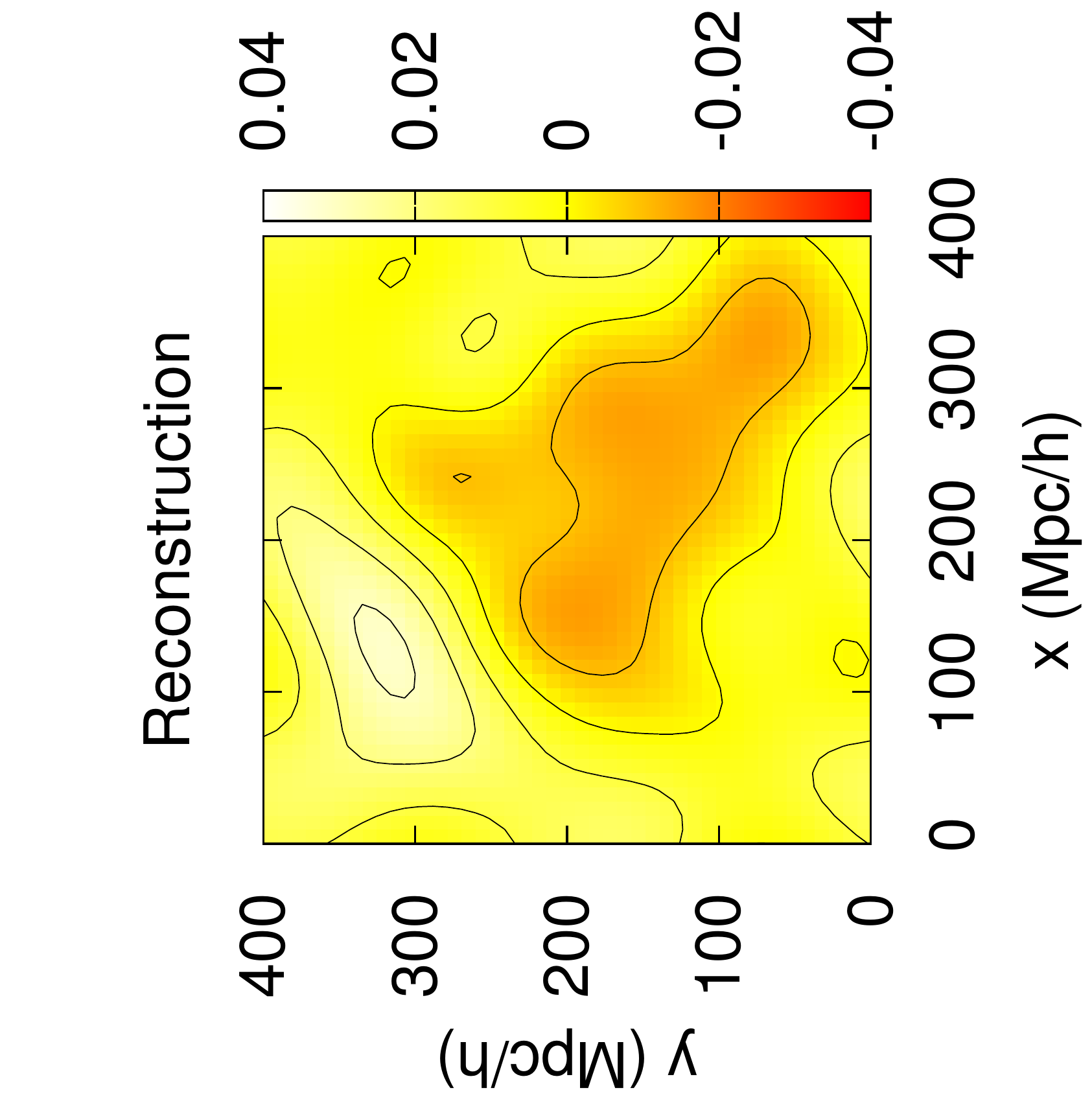}
\captionsetup{justification=centering}
\caption{$z=2$, $N_{\rm LOS}=200$, parallel to LOS}
\end{subfigure}
\caption{Slices extracted from the middle planes of the simulation cube are shown at $z=2$ with $N_{\rm LOS} = 200$, without pixel noise. 
The color scale indicates flux contrast, $\delta_{f}$.
The top row shows slices perpendicular to LOSs, whereas the bottom row shows slices in the parallel direction. True field slices are given in (a) and (c), while (b) and (d) show reconstructed field slices. The smoothed reconstructed field recovers the general features of the simulation.}
\label{Fig:Slices_z2_nonoise_200LOS}
\end{figure}

\begin{figure}
\centering
\begin{subfigure}{0.23\textwidth}
\includegraphics[scale=0.24, angle=-90]
{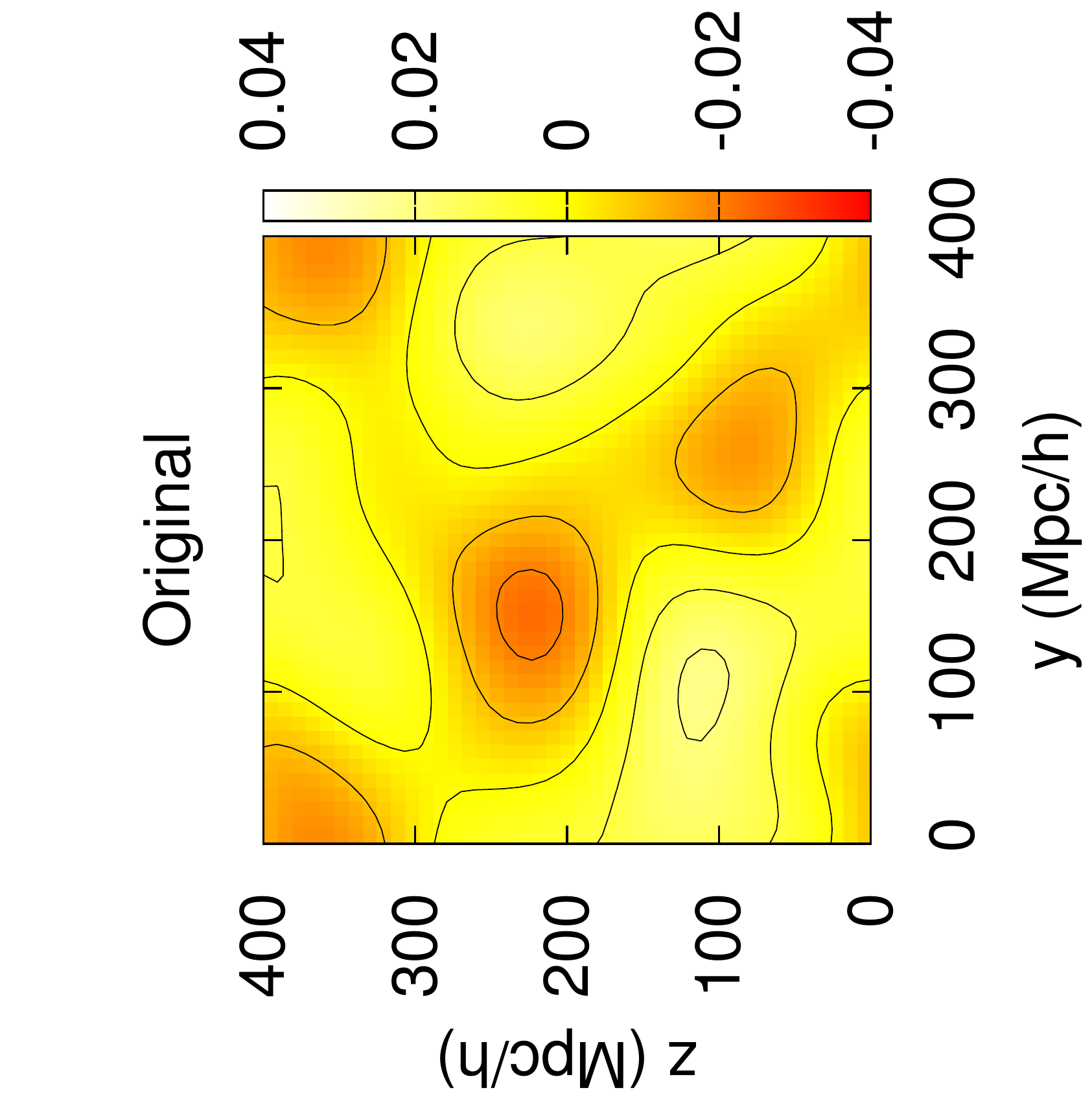}
\captionsetup{justification=centering}
\caption{$z=2$, $N_{\rm LOS}=200$, S/N=2, perpendicular to LOS}
\end{subfigure}
\begin{subfigure}{0.23\textwidth}
\includegraphics[scale=0.24, angle=-90]
{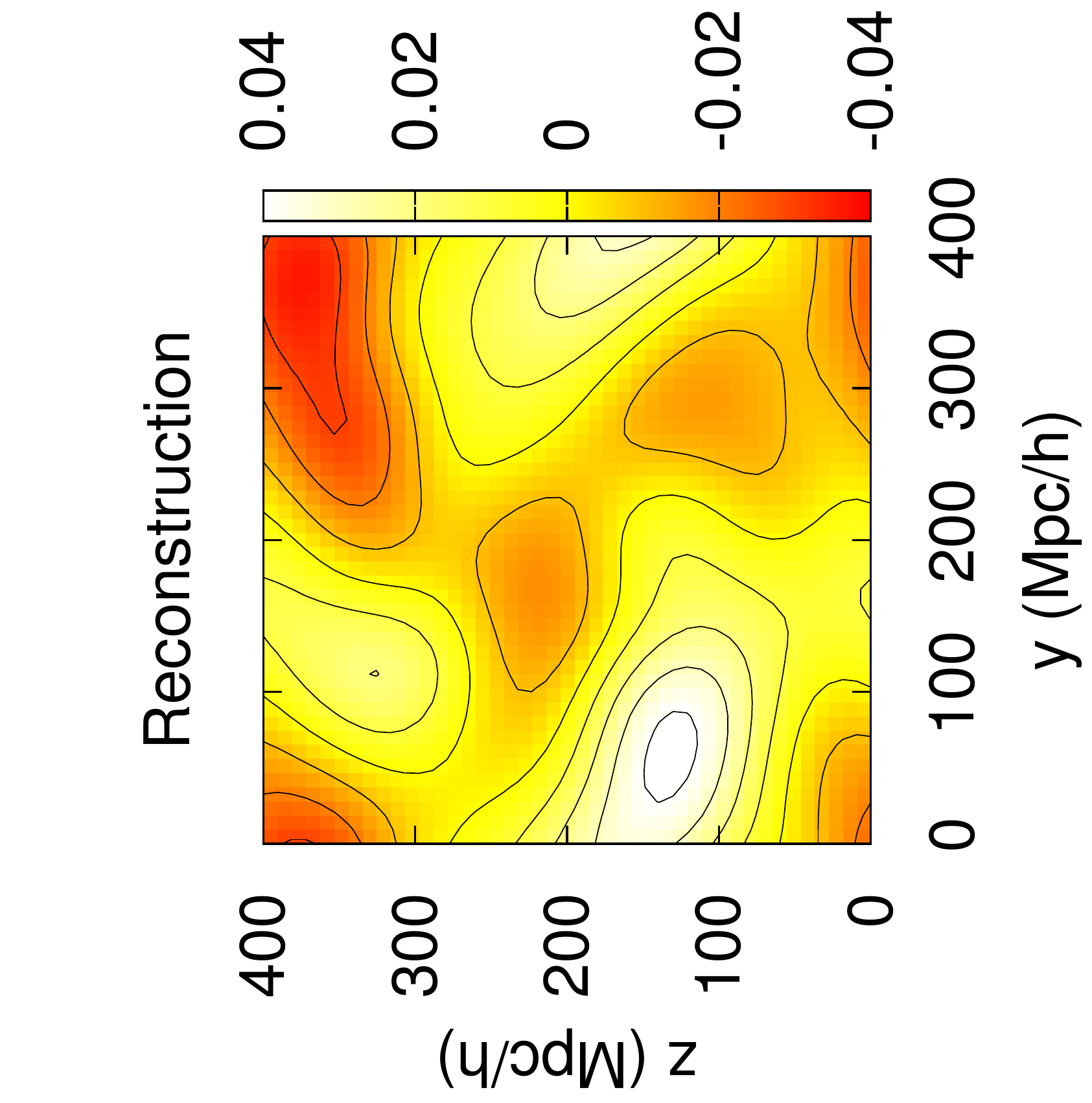}
\captionsetup{justification=centering}
\caption{$z=2$, $N_{\rm LOS}=200$, S/N=2, perpendicular to LOS}
\end{subfigure}

\vspace{4mm}

\begin{subfigure}{0.23\textwidth}
\includegraphics[scale=0.24, angle=-90]
{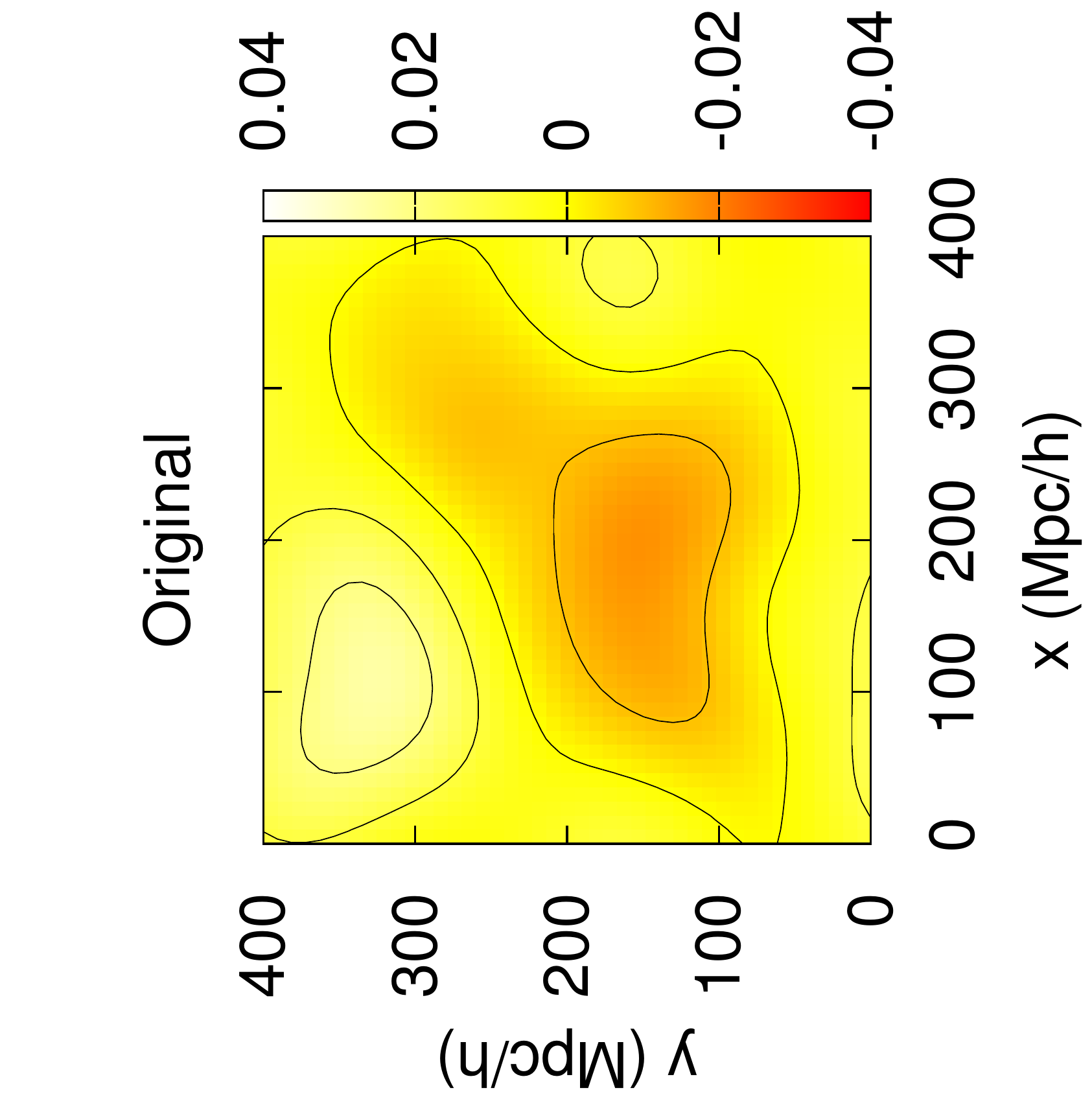}
\captionsetup{justification=centering}
\caption{$z=2$, $N_{\rm LOS}=200$, S/N=2, parallel to LOS}
\end{subfigure}
\begin{subfigure}{0.23\textwidth}
\includegraphics[scale=0.24, angle=-90]
{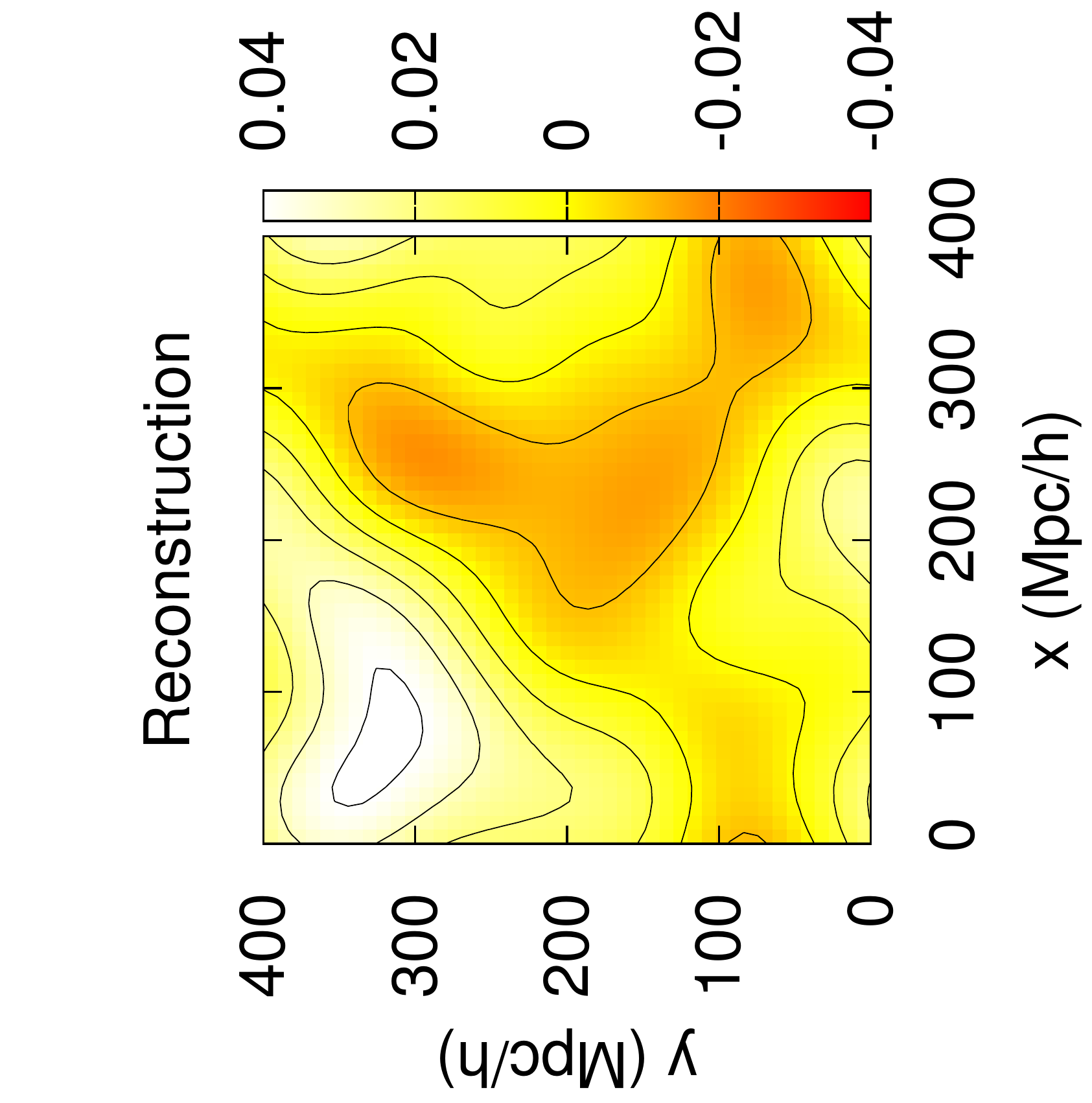}
\captionsetup{justification=centering}
\caption{$z=2$, $N_{\rm LOS}=200$, S/N=2, parallel to LOS}
\end{subfigure}

\vspace{1mm}

\caption{Slices extracted from the middle planes of the simulation cube are shown at $z=2$ with $N_{\rm LOS} = 200$, with Gaussian pixel noise added (S/N = 2). The top row shows slices perpendicular to LOSs, whereas the bottom row shows slices in the parallel direction. True field slices are given in (a) and (c), while (b) and (d) show reconstructed field slices. The smoothed reconstructed field recovers the general features of the simulation.}
\label{Fig:Slices_z2_noisySNR2_200LOS}
\end{figure}

\begin{figure}
\centering
\begin{subfigure}{0.23\textwidth}
\includegraphics[scale=0.24, angle=-90]
{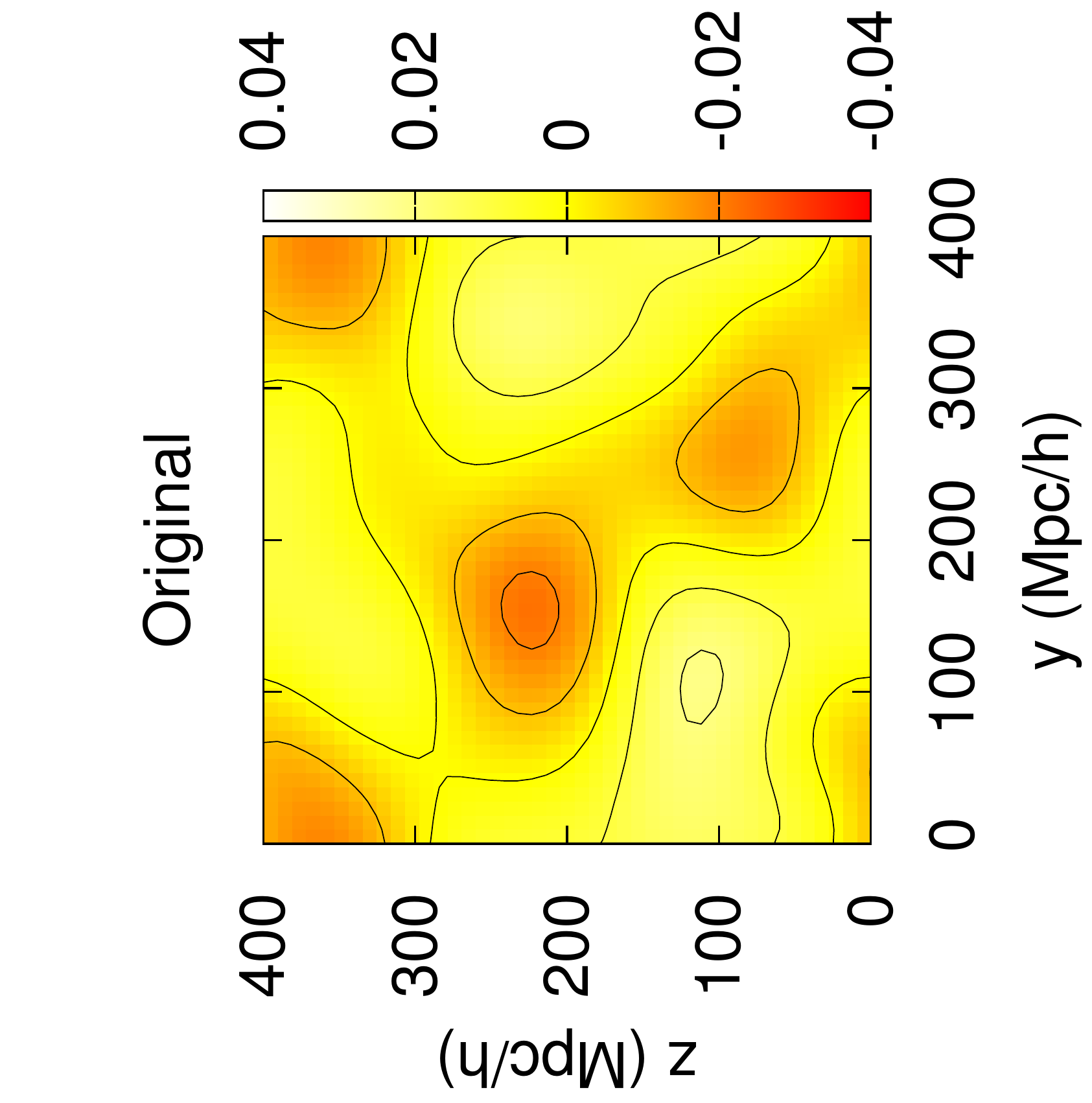}
\captionsetup{justification=centering}
\caption{$z=2$, $N_{\rm LOS}=200$, S/N=1, perpendicular to LOS}
\end{subfigure}
\begin{subfigure}{0.23\textwidth}
\includegraphics[scale=0.24, angle=-90]
{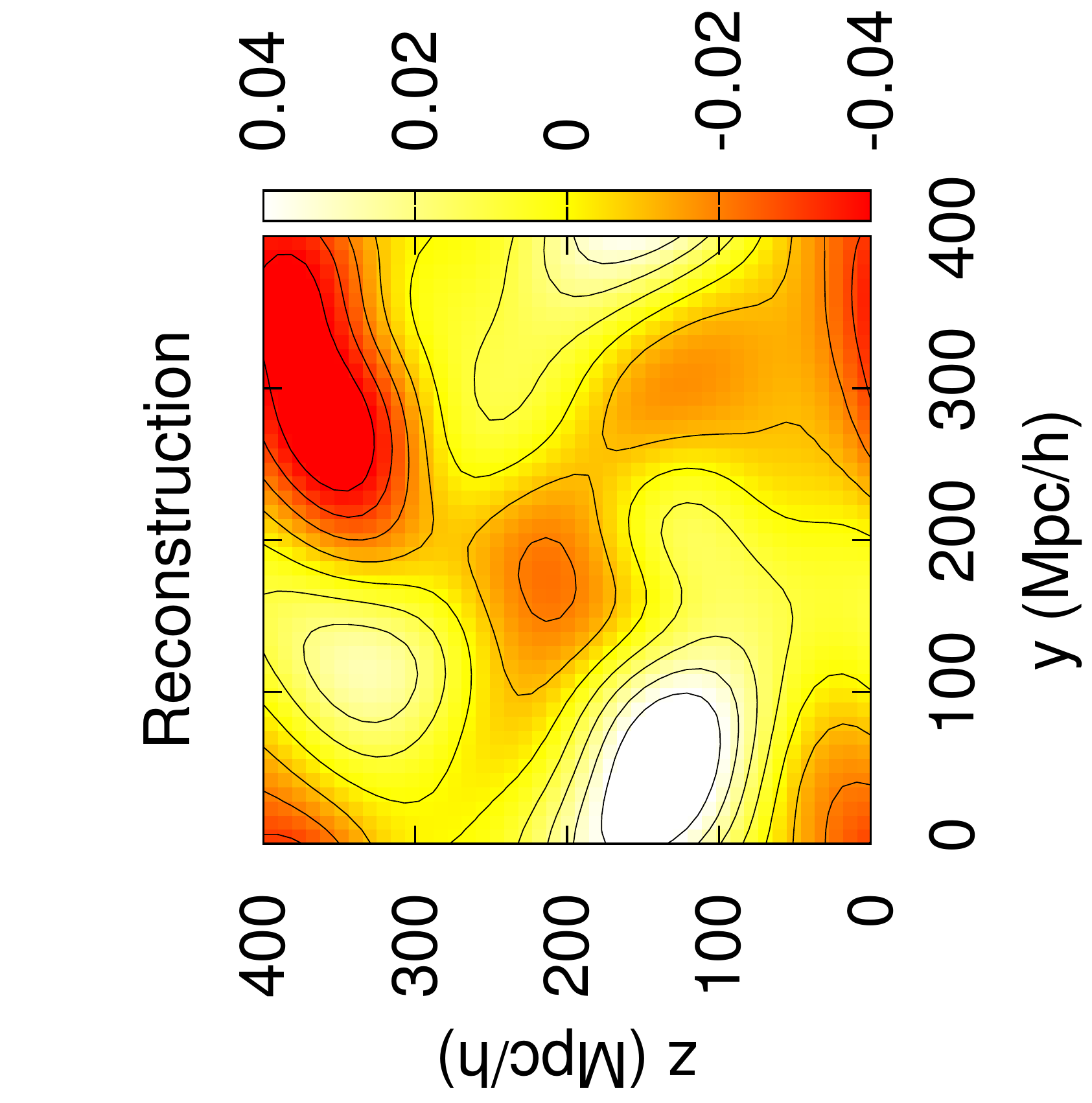}
\captionsetup{justification=centering}
\caption{$z=2$, $N_{\rm LOS}=200$, S/N=1, perpendicular to LOS}
\end{subfigure}

\vspace{4mm}

\begin{subfigure}{0.23\textwidth}
\includegraphics[scale=0.24, angle=-90]
{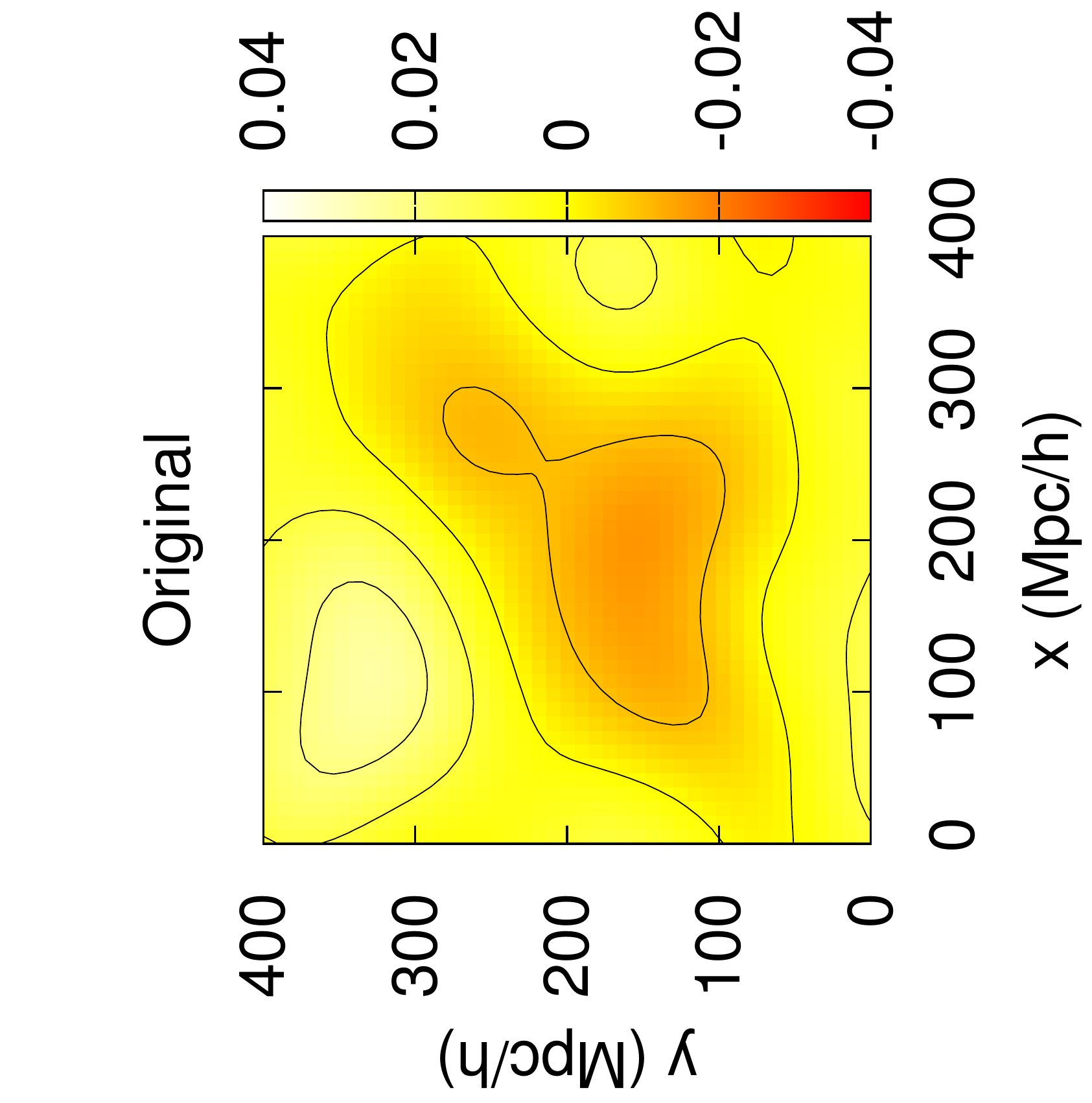}
\captionsetup{justification=centering}
\caption{$z=2$, $N_{\rm LOS}=200$, S/N=1, parallel to LOS}
\end{subfigure}
\begin{subfigure}{0.23\textwidth}
\includegraphics[scale=0.24, angle=-90]
{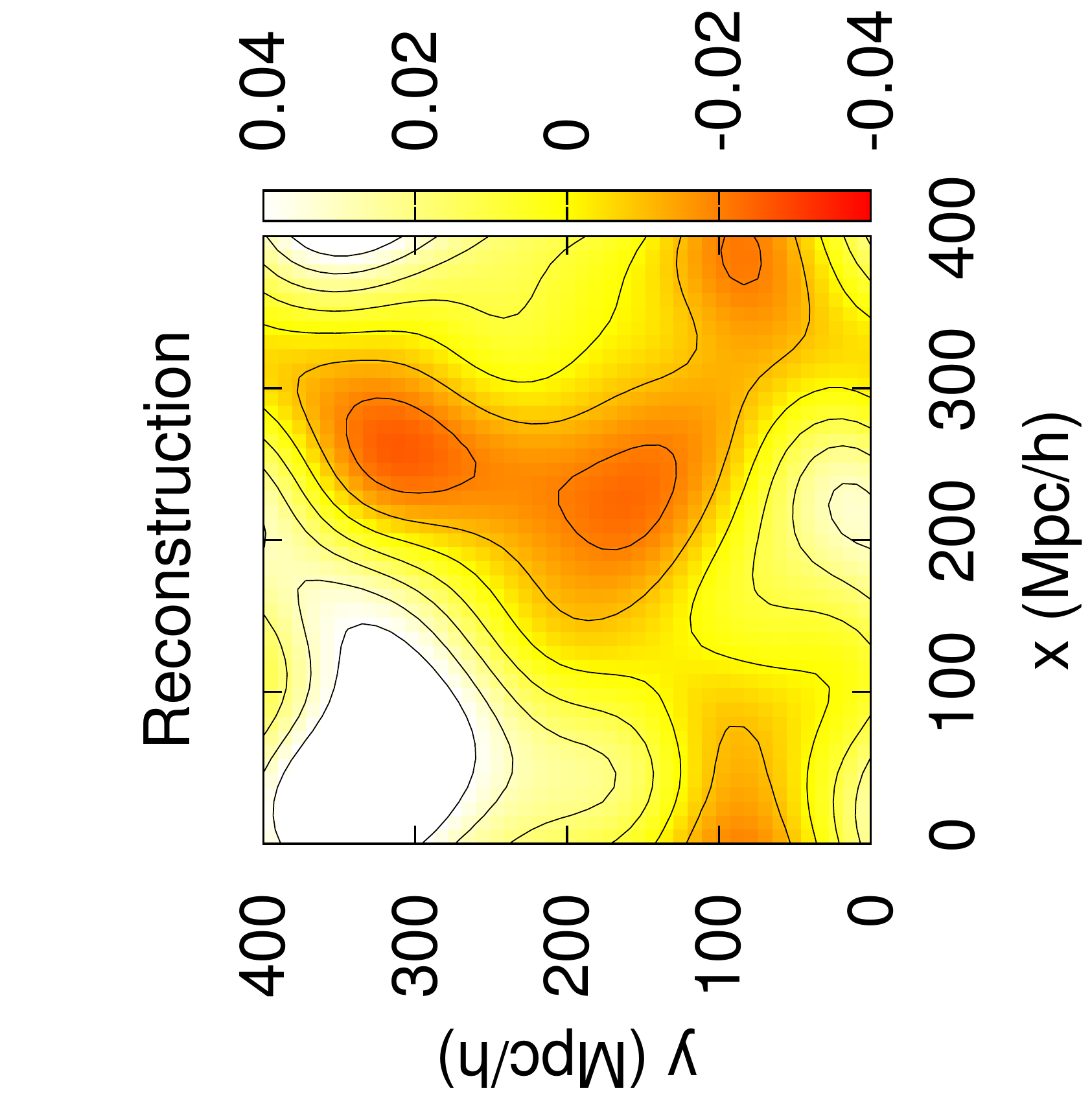}
\captionsetup{justification=centering}
\caption{$z=2$, $N_{\rm LOS}=200$, S/N=1, parallel to LOS}
\end{subfigure}

\vspace{1mm}

\caption{Slices extracted from the middle planes of the simulation cube are shown at $z=2$ with $N_{\rm LOS} = 200$, with Gaussian pixel noise added (S/N = 1). The top row shows slices perpendicular to LOSs, whereas the bottom row shows slices in the parallel direction. True field slices are given in (a) and (c), while (b) and (d) show reconstructed field slices. The smoothed reconstructed field recovers the general features of the simulation.}
\label{Fig:Slices_z2_noisySNR1_200LOS}
\end{figure}

\begin{figure}
\centering
\begin{subfigure}{0.23\textwidth}
\includegraphics[scale=0.24, angle=-90]
{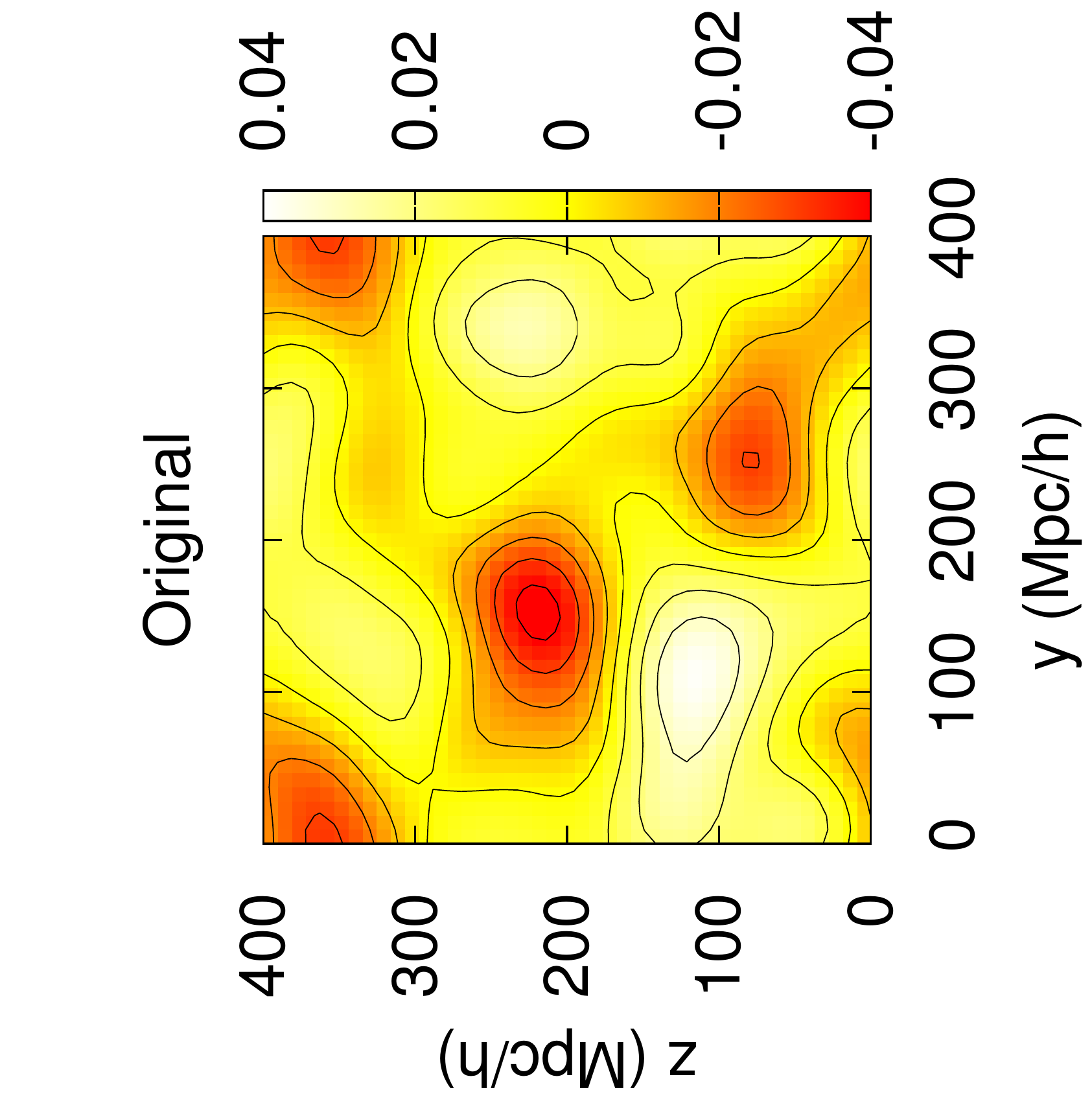}
\captionsetup{justification=centering}
\caption{$z=2$, $N_{\rm LOS}=400$, perpendicular to LOS}
\end{subfigure}
\begin{subfigure}{0.23\textwidth}
\includegraphics[scale=0.24, angle=-90]
{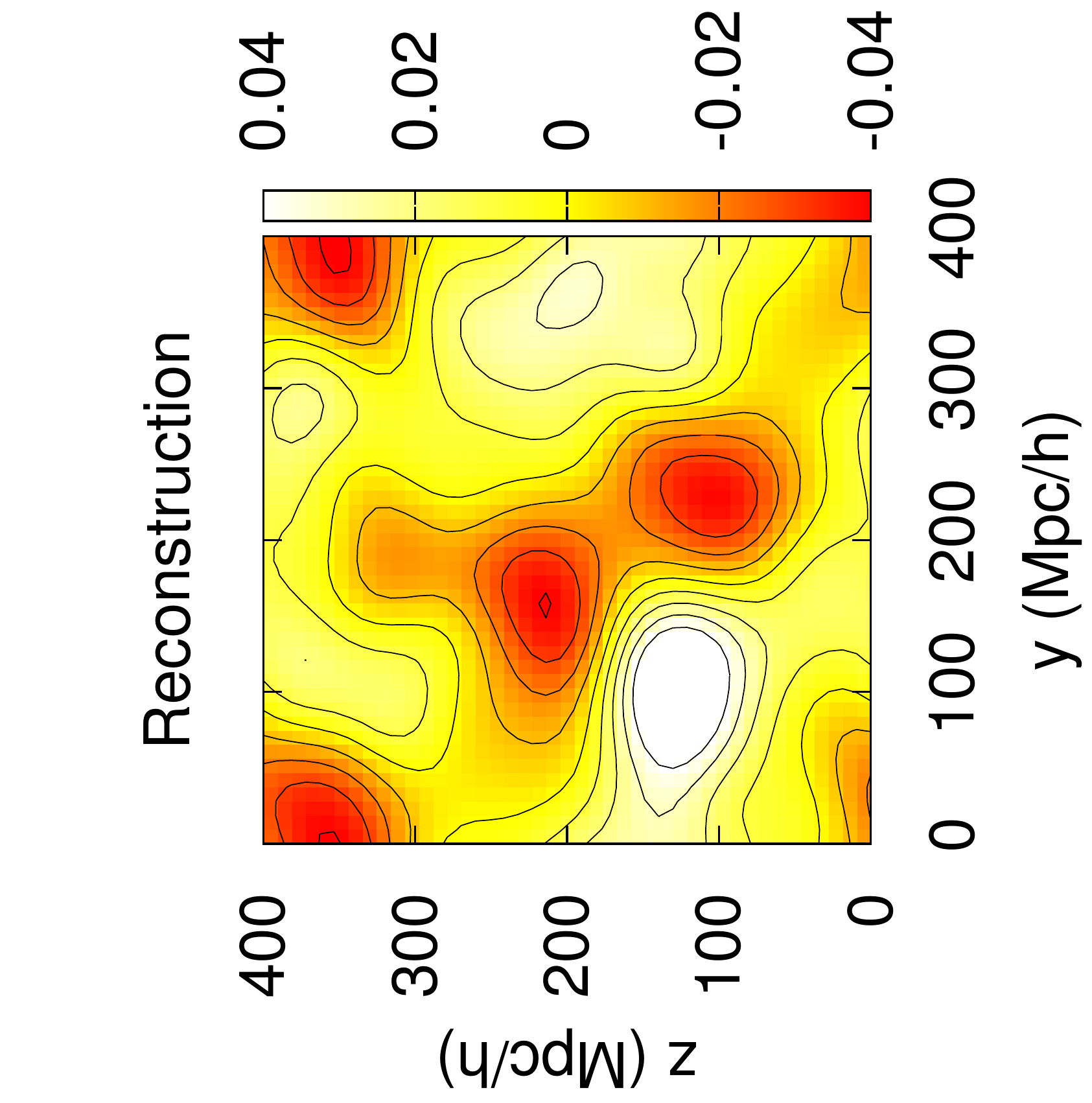}
\captionsetup{justification=centering}
\caption{$z=2$, $N_{\rm LOS}=400$, perpendicular to LOS}
\end{subfigure}

\vspace{4mm}

\begin{subfigure}{0.23\textwidth}
\includegraphics[scale=0.24, angle=-90]
{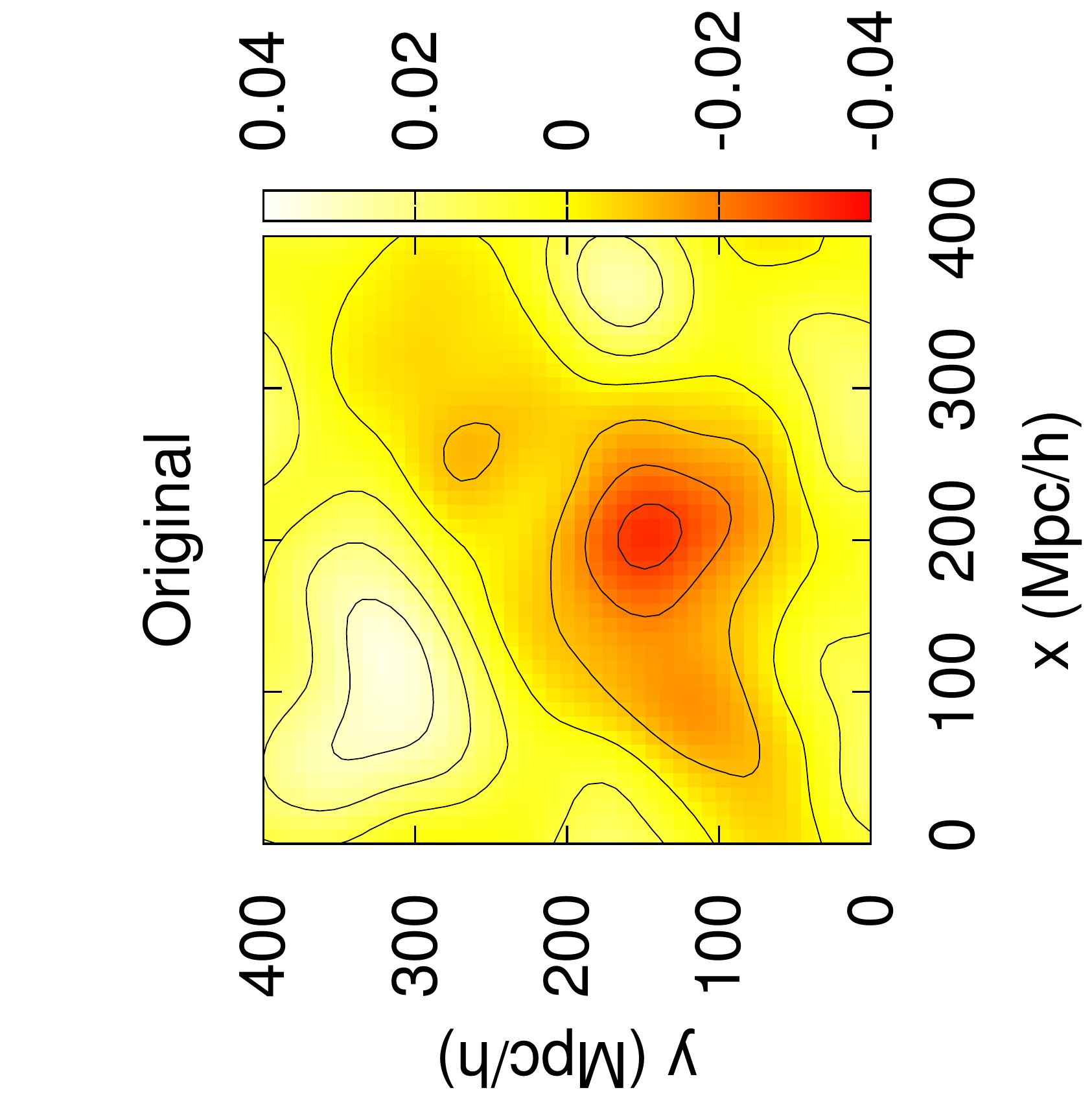}
\captionsetup{justification=centering}
\caption{$z=2$, $N_{\rm LOS}=400$, parallel to LOS}
\end{subfigure}
\begin{subfigure}{0.23\textwidth}
\includegraphics[scale=0.24, angle=-90]
{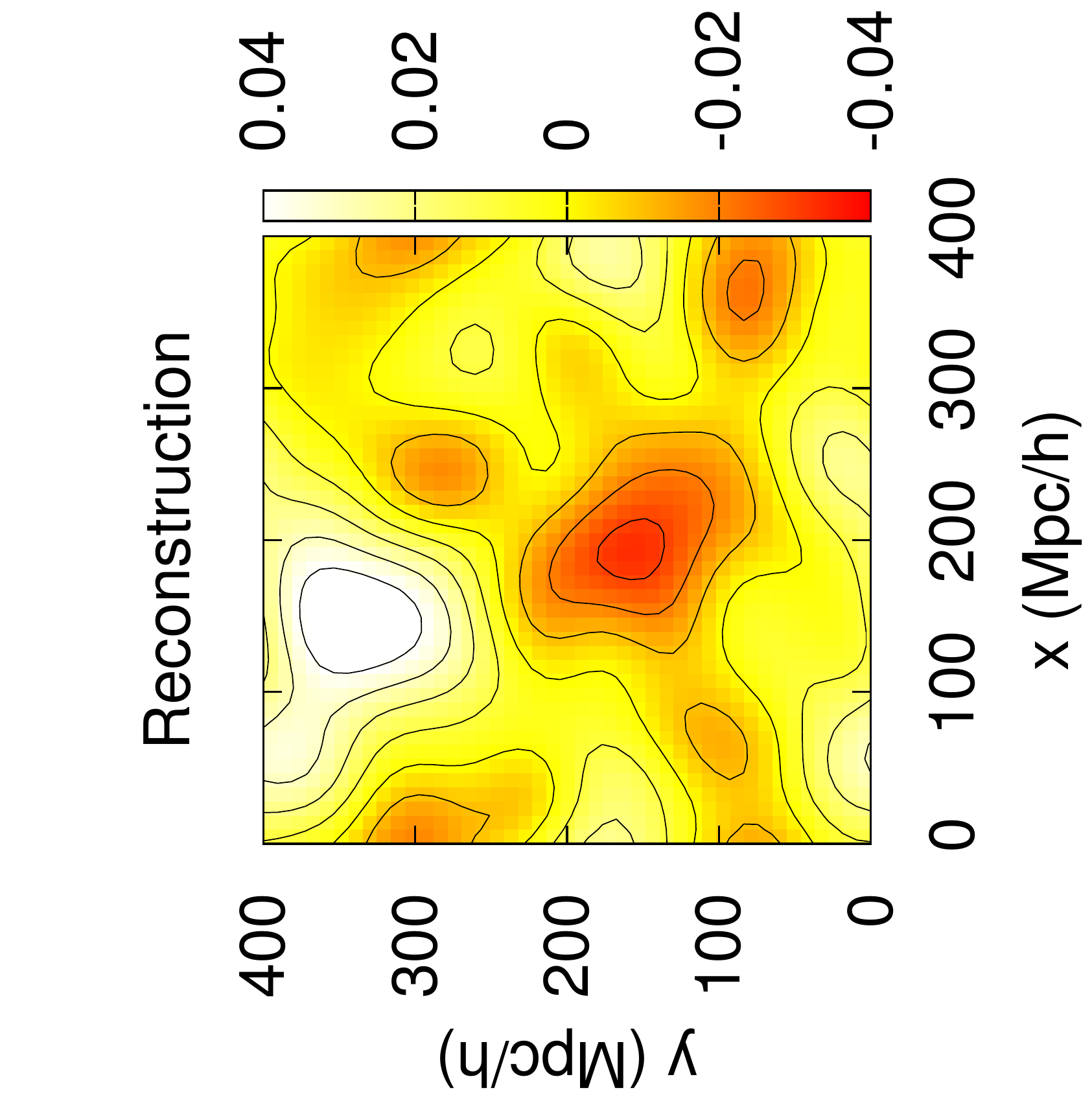}
\captionsetup{justification=centering}
\caption{$z=2$, $N_{\rm LOS}=400$, parallel to LOS}
\end{subfigure}

\vspace{1mm}

\caption{Slices extracted from the middle planes of the simulation cube are shown at $z=2$ with $N_{\rm LOS} = 400$, without pixel noise. The top row shows slices perpendicular to LOSs, whereas the bottom row shows slices in the parallel direction. True field slices are given in (a) and (c), while (b) and (d) show reconstructed field slices. The smoothed reconstructed field recovers the general features of the simulation.}
\label{Fig:Slices_z2_nonoise_400LOS}
\end{figure}

\begin{figure}
\centering
\begin{subfigure}{0.23\textwidth}
\includegraphics[scale=0.24, angle=-90]
{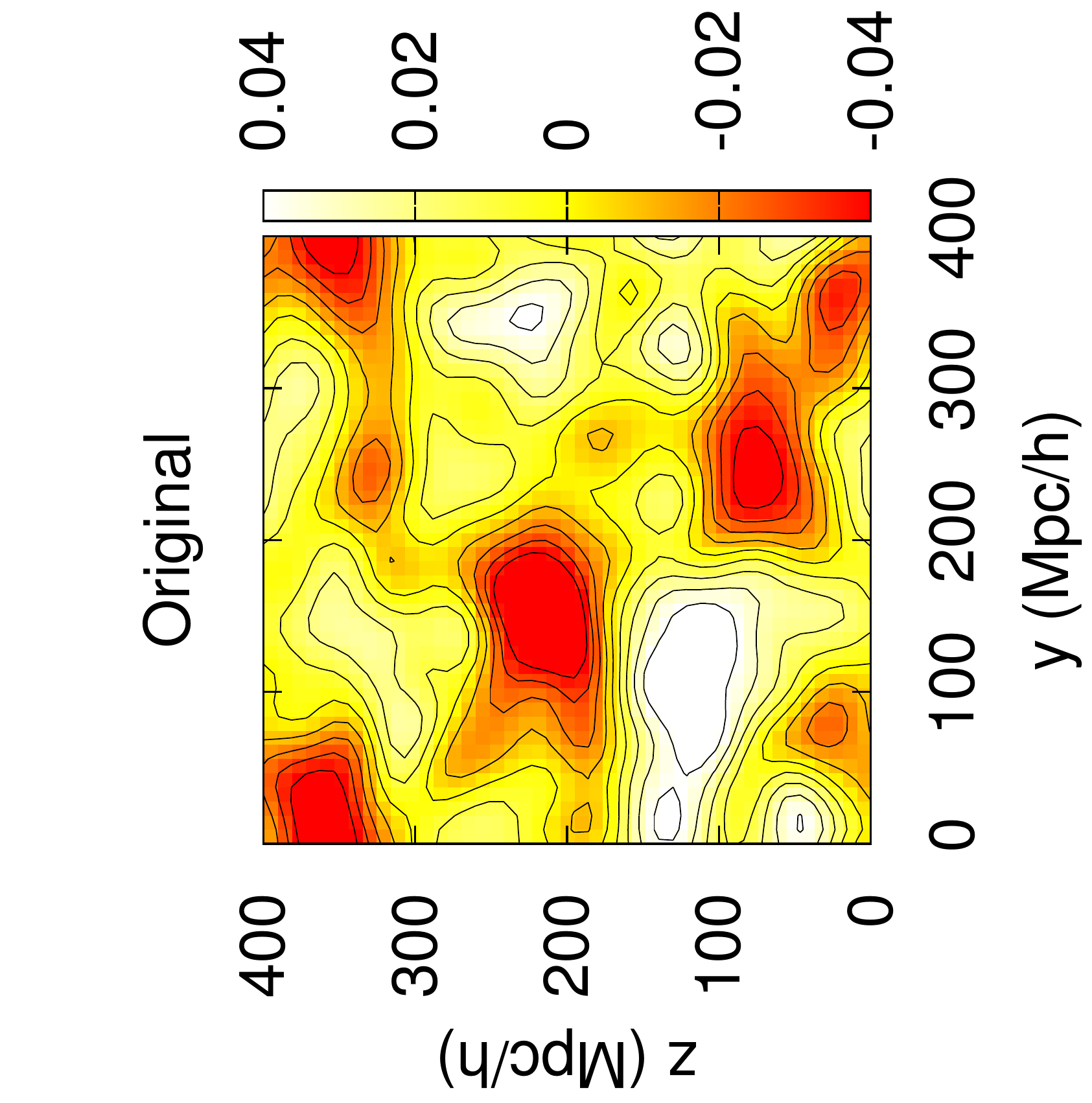}
\captionsetup{justification=centering}
\caption{$z=2$, $N_{\rm LOS}=1000$, perpendicular to LOS}
\end{subfigure}
\begin{subfigure}{0.23\textwidth}
\includegraphics[scale=0.24, angle=-90]
{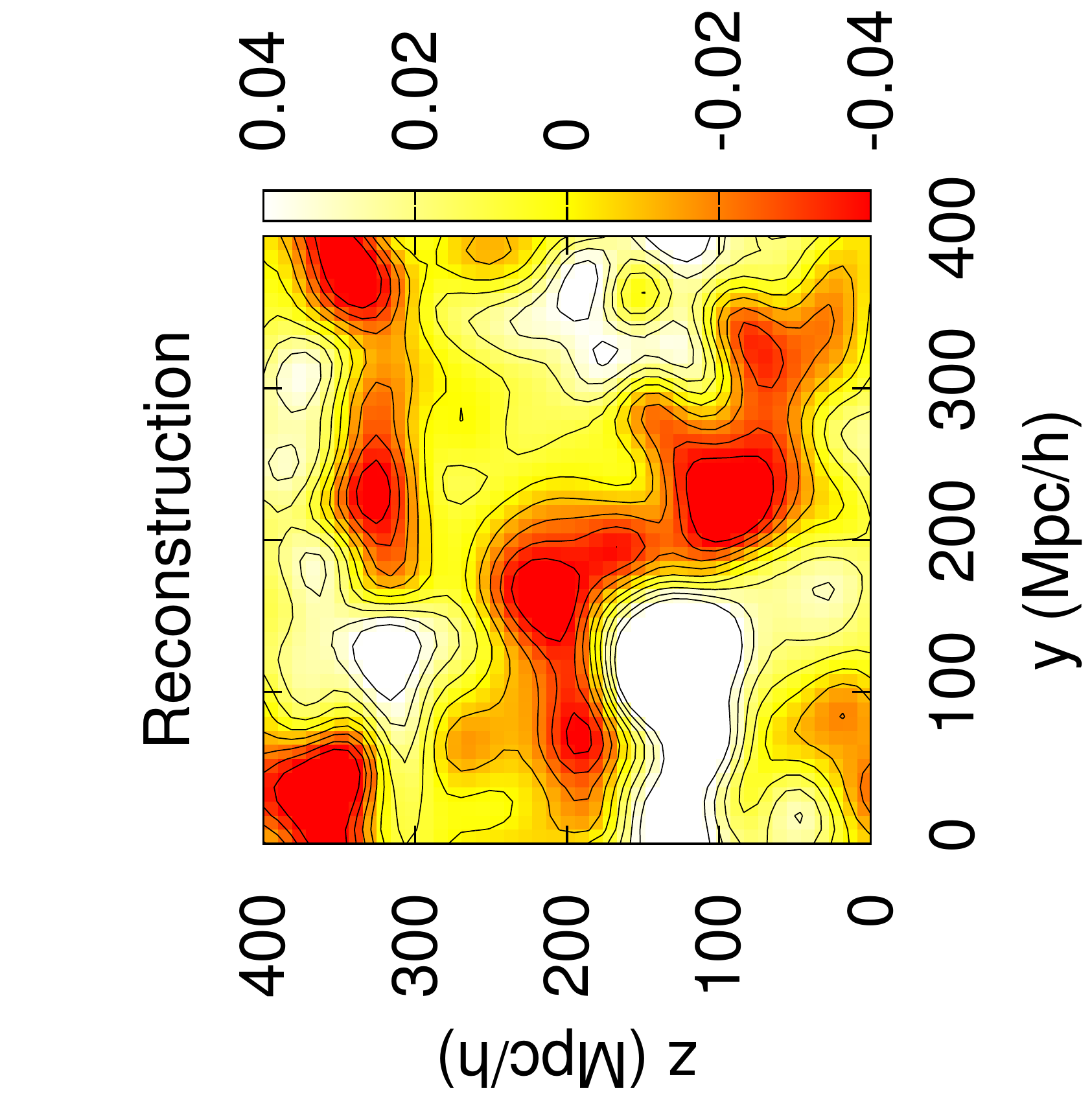}
\captionsetup{justification=centering}
\caption{$z=2$, $N_{\rm LOS}=1000$, perpendicular to LOS}
\end{subfigure}

\vspace{4mm}

\begin{subfigure}{0.23\textwidth}
\includegraphics[scale=0.24, angle=-90]
{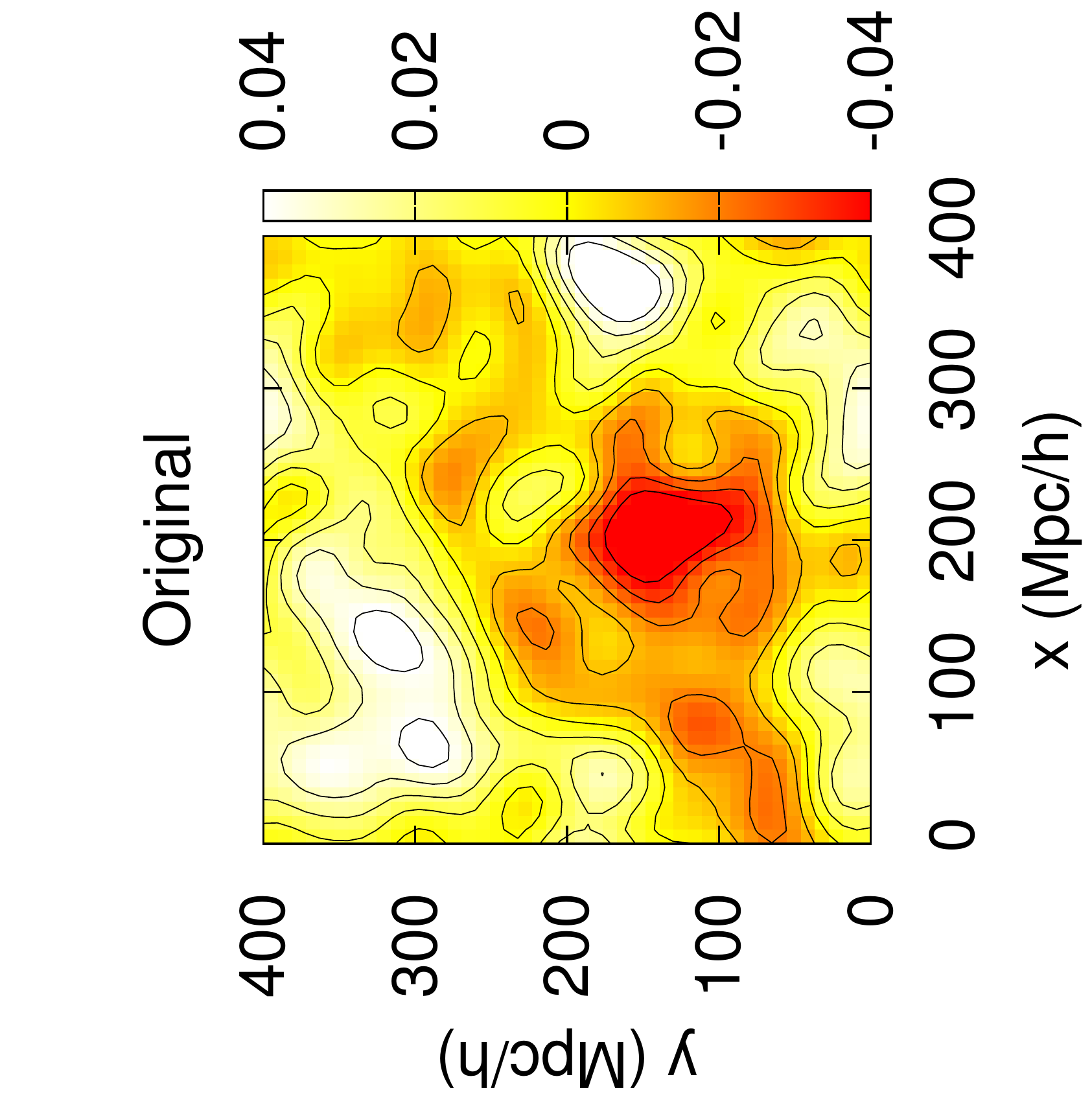}
\captionsetup{justification=centering}
\caption{$z=2$, $N_{\rm LOS}=1000$, parallel to LOS}
\end{subfigure}
\begin{subfigure}{0.23\textwidth}
\includegraphics[scale=0.24, angle=-90]
{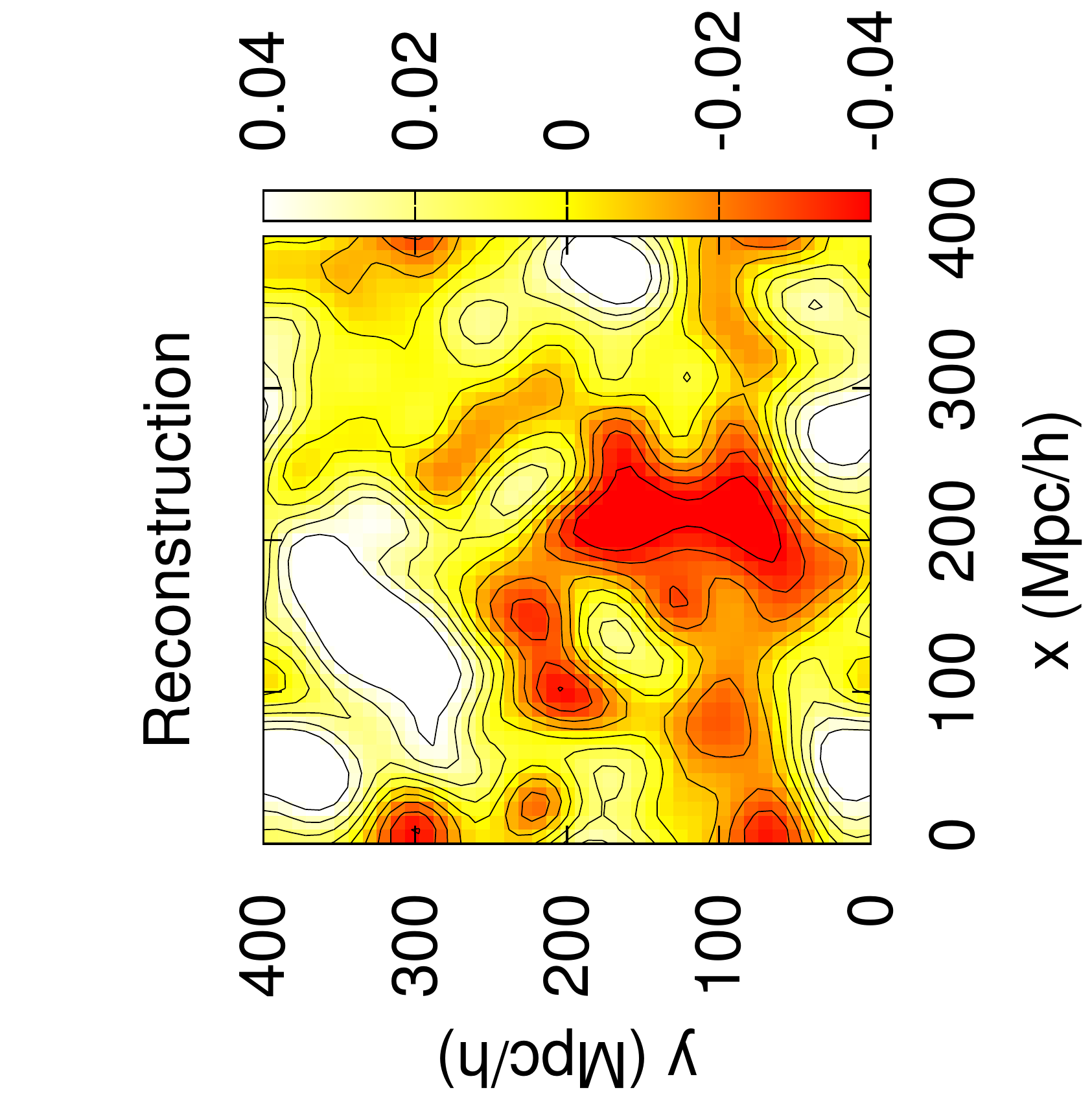}
\captionsetup{justification=centering}
\caption{$z=2$, $N_{\rm LOS}=1000$, parallel to LOS}
\end{subfigure}

\vspace{1mm}

\caption{Slices extracted from the middle planes of the simulation cube are shown at $z=2$ with $N_{\rm LOS} = 1000$, without pixel noise. The top row shows slices perpendicular to LOSs, whereas the bottom row shows slices in the parallel direction. True field slices are given in (a) and (c), while (b) and (d) show reconstructed field slices. The smoothed reconstructed field recovers the general features of the simulation.}
\label{Fig:Slices_z2_nonoise_1000LOS}
\end{figure}


\begin{figure}
\centering
\begin{subfigure}{0.23\textwidth}
\includegraphics[scale=0.24, angle=-90]
{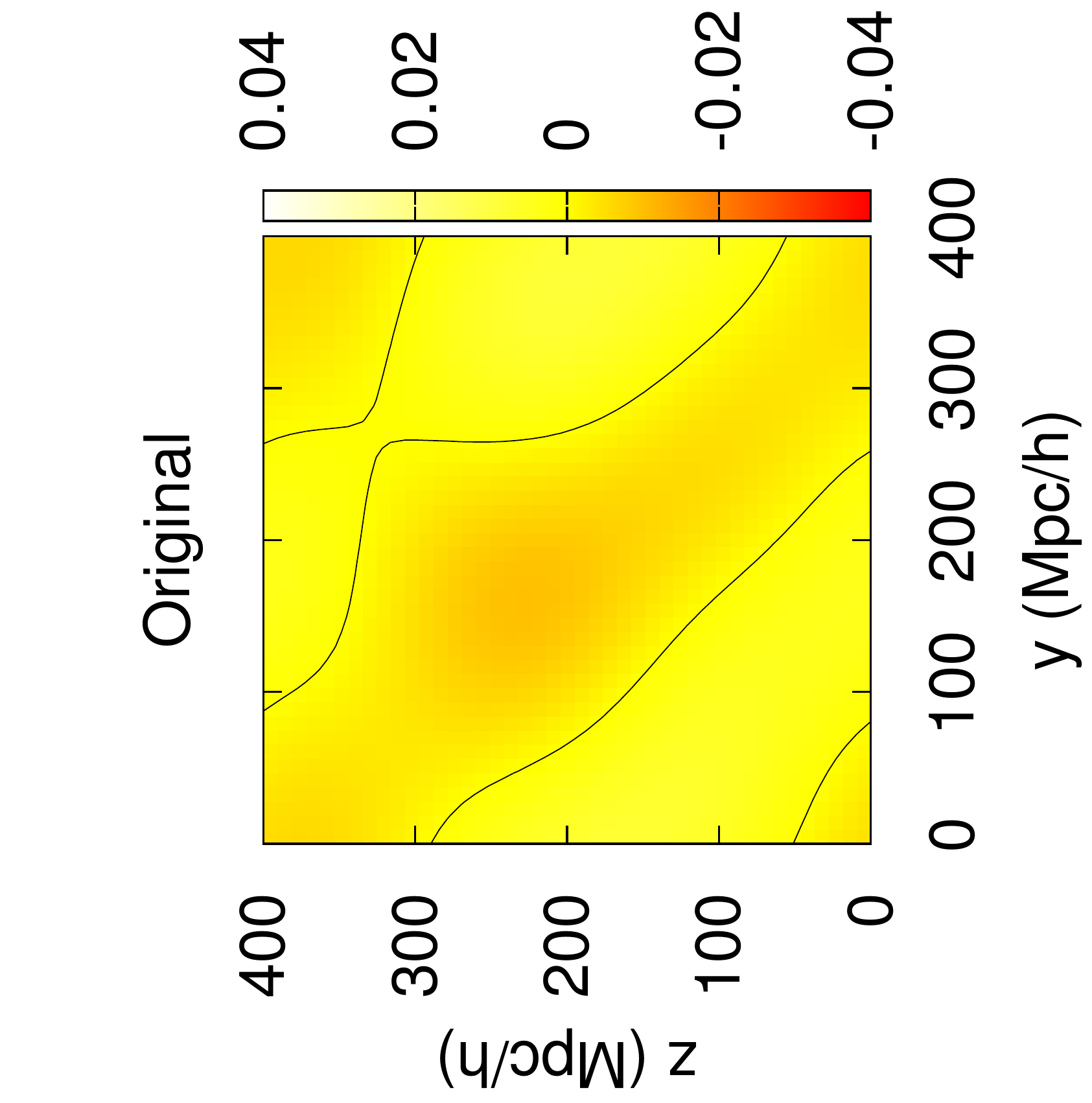}
\captionsetup{justification=centering}
\caption{$z=3$, $N_{\rm LOS}=60$, perpendicular to LOS}
\end{subfigure}
\begin{subfigure}{0.23\textwidth}
\includegraphics[scale=0.24, angle=-90]
{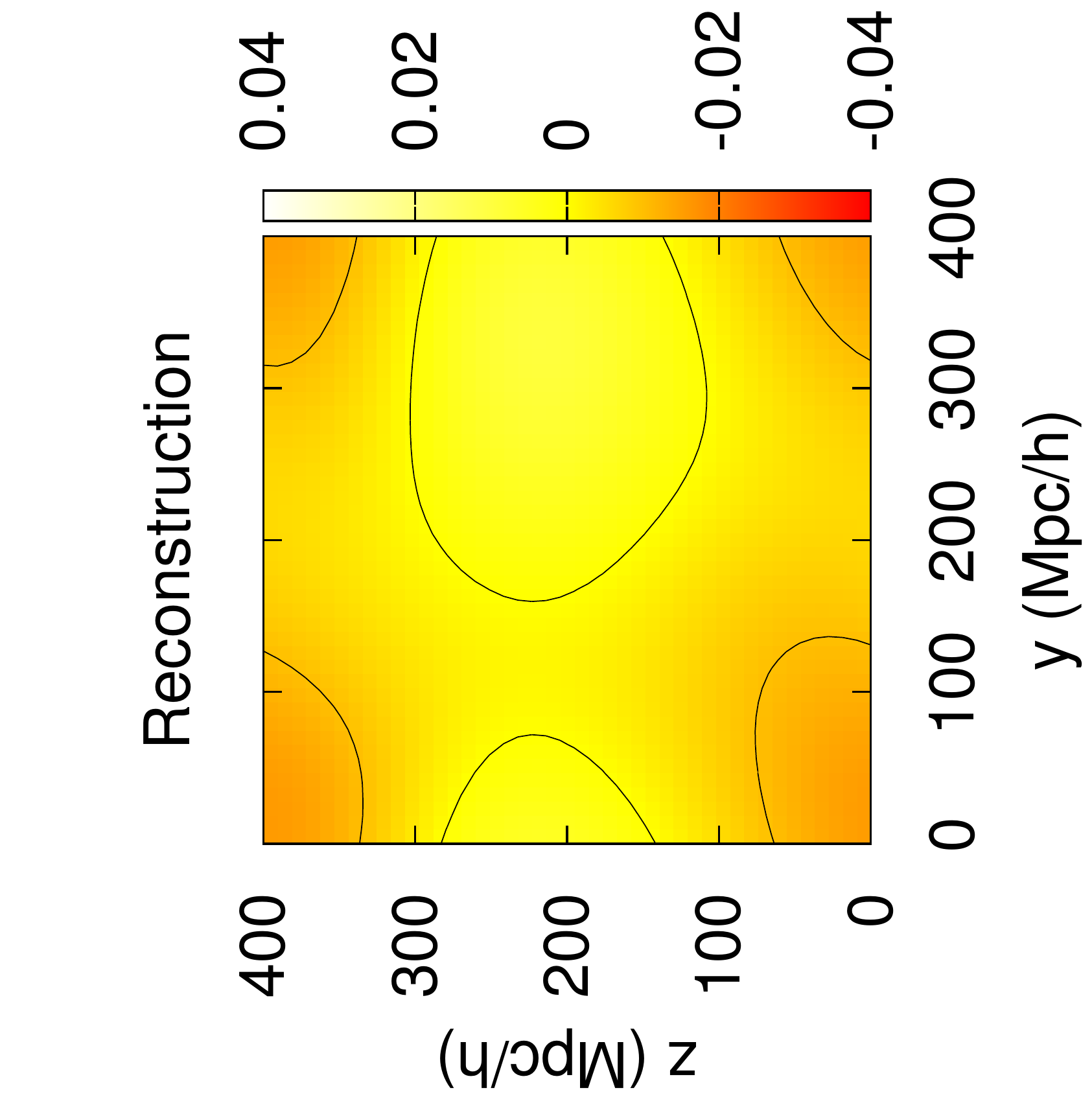}
\captionsetup{justification=centering}
\caption{$z=3$, $N_{\rm LOS}=60$, perpendicular to LOS}
\end{subfigure}

\vspace{4mm}

\begin{subfigure}{0.23\textwidth}
\includegraphics[scale=0.24, angle=-90]
{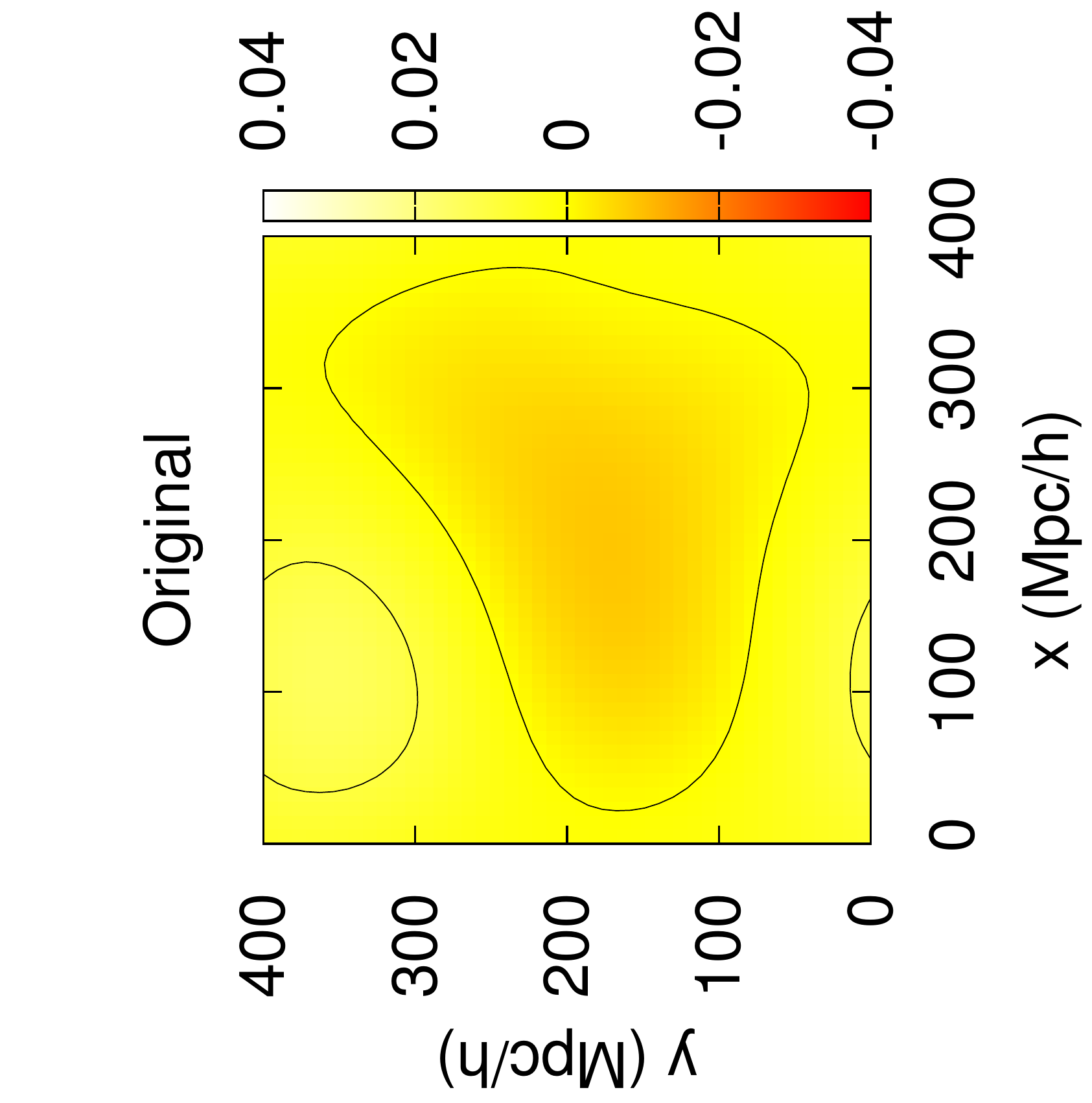}
\captionsetup{justification=centering}
\caption{$z=3$, $N_{\rm LOS}=60$, parallel to LOS}
\end{subfigure}
\begin{subfigure}{0.23\textwidth}
\includegraphics[scale=0.24, angle=-90]
{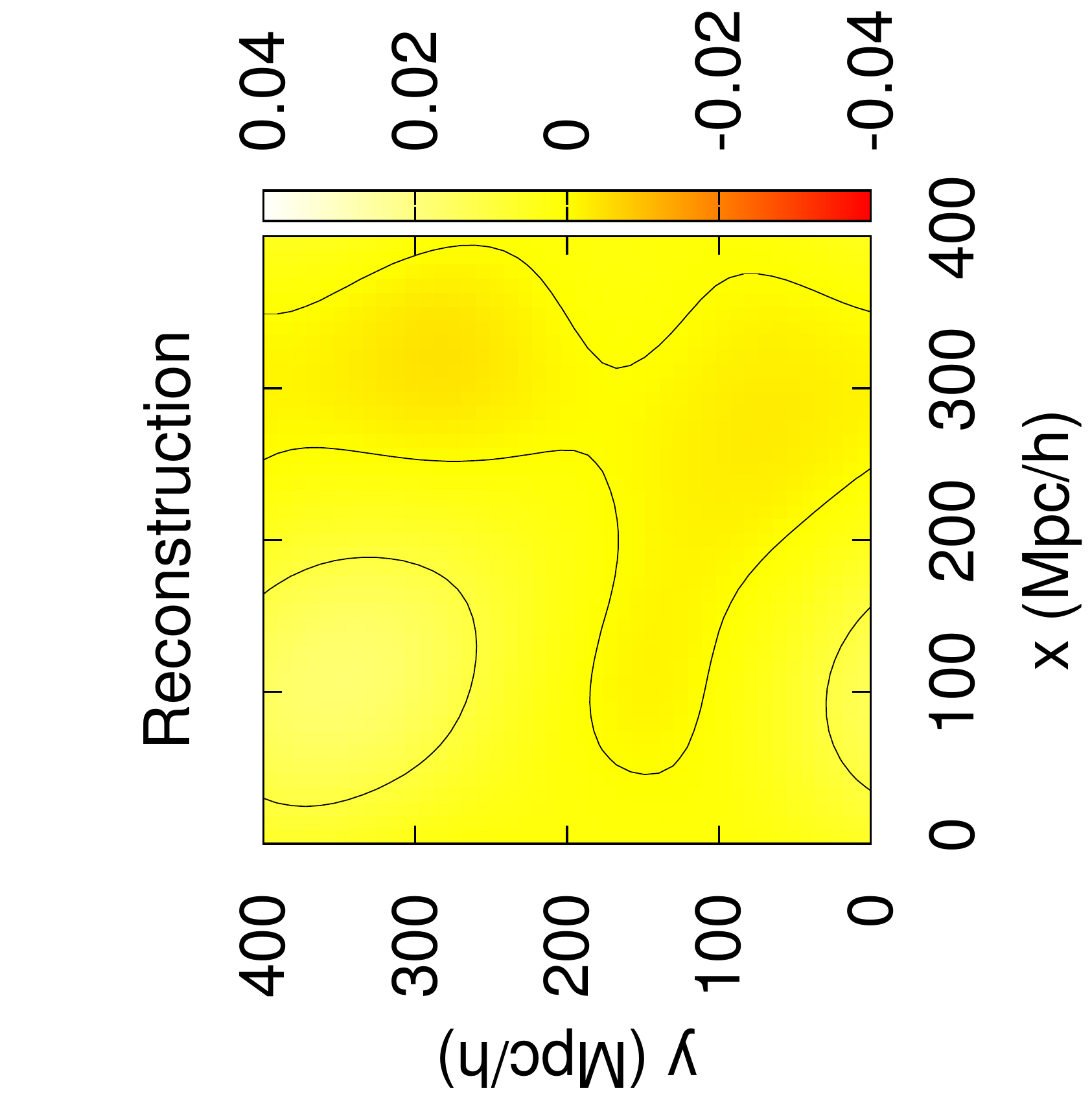}
\captionsetup{justification=centering}
\caption{$z=3$, $N_{\rm LOS}=60$, parallel to LOS}
\end{subfigure}

\vspace{1mm}

\caption{Slices extracted from the middle planes of the simulation cube are shown at $z=3$ with $N_{\rm LOS} = 60$, without pixel noise. The top row shows slices perpendicular to LOSs, whereas the bottom row shows slices in the parallel direction. True field slices are given in (a) and (c), while (b) and (d) show reconstructed field slices. The smoothed reconstructed field cannot recover the general features of the simulation well when the areal density of the absorption skewers is low.}
\label{Fig:Slices_z3_nonoise_60LOS}
\end{figure}

\begin{figure}
\centering
\begin{subfigure}{0.23\textwidth}
\includegraphics[scale=0.24, angle=-90]
{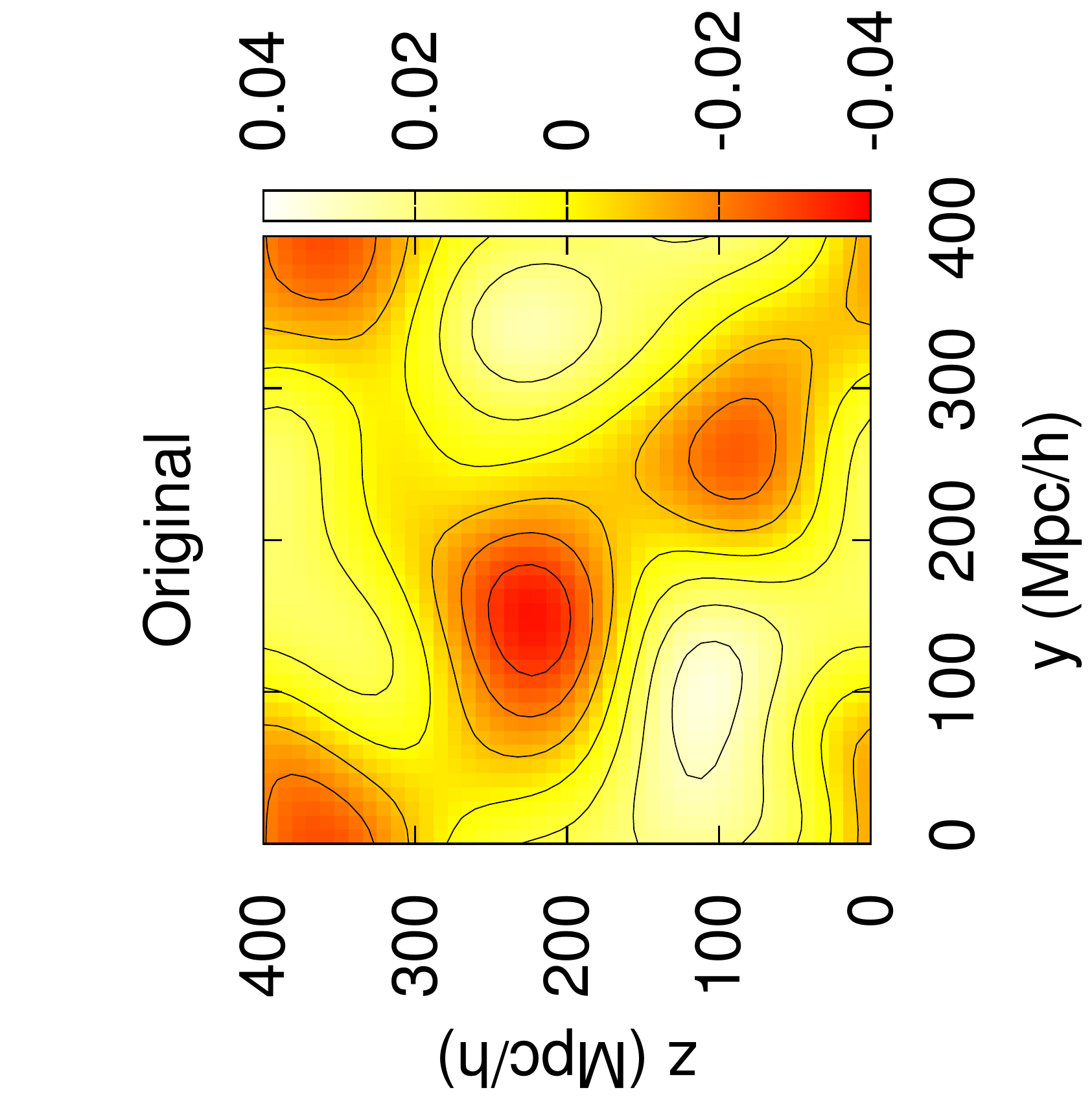}
\captionsetup{justification=centering}
\caption{$z=3$, $N_{\rm LOS}=200$, perpendicular to LOS}
\end{subfigure}
\begin{subfigure}{0.23\textwidth}
\includegraphics[scale=0.24, angle=-90]
{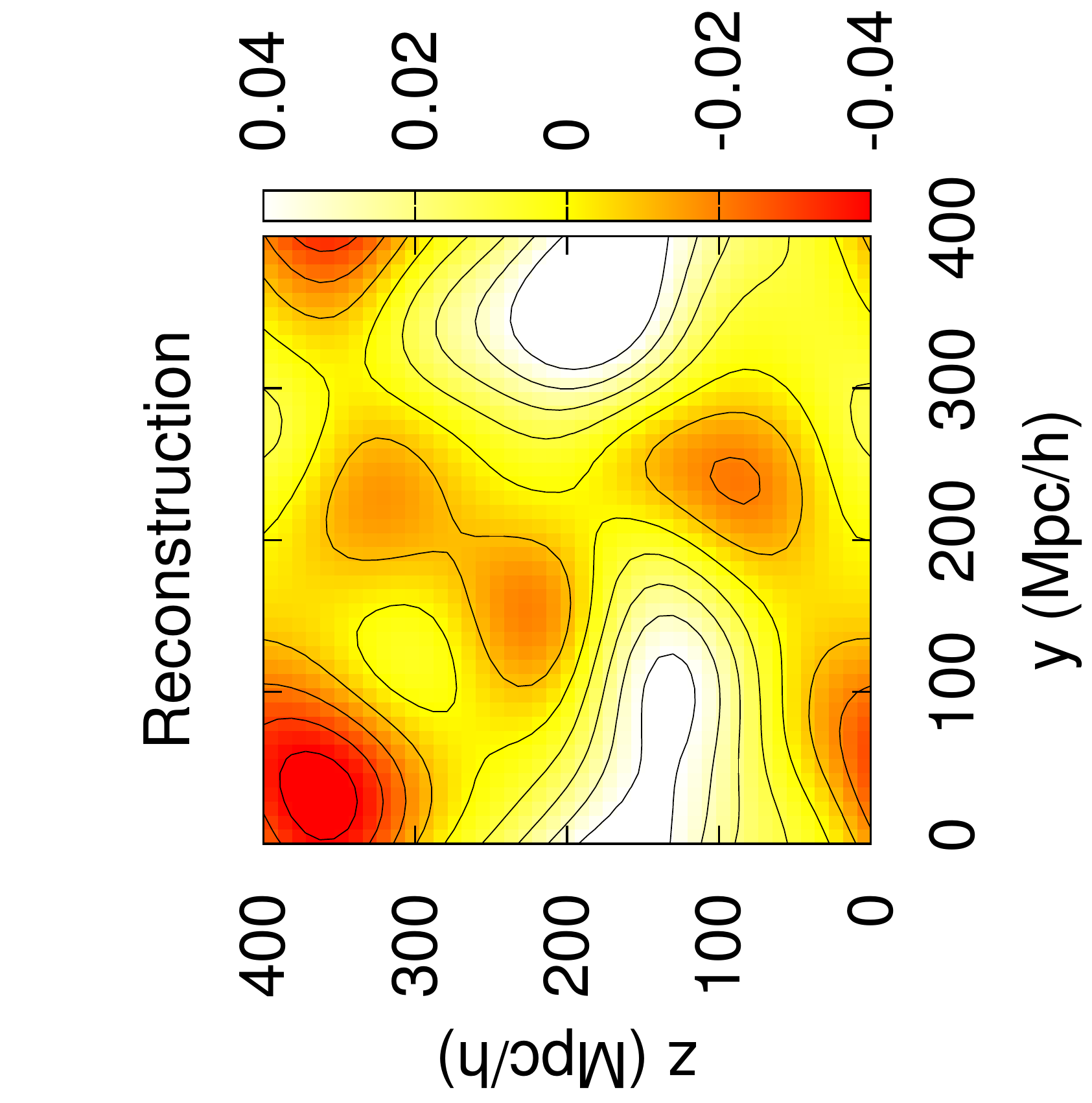}
\captionsetup{justification=centering}
\caption{$z=3$, $N_{\rm LOS}=200$, perpendicular to LOS}
\end{subfigure}

\vspace{4mm}

\begin{subfigure}{0.23\textwidth}
\includegraphics[scale=0.24, angle=-90]
{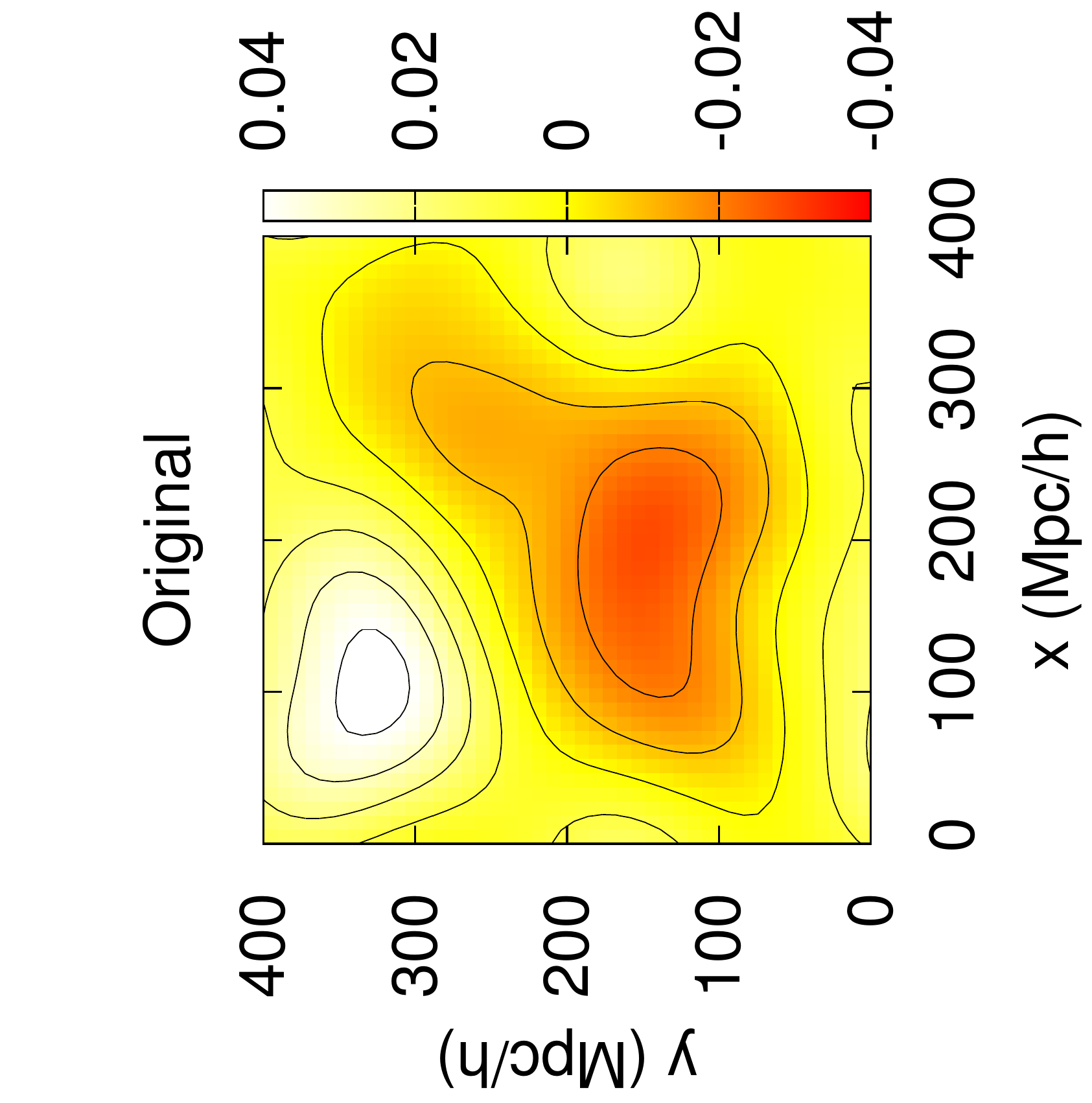}
\captionsetup{justification=centering}
\caption{$z=3$, $N_{\rm LOS}=200$, parallel to LOS}
\end{subfigure}
\begin{subfigure}{0.23\textwidth}
\includegraphics[scale=0.24, angle=-90]
{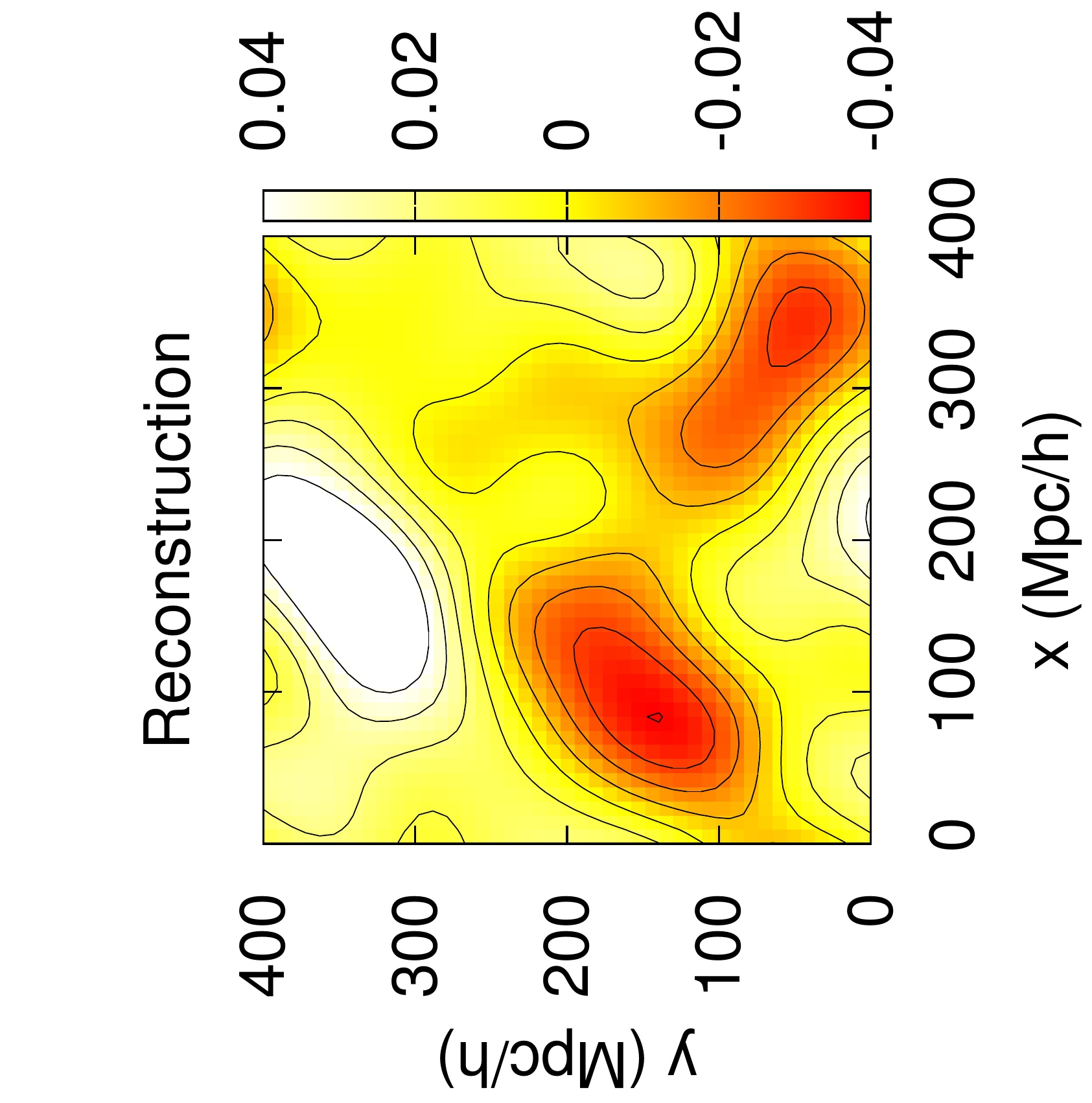}
\captionsetup{justification=centering}
\caption{$z=3$, $N_{\rm LOS}=200$, parallel to LOS}
\end{subfigure}

\vspace{1mm}

\caption{Slices extracted from the middle planes of the simulation cube are shown at $z=3$ with $N_{\rm LOS} = 200$, without pixel noise. The top row shows slices perpendicular to LOSs, whereas the bottom row shows slices in the parallel direction. True field slices are given in (a) and (c), while (b) and (d) show reconstructed field slices. The smoothed reconstructed field recovers the general features of the simulation, although the quality of the reconstruction is lower than that with the $z=2$ data set.}
\label{Fig:Slices_z3_nonoise_200LOS}
\end{figure}

We now turn to a visual comparison of the structures in the real and 
reconstructed maps. The three-dimensional datacubes have $z$ axes oriented
parallel to the line of sight, and $x$ and $y$ axes perpendicular to it.
The sampling of pixels in a mock data set is 
therefore different depending on the plot orientation, and this could
influence the recovery of structure. We therefore show two orientations
for each plot, one in the $y-z$ plane (an ''x'' slice) and one
in the $x-y$ plane (a ''z'' slice). In our plots we show the flux contrast
in a slice of thickness one grid cell. As our volumes are
44 cells on a side, this corresponds to a thickness $400/43 = 9 \; \hmpc$.

We have also seen in Section 4.1 that there is a bias in the reconstructed
field which leads it to have higher contrast. Changes in $N_{\rm LOS}$ and 
adjusting the correlation lengths do not alter this and so we follow
\citep{lee2014a} in applying a bias correction before visualizing the
fields.

In Figs \ref{Fig:Slices_z2_nonoise_200LOS} through \ref{Fig:Slices_z3_nonoise_200LOS}
we present image slices through the simulation volume.
The images show the flux contrast $\delta_{F}= (F/\left<F\right>)-1$, which means
that low values correspond to high values of the matter density.  
The red colour shows these higher density regions and white those of lower density.
The image slices are taken from the centre of the cube in directions 
parallel and perpendicular to the LOSs. 
As explained in Section 3, the motivation for choosing to display slices
through the centre of the cube is because we
are not using information about the periodic boundary conditions in
the simulation when carrying out the reconstruction. The edges of each
image slice will therefore give an idea of how well the reconstruction
would succeed at the edges of a survey volume. We have checked other
random slices and verified that the reconstruction recovers the general features
of the original field, even when close to edges of the simulation volume.

In Fig. \ref{Fig:Slices_z2_nonoise_200LOS} we can see the
results for our lowest density of sightlines (we have $N_{\rm LOS}=200$),
at redshift $z=2$. We can see that the general morphology of the field is
recognizably similar in the true and reconstructed maps. In detail, 
the maps have some differences, but the maxima, minima and their gross
shapes are fairly well reproduced, and one could therefore expect that 
observational data from the BOSS survey (which has 
approximately this number density of quasars) would yield visually quite
accurate maps of the large scale structure, at least when smoothed on
the relevant filter scale (a filter of 39.6 $\hmpc$ was used here.).

The top and bottom rows of Fig. \ref{Fig:Slices_z2_nonoise_200LOS} show
results for slices parallel and perpendicular to the line of sight. We see
no obvious difference in the fidelity of reconstruction for each, 
and there is no obvious sign of the discrete sampling of the
field by pixels and sightlines (which is
different for the top and bottom rows). General features of the field
are recovered well, especially for mildly dense regions.

Figs \ref{Fig:Slices_z2_noisySNR2_200LOS} and \ref{Fig:Slices_z2_noisySNR1_200LOS}
demonstrate the effect of adding uncorrelated Gaussian noise to the flux field in order to better mimic 
observational data, and how the fidelity of the reconstruction changes when noise is introduced.
Noise levels (S/N = 1 or 2) indicate the amount of noise for a simulation pixel ($\sim$ 0.76 $\hmpc$) wide.
Since our pixels are rebinned to $\sim$ 9 $\hmpc$, the added noise is reduced by a factor of 3.4. Hence, the difference 
between the true fields before and after adding noise is small. However, the reconstruction is sensitive to the 
amount of noise, therefore the fidelity of the noisy reconstruction is noticeably worse, especially for overdense and underdense regions.

We increase the density of the sightlines in Figs \ref{Fig:Slices_z2_nonoise_400LOS} and \ref{Fig:Slices_z2_nonoise_1000LOS}. As a result, the quality of the reconstruction is visually better. Although the reconstruction
code does not take into account the periodic boundary conditions of the simulation, the fields are comparable even at the edges. This is likely due to lower smoothing levels, as the smoothing level scales inversely with $N_{\rm LOS}$.

Fluctuations in the flux field are greater at higher redshifts. As Figs \ref{Fig:Slices_z3_nonoise_60LOS} and \ref{Fig:Slices_z3_nonoise_200LOS} clearly show, this results in a decrease in the fidelity of the reconstruction. It is obvious that at redshift $z=3$, the LOS density of $N_{\rm LOS} = 60$ is not good enough to yield a comparable reconstructed field. For observational data, we naturally expect a better map at $z=2$ than at $z=3$, as the LOS density is higher at $z=2$. In this study, although the LOS density is set to be the same at both redshifts, we get a better recovery of the field at $z=2$. 

\begin{figure*}
\centering
\begin{subfigure}{0.49\textwidth}
\includegraphics[scale=0.35]{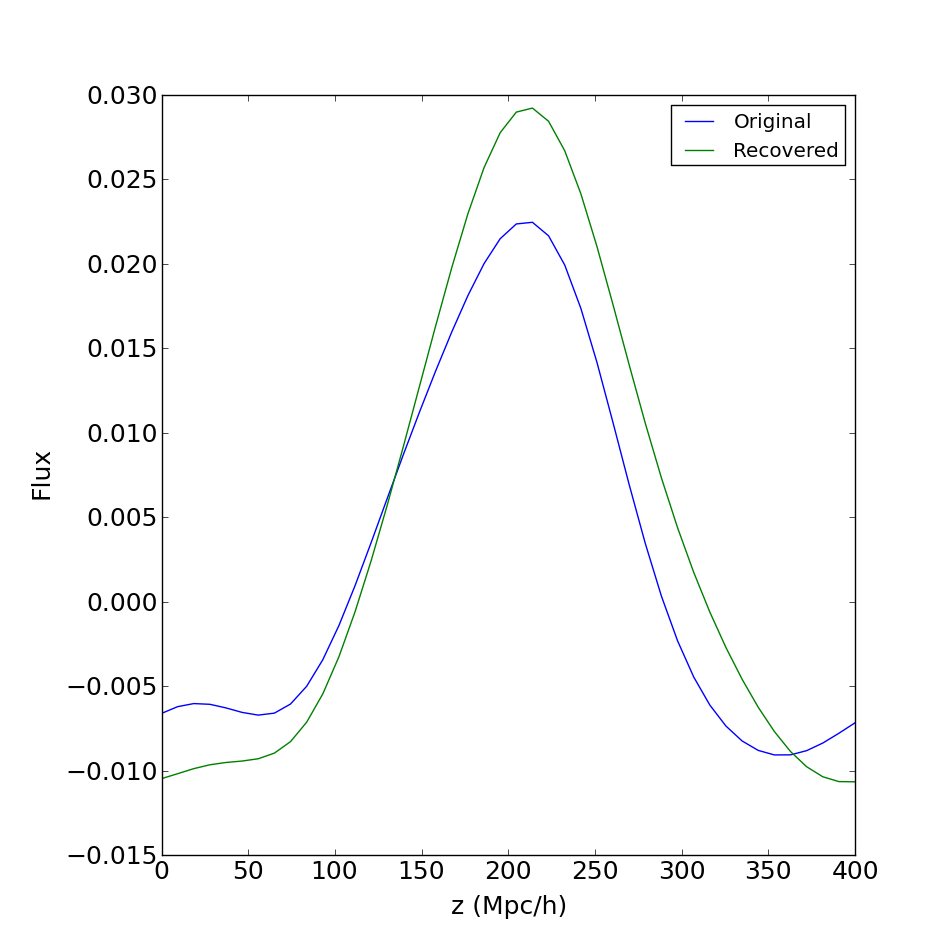}
\caption{Spectrum at (130.2, 297.7) $\hmpc$.}
\end{subfigure}
\begin{subfigure}{0.49\textwidth}
\includegraphics[scale=0.35]{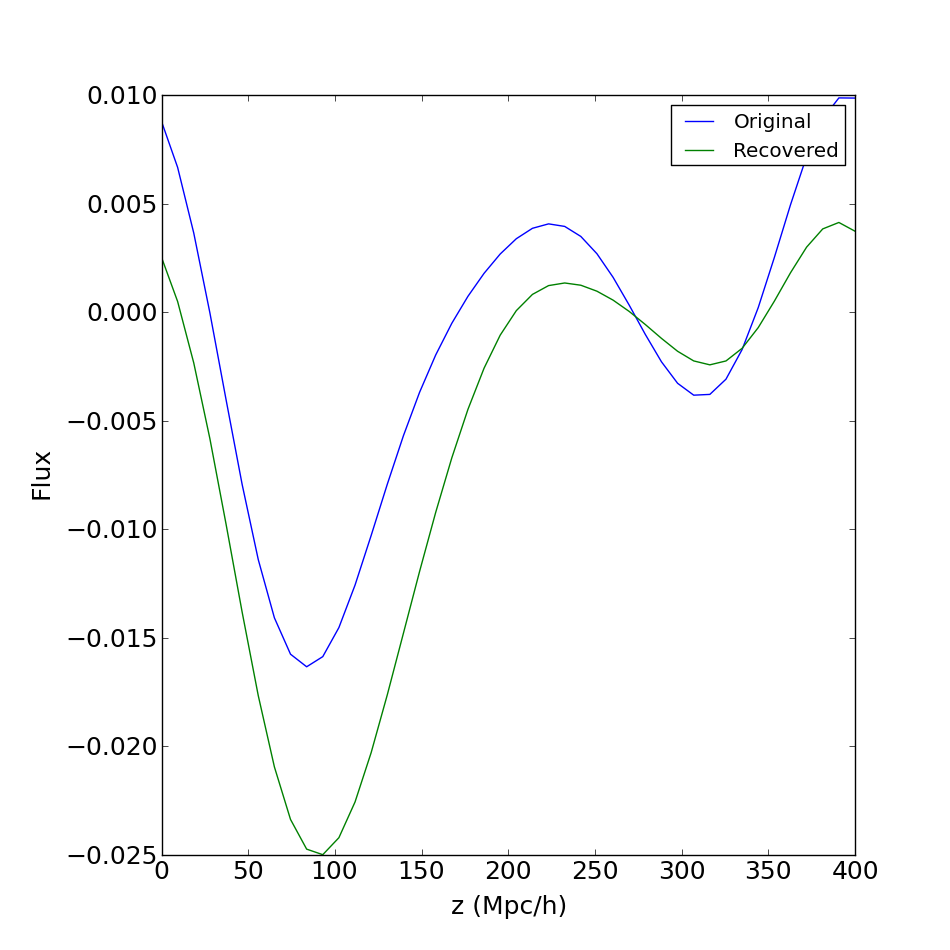}
\caption{Spectrum at (176.7, 260.5) $\hmpc$.}
\end{subfigure}
\begin{subfigure}{0.49\textwidth}
\includegraphics[scale=0.35]{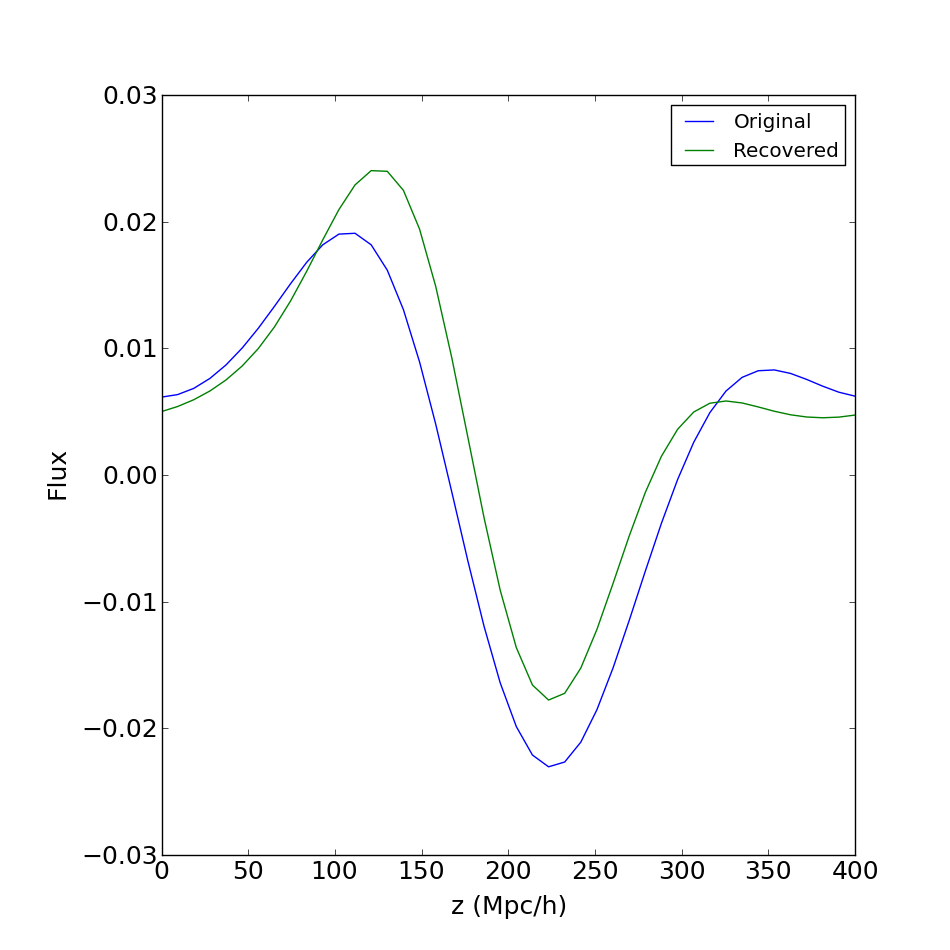}
\caption{Spectrum at (186.0, 130.2) $\hmpc$.}
\end{subfigure}
\begin{subfigure}{0.49\textwidth}
\includegraphics[scale=0.35]{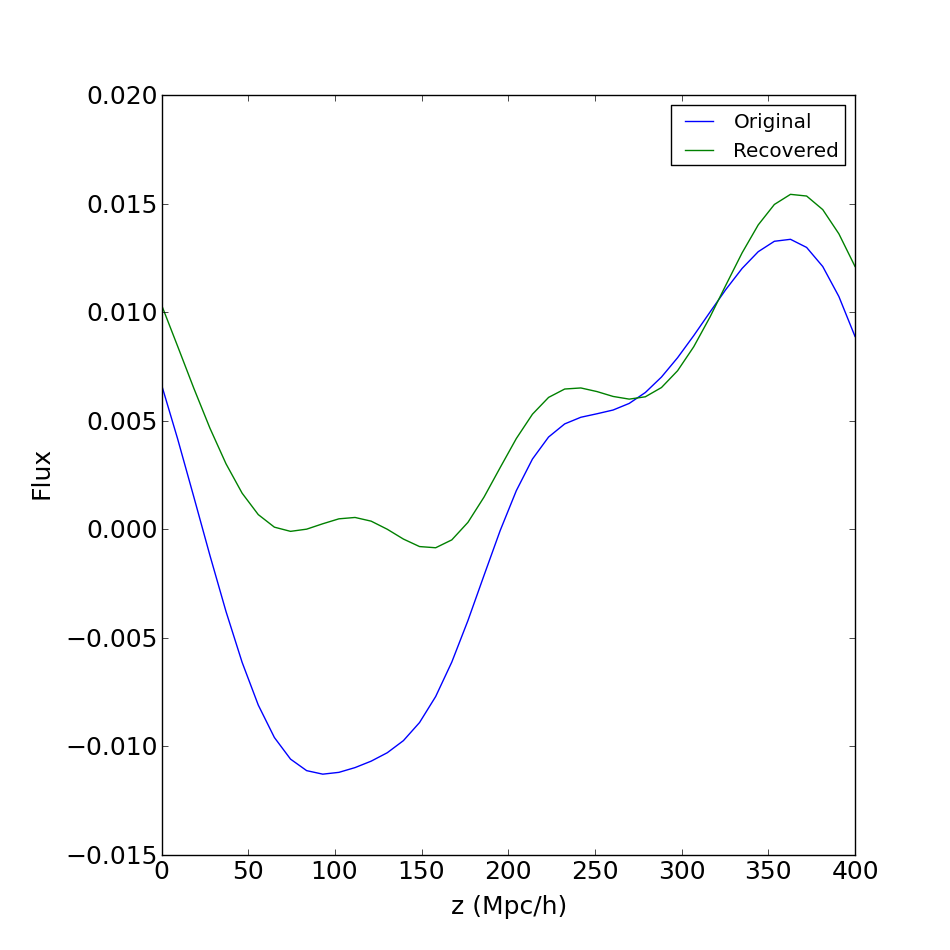}
\caption{Spectrum at (316.3, 353.5) $\hmpc$.}
\end{subfigure}
\vspace*{4mm}
\caption{Random LOS comparisons are shown above from a data sample at redshift $z=2$ with $N_{LOS} = 200$. Pixel values are compared along a single LOS at the coordinates given in the captions. General features of individual spectra are captured by the recovered field.}
\label{Fig:RandomLOS_z2_nonoise200LOS}
\end{figure*}

Furthermore, in Fig. \ref{Fig:RandomLOS_z2_nonoise200LOS}, we show one--dimensional visual comparisons along four lines of sight chosen at random using the data set $z2\textunderscore N200$, whose source density matches the areal LOS density of BOSS, and observe that the recovered skewers capture general features of the original ones. 

 The recovery is accurate for scales larger than $\sim$ $1.4 \dlos$, as found in \citep{caucci2008}, especially for 
mildly dense regions (standardized correlation plots). Due to the isotropic 
nature of the recovery and the smoothing, we do not notice any significant 
statistical difference between the directions parallel to and perpendicular 
to LOSs. As $N_{\rm LOS}$ increases, naturally, the recovery gets substantially 
better. This means that with future experiments like eBOSS and MS--DESI, 
which have higher areal density of LOSs, a very accurate large map of the 
IGM can be generated.

Adding noise to our data (pixel by pixel) and carrying out the recovery 
is an important step in order to better simulate real data from 
experiments. It is clearly seen from the figures that adding noise makes 
the recovery of overdense and underdense regions significantly worse. 
Furthermore, our results with the data set at $z=3$ are significantly worse 
than the other data set at $z=2$.

\FloatBarrier
\subsection{Using Observed Correlations in the Covariance Matrix}

\begin{figure*}
\centering
\begin{subfigure}{0.32\textwidth}
\includegraphics[scale=0.20, angle=-90]{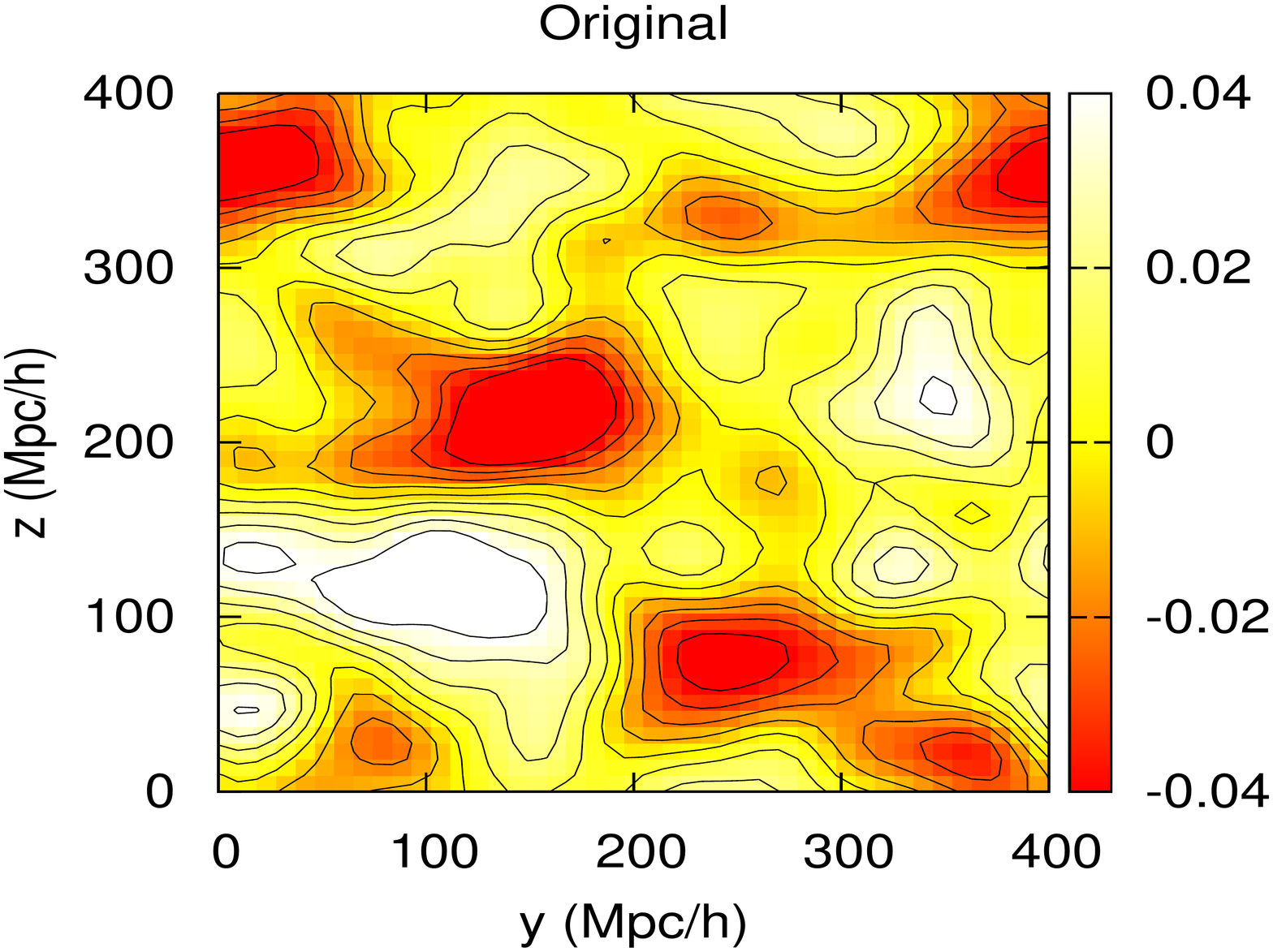}
\caption{True Field}
\end{subfigure}
\begin{subfigure}{0.32\textwidth}
\includegraphics[scale=0.20, angle=-90]{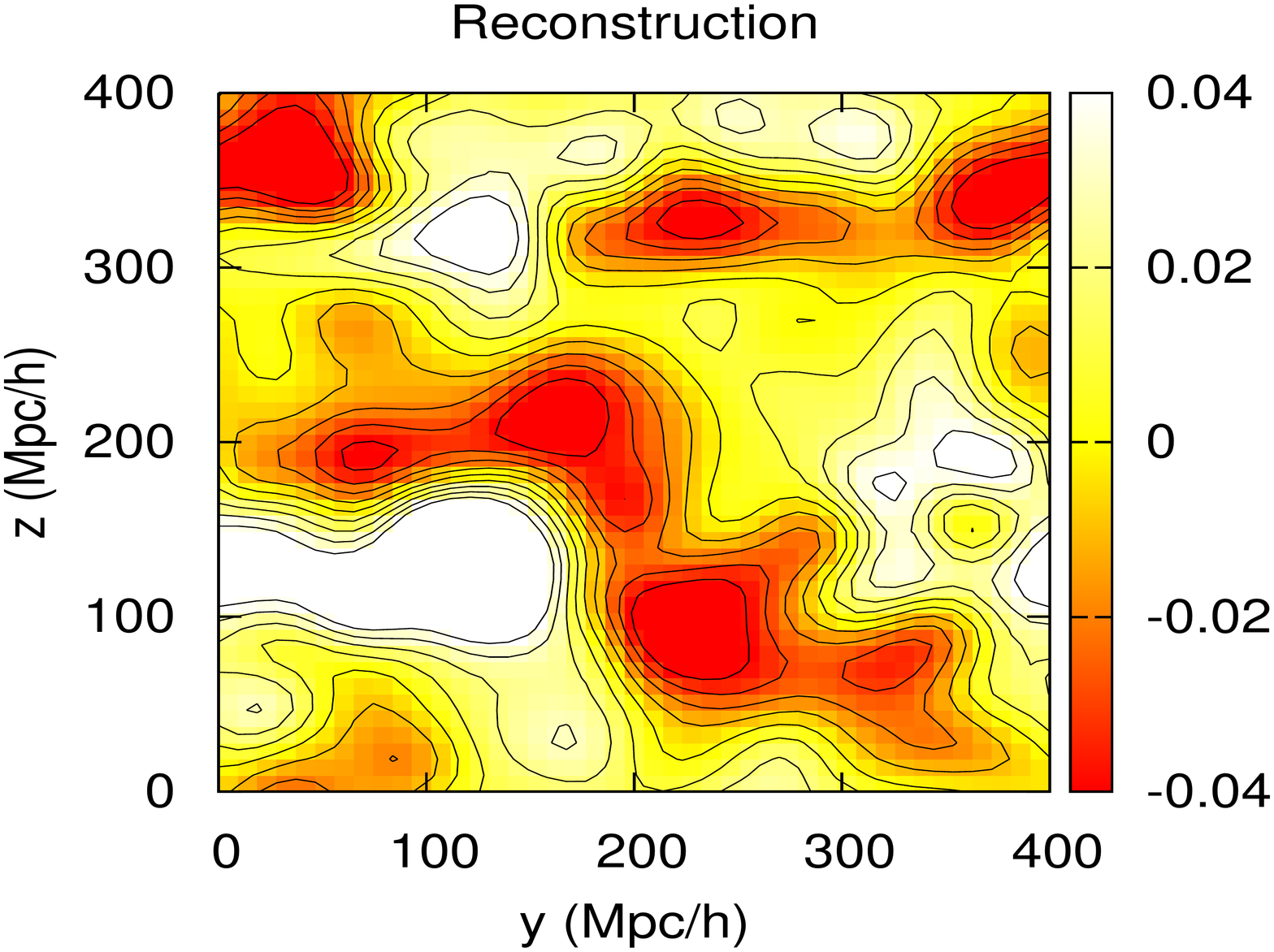}
\caption{Gaussian Correlation Function}
\end{subfigure}
\begin{subfigure}{0.32\textwidth}
\includegraphics[scale=0.20, angle=-90]{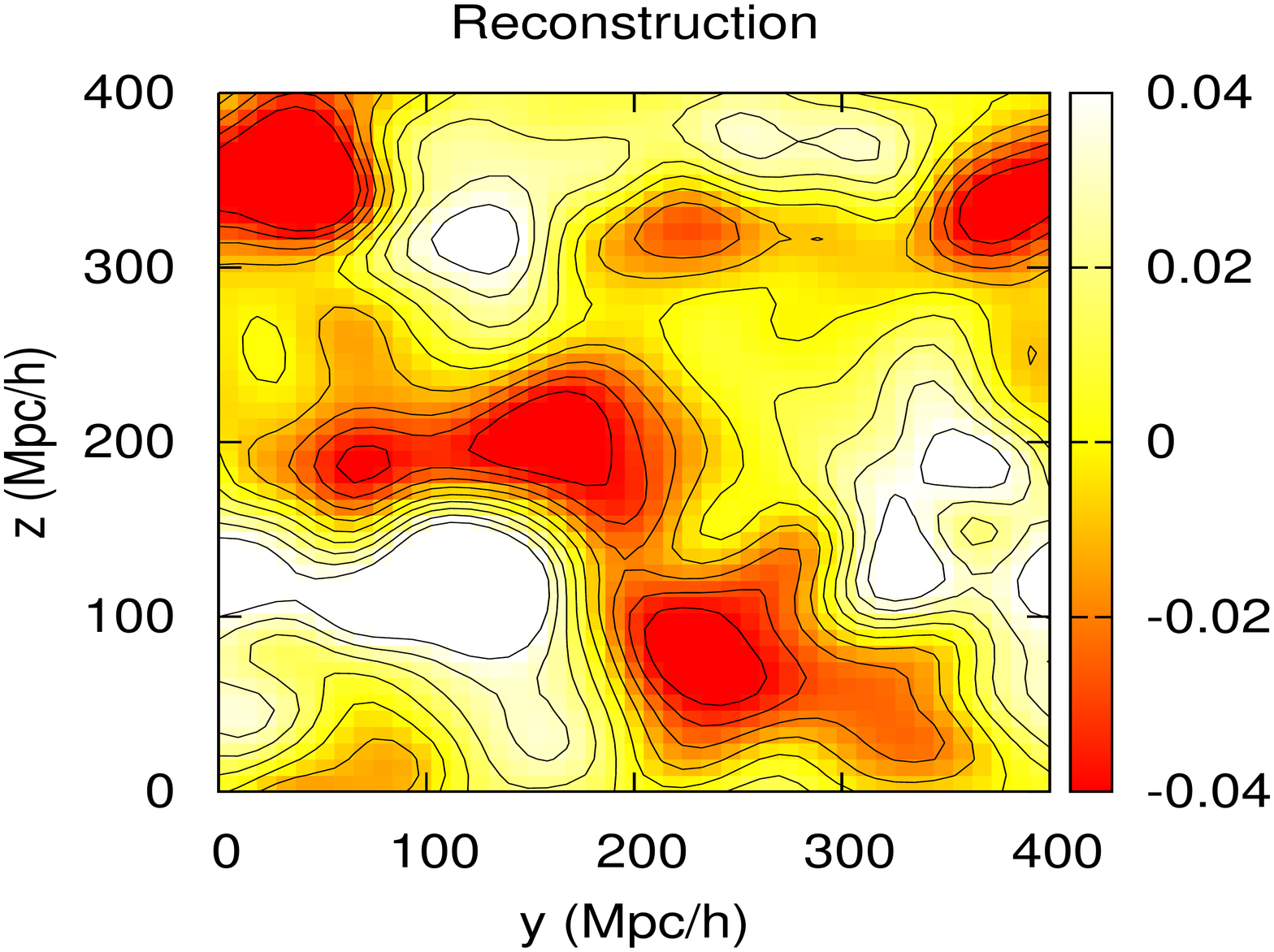}
\caption{CDM Correlation Function}
\end{subfigure}

\vspace*{4mm}

\begin{subfigure}{0.32\textwidth}
\includegraphics[scale=0.20, angle=-90]{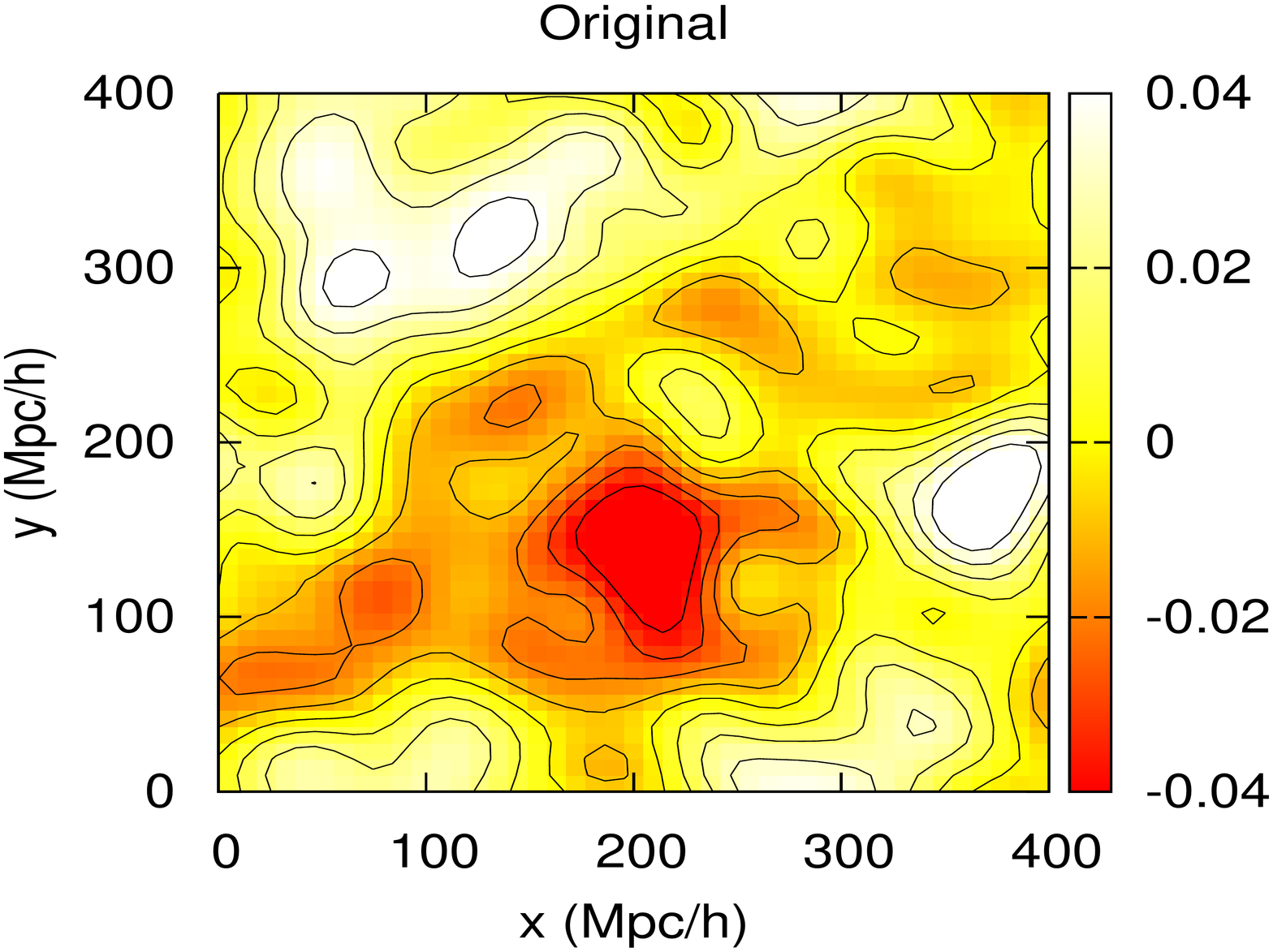}
\caption{True Field}
\end{subfigure}
\begin{subfigure}{0.32\textwidth}
\includegraphics[scale=0.20, angle=-90]{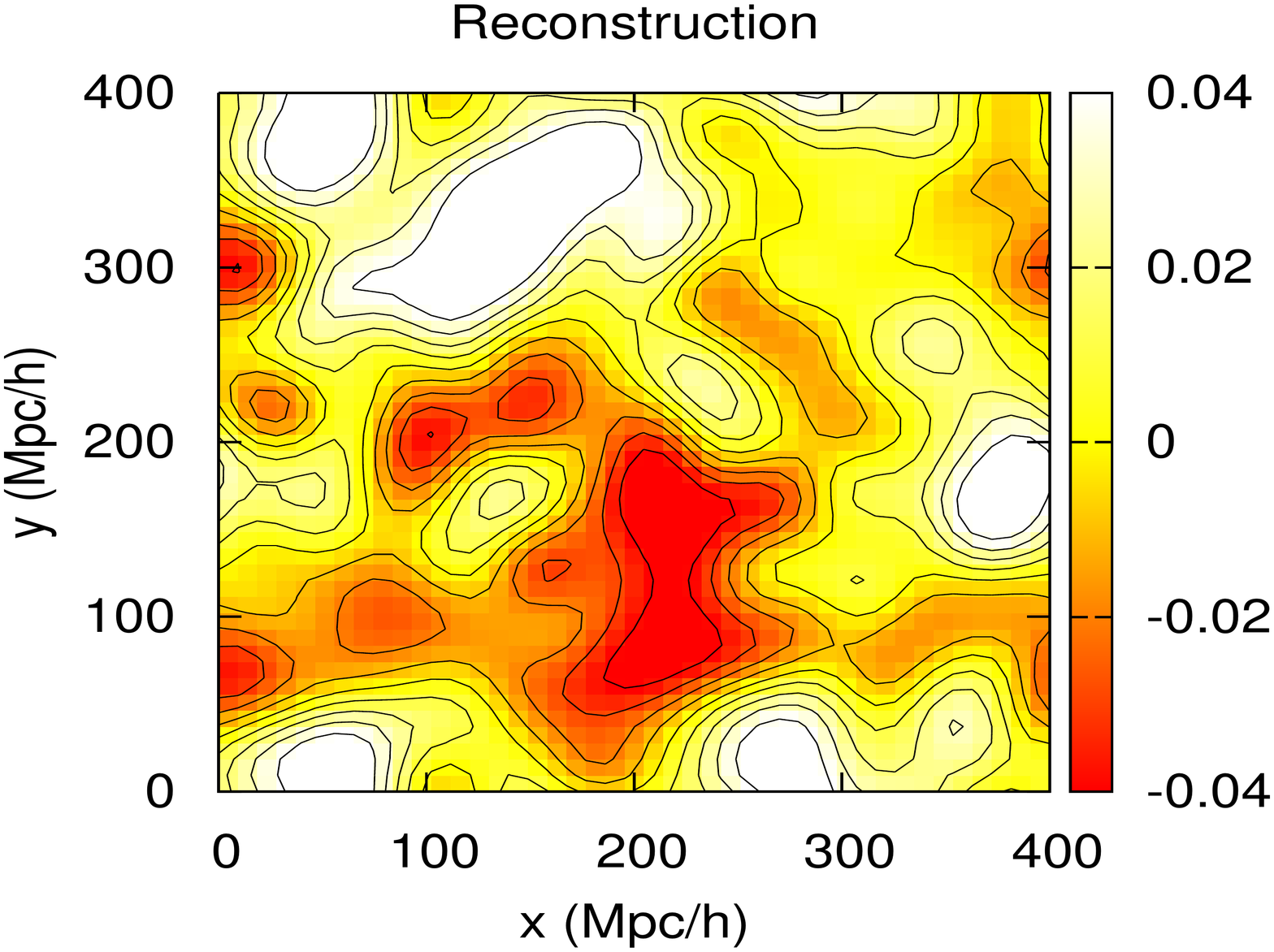}
\caption{Gaussian Correlation Function}
\end{subfigure}
\begin{subfigure}{0.32\textwidth}
\includegraphics[scale=0.20, angle=-90]{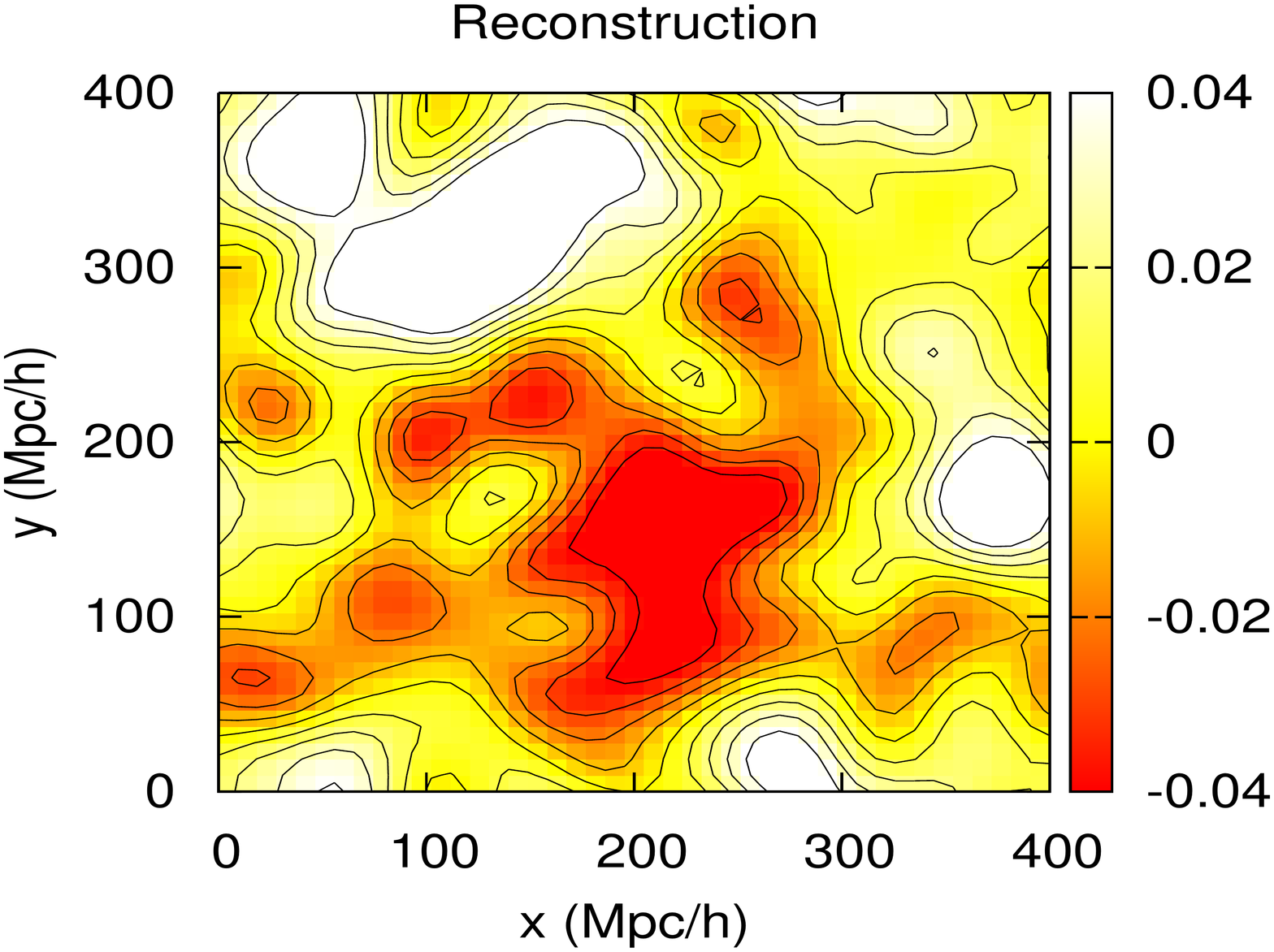}
\caption{CDM Correlation Function}
\end{subfigure}

\vspace*{4mm}
\caption{Slice plots from the middle of the cube are shown above from a data sample at redshift $z=2$ with $N_{LOS} = 1000$. We use the original Gaussian correlation function as well as a correlation function measured from observations to recover the field. The top row shows results for the middle of the cube, perpendicular to LOSs. The bottom row has plots in the middle of the cube, parallel to LOSs. The recovery of the field using the Gaussian correlation function yields better results than using the CDM correlation function obtained from observations.}
\label{Fig:Camb slice images}
\end{figure*}

Along with Caucci et al. 2008, and Lee et al., 2014, we use a simple
Gaussian form for the correlation function which appears in the 
Wiener interpolation covariance matrix (Equation \ref{Eq:correl_matrix}).
This is motivated by simplicity, and the fact that it is well behaved 
numerically at large separations. One might expect covariance matrices 
computed from the actual correlation functions of the field to give more
accurate reconstruction results, however, and we now test this.

In \citep{slosar2011}, the three-dimensional correlation function of the absorption in the Lyman-$\alpha$ forest was measured for the first time. The measurement was extended to greater than $100 \, \hmpc$ scales by \citep{delubac2013baryon} and \citep{slosar2013measurement}. We use this measurement  
of correlation function to construct a correlation matrix instead
of the Gaussian covariances we have used (Equation \ref{Eq:correl_matrix}).

The correlation function measured from the observational \lya forest
data is anisotropic because of redshift distortions. We construct the
correlation matrix not from the observational data results of 
\citep{slosar2011}, but from the linear theory CDM model 
consistent with the data. 
This redshift space model fit is given by Equations 4.5 -- 4.13
of \citep{slosar2011}. We use these equations, along with the linear 
theory correlation 
function from Section 2, and the following parameters:
bias factor $b=0.2$, and redshift
distortion factor $\beta=1.5$ to compute $\xi_{F}({r_{\perp},r_{\parallel})}$,
the flux correlation function for line of sight separation
$r_{\parallel}$ and transverse separation $r_{\perp}$. The Wiener covariance
(replacing Equation \ref{Eq:correl_matrix}) is then given by 

\begin{equation}
    \textbf{\textsf{C}}(x_1,x_2,{\bf x_{1\perp}},{\bf x_{2\perp}}) =
    \xi_{F}(r_{\perp},r_{\parallel}),
\end{equation}
where 
  $r_{\parallel}=(x_1-x_2)$ and $r_{\perp}=|{\bf x_{1\perp}}-{\bf x_{2\perp}}|$.

After reconstructing the simulation field using the CDM fit to the
\citep{slosar2011} results in the covariance matrix, we compare the
results to our fiducial reconstruction technique
(Fig. \ref{Fig:Camb slice images}).
We find that the recovery of the field with the fiducial
(Gaussian) correlation functions yields slightly better results than with 
the CDM correlation function. Although the dynamic range with the
Gaussian correlation seems to be slightly higher, general features of 
the original field are recovered better. For example, for the data set 
z2\textunderscore N1000, instead of our original result of the RMS percentage error 17.2, we find 20.3 with the CDM correlation function, 
which is significantly worse. It is worth noting that the actual correlation function in the simulations is probably not the same as it is estimated by \citet{slosar2011}.

\section{SUMMARY AND CONCLUSIONS}

Using Wiener interpolation, we reconstruct the entire simulation box with a subset of the Lyman--alpha absorption skewers chosen randomly. This subset of skewers, $N_{\rm LOS}$, sets a natural resolution of our maps. The number of the skewers chosen at random is decided by matching it with the areal LOS density of current and future spectroscopic surveys such as BOSS and MS--DESI. Using the Lya forest with this method, one can make maps of the large scale structure at high redshifts ($2 < z < 3.5$).

The standardized cross correlation plot (Fig. \ref{Fig:Correlation Plots}, panel(c)) indicates that the reconstruction is much better at $z = 2$ than at $z = 3$ using BOSS areal LOS densities. Naturally, the fidelity of the reconstruction is better as $N_{\rm LOS}$ is increased. Truncating the cube 50 $\hmpc$ from each edge, in order to remove the edge artefacts resulting from periodic boundary conditions of the simulation and to better mimic observational data, yields significantly better reconstruction (Fig. \ref{Fig:Correlation Plots}, panel(d)).

We find that the data set at $z = 2$ yields clearly better results than the one at $z = 3$ for the simulation, even with the same $N_{\rm LOS}$.
This is most easily understood in terms of the growth of structure through gravitational instability between $z = 3$ and $z = 2$. For observational surveys, in view of the fact that the areal LOS density is also much greater at $z = 2$ than at $z = 3$, one naturally expects that the large scale structure map will be significantly better at lower redshifts.

The overall bias seen in point to point flux values in real and reconstructed fields is an issue which does not have an easy explanation. Adjusting $N_{\rm LOS}$, the correlation lengths and the buffer length does not change the situation, but using an empirical bias correction allows the fields to be well-reconstructed.

In the high redshift range covered by the Lya forest, the IGM
density field is expected to be in the mildly non-linear regime,
therefore, we look for non-Gaussianity in the probability density
functions of our reconstructed maps. While its behaviour is nearly Gaussian for noiseless data samples, it becomes less Gaussian as the noise level is increased, as Table \ref{Table:Kolmogorov-Smirnov} indicates.

Since the smoothing levels used in this study are greater than
10 $\hmpc$, it is not possible to see the filamentary structure in the IGM topology. We remind the reader that there are no wide-field galaxy surveys that can detect the topology of the IGM at $z > 2$, as it is increasingly expensive to detect galaxies at higher redshifts to reach a high source density, even with 8--10 m telescopes \citep{le2013vimos}. However, searching for local peaks allows us
to discover the potential locations of the superclusters, which can be cross--correlated with galaxy surveys. Both the number of the local peaks and their locations are reproduced reasonably well by the interpolation.

Slice images allow a visual comparison between the original
and the reconstructed fields. General features of the flux field are well reproduced by the interpolation, especially for mildly overdense regions.

Using a correlation matrix derived from a CDM--fit to observational data instead of the simple Gaussian correlation matrix used in our fiducial Wiener filtering leads to a slightly worse recovery of the field.

As an improvement, the isotropic smoothing of both fields can be altered, as one does not necessarily expect the same statistics parallel and perpendicular to the LOSs. Furthermore, as future surveys like eBOSS and MS--DESI discover more quasars, the fidelity of the large scale structure maps will improve. For example, in order to reach resolutions in the sub 10 $\hmpc$ regime at $z\sim 2$ to study the IGM filamentary structure, QSO densities of over 100 $deg^{-2}$ will be necessary.

\footnotesize{
 \bibliographystyle{mn2e}
 \bibliography{largescale3dmapping_arxiv}

\begin{thebibliography}{}
 \providecommand{\href}[2]{#2}
  \providecommand{\doi}[1]{\href{http://dx.doi.org/#1}{doi:#1}}
  \providecommand{\eprint}[1]{\href{http://arxiv.org/abs/#1}{arXiv:#1}}

\bibitem[\protect\citeauthoryear{Adelberger, Steidel, Pettini, Shapley, Reddy
  \& Erb}{Adelberger et~al.}{2005}]{adelberger2005spatial}
Adelberger K.~L.,  Steidel C.~C.,  Pettini M.,  Shapley A.~E.,  Reddy N.~A.,
  Erb D.~K.,  2005, The Astrophysical Journal, 619, 697

\bibitem[\protect\citeauthoryear{Alam et~al.,}{Alam
  et~al.}{2015}]{alam2015eleventh}
Alam S.  et~al., 2015, arXiv preprint arXiv:1501.00963

\bibitem[\protect\citeauthoryear{Aubourg et~al.,}{Aubourg
  et~al.}{2014}]{aubourg2014cosmological}
Aubourg {\'E}.  et~al., 2014, arXiv preprint arXiv:1411.1074

\bibitem[\protect\citeauthoryear{Bardeen, Bond, Kaiser \& Szalay}{Bardeen
  et~al.}{1986}]{bardeen1986statistics}
Bardeen J.~M.,  Bond J.,  Kaiser N.,    Szalay A.,  1986, The Astrophysical
  Journal, 304, 15

\bibitem[\protect\citeauthoryear{Bi}{Bi}{1993}]{bi1993lyman}
Bi H.,  1993, The Astrophysical Journal, 405, 479

\bibitem[\protect\citeauthoryear{Busca, Delubac, Rich, Bailey, Font-Ribera,
  Kirkby et~al.,}{Busca et~al.}{2012}]{boss2616ng}
Busca N.,  Delubac T.,  Rich J.,  Bailey S.,  Font-Ribera A.,  Kirkby D.,
  et~al., 2012, arXiv preprint arXiv:1211.2616

\bibitem[\protect\citeauthoryear{Busca et~al.,}{Busca
  et~al.}{2013}]{delubac2013baryon}
Busca N.~G.  et~al., 2013, Astronomy \& Astrophysics, 552, A96

\bibitem[\protect\citeauthoryear{Caucci, Colombi, Pichon, Rollinde, Petitjean
  \& Sousbie}{Caucci et~al.}{2008}]{caucci2008}
Caucci S.,  Colombi S.,  Pichon C.,  Rollinde E.,  Petitjean P.,    Sousbie T.,
   2008, Monthly Notices of the Royal Astronomical Society, 386, 211

\bibitem[\protect\citeauthoryear{Cisewski, Croft, Freeman, Genovese, Khandai,
  Ozbek \& Wasserman}{Cisewski et~al.}{2014}]{cisewski2014}
Cisewski J.,  Croft R.~A.,  Freeman P.~E.,  Genovese C.~R.,  Khandai N.,  Ozbek
  M.,    Wasserman L.,  2014, arXiv preprint arXiv:1401.1867

\bibitem[\protect\citeauthoryear{Croft \& Gaztanaga}{Croft \&
  Gaztanaga}{1998}]{croft1998space}
Croft R.~A.,  Gaztanaga E.,  1998, The Astrophysical Journal, 495, 554

\bibitem[\protect\citeauthoryear{Croom, Smith, Boyle, Shanks, Miller, Outram \&
  Loaring}{Croom et~al.}{2004}]{croom20042df}
Croom S.~M.,  Smith R.,  Boyle B.,  Shanks T.,  Miller L.,  Outram P.,
  Loaring N.,  2004, Monthly Notices of the Royal Astronomical Society, 349,
  1397

\bibitem[\protect\citeauthoryear{Dawson et~al.,}{Dawson
  et~al.}{2013}]{dawson2013baryon}
Dawson K.~S.  et~al., 2013, The Astronomical Journal, 145, 10

\bibitem[\protect\citeauthoryear{De \& Croft}{De \& Croft}{2007}]{de2007peaks}
De S.,  Croft R.~A.,  2007, Monthly Notices of the Royal Astronomical Society,
  382, 1591

\bibitem[\protect\citeauthoryear{De \& Croft}{De \& Croft}{2010}]{de2010peaks}
De S.,  Croft R.~A.,  2010, Monthly Notices of the Royal Astronomical Society,
  401, 1989

\bibitem[\protect\citeauthoryear{Delubac et~al.,}{Delubac
  et~al.}{2014}]{delubac2014baryon}
Delubac T.  et~al., 2014, arXiv preprint arXiv:1404.1801

\bibitem[\protect\citeauthoryear{Dijkstra, Lidz \&
  Hui}{\protect\mniiiauthor{dijkstra2003}{Dijkstra, Lidz \& Hui}{Dijkstra
  et~al.}}{2003}]{dijkstra2003}
Dijkstra M.,  Lidz A.,    Hui L.,  2003, arXiv preprint astro-ph/0305498

\bibitem[\protect\citeauthoryear{Dinshaw, Impey, Foltz, Weymann \&
  Chaffee}{Dinshaw et~al.}{1994}]{dinshaw1994common}
Dinshaw N.,  Impey C.~D.,  Foltz C.~B.,  Weymann R.~J.,    Chaffee F.~H.,
  1994, The Astrophysical Journal, 437, L87

\bibitem[\protect\citeauthoryear{Dinshaw, Foltz, Impey, Weymann \&
  Morris}{Dinshaw et~al.}{1995}]{dinshaw1995large}
Dinshaw N.,  Foltz C.~B.,  Impey C.~D.,  Weymann R.~J.,    Morris S.~L.,  1995,
  Nature, 373, 223

\bibitem[\protect\citeauthoryear{Di~Matteo, Khandai, DeGraf, Feng, Croft, Lopez
  \& Springel}{Di~Matteo et~al.}{2012}]{di2012cold}
Di~Matteo T.,  Khandai N.,  DeGraf C.,  Feng Y.,  Croft R.,  Lopez J.,
  Springel V.,  2012, The Astrophysical Journal Letters, 745, L29

\bibitem[\protect\citeauthoryear{Fang, Duncan, Crotts \& Bechtold}{Fang
  et~al.}{1995}]{fang1995size}
Fang Y.,  Duncan R.~C.,  Crotts A.~P.,    Bechtold J.,  1995, arXiv preprint
  astro-ph/9510112

\bibitem[\protect\citeauthoryear{Haardt \& Madau}{Haardt \&
  Madau}{1995}]{haardt1995radiative}
Haardt F.,  Madau P.,  1995, arXiv preprint astro-ph/9509093

\bibitem[\protect\citeauthoryear{Hennawi \& Prochaska}{Hennawi \&
  Prochaska}{2007}]{hennawi2007}
Hennawi J.~F.,  Prochaska J.~X.,  2007, The Astrophysical Journal, 655, 735

\bibitem[\protect\citeauthoryear{Hernquist, Katz, Weinberg \&
  Miralda-Escude}{Hernquist et~al.}{1996}]{hernquist1996lyman}
Hernquist L.,  Katz N.,  Weinberg D.~H.,    Miralda-Escude J.,  1996, The
  Astrophysical Journal Letters, 457, L51

\bibitem[\protect\citeauthoryear{Hui \& Gnedin}{Hui \&
  Gnedin}{1997}]{hui1997equation}
Hui L.,  Gnedin N.~Y.,  1997, Monthly Notices of the Royal Astronomical
  Society, 292, 27

\bibitem[\protect\citeauthoryear{Ikeuchi}{Ikeuchi}{1986}]{ikeuchi1986baryon}
Ikeuchi S.,  1986, Astrophysics and space science, 118, 509

\bibitem[\protect\citeauthoryear{Ir{\v{s}}i{\v{c}} et~al.,}{Ir{\v{s}}i{\v{c}}
  et~al.}{2013}]{irvsivc2013detection}
Ir{\v{s}}i{\v{c}} V.  et~al., 2013, Journal of Cosmology and Astroparticle
  Physics, 2013, 016

\bibitem[\protect\citeauthoryear{Kim, Viel, Haehnelt, Carswell \&
  Cristiani}{Kim et~al.}{2004}]{kim2004power}
Kim T.-S.,  Viel M.,  Haehnelt M.,  Carswell R.,    Cristiani S.,  2004,
  Monthly Notices of the Royal Astronomical Society, 347, 355

\bibitem[\protect\citeauthoryear{Lee et~al.,}{Lee et~al.}{2013}]{lee2013boss}
Lee K.-G.  et~al., 2013, The Astronomical Journal, 145, 69

\bibitem[\protect\citeauthoryear{Lee, Hennawi, Stark, Prochaska \& White}{Lee
  et~al.}{2014a}]{lee2014a}
Lee K.-G.,  Hennawi J.~F.,  Stark C.,  Prochaska J.~X.,    White M.,  2014a,
  The Astrophysical Journal, 788, 8

\bibitem[\protect\citeauthoryear{Lee, Hennawi, White, Croft \& Ozbek}{Lee
  et~al.}{2014b}]{lee2014b}
Lee K.-G.,  Hennawi J.~F.,  White M.,  Croft R.~A.,    Ozbek M.,  2014b, The
  Astrophysical Journal, 788, 49

\bibitem[\protect\citeauthoryear{Lee et~al.,}{Lee et~al.}{2015}]{lee2015igm}
Lee K.-G.  et~al., 2015, The Astrophysical Journal, 799, 196

\bibitem[\protect\citeauthoryear{Levi et~al.,}{Levi
  et~al.}{2013}]{levi2013desi}
Levi M.  et~al., 2013, arXiv preprint arXiv:1308.0847

\bibitem[\protect\citeauthoryear{Lewis, Challinor \&
  Lasenby}{\protect\mniiiauthor{lewis2000efficient}{Lewis, Challinor \&
  Lasenby}{Lewis et~al.}}{2000}]{lewis2000efficient}
Lewis A.,  Challinor A.,    Lasenby A.,  2000, The Astrophysical Journal, 538,
  473

\bibitem[\protect\citeauthoryear{Le~F{\`e}vre et~al.,}{Le~F{\`e}vre
  et~al.}{2013}]{le2013vimos}
Le~F{\`e}vre O.  et~al., 2013, Astronomy \& Astrophysics, 559, A14

\bibitem[\protect\citeauthoryear{Lynds}{Lynds}{1971}]{lynds1971}
Lynds R.,  1971, ApJ, L73

\bibitem[\protect\citeauthoryear{Miller, Lopes, Smith, Croom, Boyle, Shanks \&
  Outram}{Miller et~al.}{2002}]{miller2002possible}
Miller L.,  Lopes A.,  Smith R.,  Croom S.,  Boyle B.,  Shanks T.,    Outram
  P.,  2002, arXiv preprint astro-ph/0210644

\bibitem[\protect\citeauthoryear{Miller, Croom, Boyle, Loaring, Smith, Shanks
  \& Outram}{Miller et~al.}{2004}]{miller2004200}
Miller L.,  Croom S.,  Boyle B.,  Loaring N.,  Smith R.,  Shanks T.,    Outram
  P.,  2004, Monthly Notices of the Royal Astronomical Society, 355, 385

\bibitem[\protect\citeauthoryear{Nuza, S{\'a}nchez, Prada, Klypin, Schlegel,
  Gottl\"ober \& Montero-Dorta}{Nuza et~al.}{2013}]{BOSS_CMASS_DR9}
Nuza S.~E.,  S{\'a}nchez A.~G.,  Prada F.,  Klypin A.,  Schlegel D.~J.,
  Gottl\"ober S.,    Montero-Dorta A.~D.,  2013, Monthly Notices of the Royal
  Astronomical Society, 432, 743, \eprint{1202.6057},
  \doi{10.1093/mnras/stt513}

\bibitem[\protect\citeauthoryear{Outram, Hoyle, Shanks, Croom, Boyle, Miller,
  Smith \& Myers}{Outram et~al.}{2003}]{outram20032df}
Outram P.,  Hoyle F.,  Shanks T.,  Croom S.,  Boyle B.,  Miller L.,  Smith R.,
    Myers A.,  2003, Monthly Notices of the Royal Astronomical Society, 342,
  483

\bibitem[\protect\citeauthoryear{Ozbek, Croft, SDSS, SDSS-III \& BOSS}{Ozbek
  et~al.}{2015}]{ozbek2015}
Ozbek M.,  Croft A.~R.,  SDSS SDSS-III c.,    BOSS c.,  2015, in prep

\bibitem[\protect\citeauthoryear{Petry, Impey \&
  Foltz}{\protect\mniiiauthor{petry1998small}{Petry, Impey \& Foltz}{Petry
  et~al.}}{1998}]{petry1998small}
Petry C.,  Impey C.,    Foltz C.,  1998, The Astrophysical Journal, 494, 60

\bibitem[\protect\citeauthoryear{Pichon, Vergely, Rollinde, Colombi \&
  Petitjean}{Pichon et~al.}{2001}]{pichon2001}
Pichon C.,  Vergely J.,  Rollinde E.,  Colombi S.,    Petitjean P.,  2001,
  Arxiv preprint astro-ph/0105196

\bibitem[\protect\citeauthoryear{Raichoor et~al.,}{Raichoor
  et~al.}{2015}]{raichoor2015sdss}
Raichoor A.  et~al., 2015, arXiv preprint arXiv:1505.01797

\bibitem[\protect\citeauthoryear{Rees}{Rees}{1986}]{rees1986lyman}
Rees M.~J.,  1986, Monthly Notices of the Royal Astronomical Society, 218, 25P

\bibitem[\protect\citeauthoryear{Richards et~al.,}{Richards
  et~al.}{2006}]{richards2006sloan}
Richards G.~T.  et~al., 2006, The Astronomical Journal, 131, 2766

\bibitem[\protect\citeauthoryear{Rollinde, Petitjean \&
  Pichon}{\protect\mniiiauthor{rollinde2001}{Rollinde, Petitjean \&
  Pichon}{Rollinde et~al.}}{2001}]{rollinde2001}
Rollinde E.,  Petitjean P.,    Pichon C.,  2001, arXiv preprint
  astro-ph/0106198

\bibitem[\protect\citeauthoryear{Rollinde, Petitjean, Pichon, Colombi, Aracil,
  D'Odorico \& Haehnelt}{Rollinde et~al.}{2003}]{rollinde2003correlation}
Rollinde E.,  Petitjean P.,  Pichon C.,  Colombi S.,  Aracil B.,  D'Odorico V.,
     Haehnelt M.,  2003, Monthly Notices of the Royal Astronomical Society,
  341, 1279

\bibitem[\protect\citeauthoryear{Sargent, Young, Boksenberg \& Tytler}{Sargent
  et~al.}{1980}]{sargent1980distribution}
Sargent W.~L.,  Young P.~J.,  Boksenberg A.,    Tytler D.,  1980, The
  Astrophysical Journal Supplement Series, 42, 41

\bibitem[\protect\citeauthoryear{Schneider et~al.,}{Schneider
  et~al.}{2002}]{schneider2002sloan}
Schneider D.~P.  et~al., 2002, The Astronomical Journal, 123, 567

\bibitem[\protect\citeauthoryear{Schneider et~al.,}{Schneider
  et~al.}{2003}]{schneider2003sloan}
Schneider D.~P.  et~al., 2003, The Astronomical Journal, 126, 2579

\bibitem[\protect\citeauthoryear{Schneider et~al.,}{Schneider
  et~al.}{2005}]{schneider2005sloan}
Schneider D.~P.  et~al., 2005, The Astronomical Journal, 130, 367

\bibitem[\protect\citeauthoryear{Shull et~al.,}{Shull et~al.}{2000}]{shull2000}
Shull J.~M.  et~al., 2000, arXiv preprint astro-ph/0005011

\bibitem[\protect\citeauthoryear{Slosar et~al.,}{Slosar
  et~al.}{2011}]{slosar2011}
Slosar A.  et~al., 2011, Journal of Cosmology and Astroparticle Physics, 2011,
  001

\bibitem[\protect\citeauthoryear{Slosar et~al.,}{Slosar
  et~al.}{2013}]{slosar2013measurement}
Slosar A.  et~al., 2013, Journal of Cosmology and Astroparticle Physics, 2013,
  026

\bibitem[\protect\citeauthoryear{Springel}{Springel}{2005}]{springel2005cosmological}
Springel V.,  2005, Monthly Notices of the Royal Astronomical Society, 364,
  1105

\bibitem[\protect\citeauthoryear{Springel \& Hernquist}{Springel \&
  Hernquist}{2002}]{springel2002cosmological}
Springel V.,  Hernquist L.,  2002, arXiv preprint astro-ph/0206393

\bibitem[\protect\citeauthoryear{Steidel \& Hamilton}{Steidel \&
  Hamilton}{1992}]{steidel1992deep}
Steidel C.~C.,  Hamilton D.,  1992, The Astronomical Journal, 104, 941

\bibitem[\protect\citeauthoryear{Steidel, Adelberger, Dickinson, Giavalisco \&
  Pettini}{Steidel et~al.}{1998}]{steidel1998lyman}
Steidel C.,  Adelberger K.,  Dickinson M.,  Giavalisco M.,    Pettini M.,
  1998, arXiv preprint astro-ph/9812167

\bibitem[\protect\citeauthoryear{Steidel, Pettini \&
  Adelberger}{\protect\mniiiauthor{steidel2001lyman}{Steidel, Pettini \&
  Adelberger}{Steidel et~al.}}{2001}]{steidel2001lyman}
Steidel C.~C.,  Pettini M.,    Adelberger K.~L.,  2001, The Astrophysical
  Journal, 546, 665

\bibitem[\protect\citeauthoryear{Totani et~al.,}{Totani
  et~al.}{2013}]{totani2013}
Totani T.  et~al., 2013, PASJ

\bibitem[\protect\citeauthoryear{Vikas et~al.,}{Vikas
  et~al.}{2013}]{vikas2013moderate}
Vikas S.  et~al., 2013, The Astrophysical Journal, 768, 38

\bibitem[\protect\citeauthoryear{Weinberg, Croft, Hernquist, Katz \&
  Pettini}{Weinberg et~al.}{1999}]{weinberg1999closing}
Weinberg D.~H.,  Croft R.~A.,  Hernquist L.,  Katz N.,    Pettini M.,  1999,
  The Astrophysical Journal, 522, 563

\bibitem[\protect\citeauthoryear{Zhu \& M{\'e}nard}{Zhu \&
  M{\'e}nard}{2013}]{zhu2013jhu}
Zhu G.,  M{\'e}nard B.,  2013, The Astrophysical Journal, 770, 130

\bibitem[\protect\citeauthoryear{{Levi} et~al.,}{{Levi}
  et~al.}{2013}]{DESI2013}
{Levi} M.  et~al., 2013, ArXiv e-prints, \eprint{1308.0847}

\bibitem[\protect\citeauthoryear{{Stark}, {White}, {Lee} \& {Hennawi}}{{Stark}
  et~al.}{2014}]{stark2014}
{Stark} C.~W.,  {White} M.,  {Lee} K.-G.,    {Hennawi} J.~F.,  2014, ArXiv
  e-prints, \eprint{1412.1507}

\bibitem[\protect\citeauthoryear{{Stark}, {Font-Ribera}, {White} \&
  {Lee}}{{Stark} et~al.}{2015}]{stark2015}
{Stark} C.~W.,  {Font-Ribera} A.,  {White} M.,    {Lee} K.-G.,  2015, ArXiv
  e-prints, \eprint{1504.03290}

\end{thebibliography}
}

\end{document}